\documentclass[amsmath,amssymb,prr,reprint,superscriptaddress]{revtex4-2} 

\usepackage {graphicx}
\usepackage{dcolumn}
\usepackage{bm}
\usepackage[whole]{bxcjkjatype}
\usepackage{verbatim}
\usepackage{color}
\usepackage{physics}
\usepackage{caption}
\usepackage[position=top,singlelinecheck=false]{subcaption}

\newcommand{\ii}{\ensuremath{\mathrm{i}}}

\newcommand{\NP}{\ensuremath{N_\mathrm{P}}}

\begin{document}

\title{
Comparative study on compact quantum circuits of hybrid quantum--classical algorithms for quantum impurity models
}

\author{Rihito Sakurai}
\email{sakurairihito@gmail.com}
\affiliation{Department of Physics, Saitama University, Saitama 338-8570, Japan}

\author{Oliver J. Backhouse}
\affiliation{Department of Physics, King's College London, Strand, London WC2R 2LS, U.K.}

\author{George H. Booth}
\affiliation{Department of Physics, King's College London, Strand, London WC2R 2LS, U.K.}

\author{Wataru Mizukami}
\affiliation{Center for Quantum Information and Quantum Biology, Osaka University, Osaka 565-0871, Japan}

\author{Hiroshi Shinaoka}
\affiliation{Department of Physics, Saitama University, Saitama 338-8570, Japan}

\date{\today}

\begin{abstract}
Predicting the properties of strongly correlated materials is a significant challenge in condensed matter theory. 
The widely used dynamical mean-field theory faces difficulty in solving quantum impurity models numerically. 
Hybrid quantum--classical algorithms such as variational quantum eigensolver emerge as a potential solution for quantum impurity models.
A common challenge in these algorithms is the rapid growth of the number of variational parameters with the number of spin-orbitals in the impurity. 
In our approach to this problem, we develop compact ansatzes using a combination of two different strategies.  
First, we employ compact physics-inspired ansatz, $k$-unitary cluster Jastrow ansatz, developed in the field of quantum chemistry. 
Second, we eliminate largely redundant variational parameters of physics-inspired ansatzes associated with bath sites based on physical intuition.  
This is based on the fact that a quantum impurity model with a star-like geometry has no direct hopping between bath sites.
We benchmark the accuracy of these ansatzes for both ground-state energy and dynamic quantities by solving typical quantum impurity models with/without shot noise.  
The results suggest that we can maintain the accuracy of ground-state energy while we drop the number of variational parameters associated with bath sites. 
Furthermore, we demonstrate that a moment expansion, when combined with the proposed ansatzes, can calculate the imaginary-time Green's functions under the influence of shot noise.
This study demonstrates the potential for addressing complex impurity models in large-scale quantum simulations with fewer variational parameters without sacrificing accuracy. 
\end{abstract}

\maketitle

\section{Introduction}
Accurately predicting the properties of strongly correlated materials poses a significant challenge in condensed matter theory, including long-standing challenges in the field, such as the mechanism of high-temperature superconductivity~\cite{2006NatPh...2..134L, RePEc:nat:nature:v:518:y:2015:i:7538:d:10.1038_nature14165}.
Simulating these strongly correlated materials is difficult due to quantum superposition, which exponentially increases the accessible Hilbert space with the number of particles. 
Even with quantum computers with more than a hundred logical qubits, simulating solids with large numbers of degrees of freedom is still challenging.
Quantum embedding theories, such as dynamical mean-field theory (DMFT)\cite{georges1996dynamical,kotliar2006electronic} or density matrix embedding theory (DMET)~\cite{PhysRevLett.109.186404, PhysRevB.104.245114}, aims to address this issue by limiting the correlated degrees of freedom in solid materials based on local approximation.

In DMFT, widely used in condensed matter physics, the original lattice system is divided into impurities with local interactions and a dynamical environment called a bath. 
This model is called a quantum impurity model. 
A self-consistent calculation is performed to update the parameters associated with the bath until the local Green's function defined on impurity matches that of the original lattice system with the dynamical mean-field. 
DMFT allows us to compute the single-particle excitation spectrum and successfully describes transitions from metallic to Mott insulating behavior.
The biggest numerical bottleneck in DMFT calculations is solving the correlated quantum impurity models, specifically computing local Green's functions for these interacting problems.
While state-of-the-art classical algorithms have been adapted for use as impurity solvers, such as tensor networks~\cite{PhysRevB.101.041101, wolf2015imaginary, bauernfeind2017fork} or quantum Monte Carlo methods~\cite{RevModPhys.83.349}, their applications are limited to models with only a few impurity and/or bath orbitals~\cite{PhysRevB.101.041101, wolf2015imaginary, bauernfeind2017fork}.
This challenge stems from the exponential increase in quantum entanglement entropy and the notorious negative sign problem.

To exploit the growing potential for solving quantum impurity models on quantum devices, quantum algorithms based on quantum phase estimation~\cite{kitaev1995quantum, doi:10.1126/science.1113479} and adiabatic algorithms~\cite{farhi2000quantum, RevModPhys.90.015002} have been proposed~\cite{bauer2016hybrid}.
Their practical implementation, however, may take decades because it requires large-scale error correction schemes.
This led to a growing interest in variational quantum algorithms~\cite{peruzzo2014variational, mcardle2019variational} for near-term quantum computers with limited hardware resources, often dubbed `noisy intermediate-scale quantum' (NISQ) devices~\cite{preskill2018quantum}. 
A number of proof-of-principle demonstrations of solving quantum impurity models using NISQ devices have been conducted~\cite{PhysRevB.105.115108, PhysRevB.107.165155, rungger2020dynamical, greenediniz2023quantum, dhawan2023quantum}.

In near-term quantum algorithms, such as those in the NISQ era, it is crucial to utilize limited hardware resources effectively. 
Therefore, there is a need to discretize a continuous bath with fewer bath sites. 
This reduction can be achieved through the use of imaginary time formalism in DMFT~\cite{Bravyi:2017cc,Nusspickel2020,sparsebathfitting2021, PhysRevB.92.155126, PhysRevB.101.035143}.
For example, a recent estimate for 20-orbital impurity models for iron-based superconductors indicates that about 300 bath sites are sufficient for accurate discretization in the imaginary-time formalism~\cite{sparsebathfitting2021}.

Once this finite Hamiltonian representation of the quantum impurity model has been found, it is now in principle amenable to solution on a quantum device. For variational quantum algorithms, the first challenge is to define an appropriate ansatz which is flexible enough to span the solution to the problem, able to be efficiently evaluated via unitary quantum gates, and where the number of variational parameters, $\NP$, does not grow prohibitively as the number of spin-orbitals $N_{\mathrm{SO}}$ increases.
Physics-inspired ansatzes based on unitary coupled cluster (UCC) methods~\cite{BARTLETT1989133, https://doi.org/10.1002/qua.21198, doi:10.1021/acs.jpca.3c02781} are widely used in previous studies for quantum impurity models~\cite{PhysRevResearch.3.013184, Mukherjee_2023, dhawan2023quantum}.
Among the family of UCC methods, for the unitary coupled cluster with generalized singles and doubles (UCCGSD)~\cite{lee2018generalized},  $\NP$ grows as $O(N_{\mathrm{SO}}^4)$. The computational times for computing imaginary-time Green's function grow even more rapidly, e.g., as $O(N_{\mathrm{depth}} \NP^{2})$~\cite{PhysRevResearch.4.023219} using the UCCGSD~\cite{lee2018generalized} and the variational quantum simulation (VQS)~\cite{mcardle2019variational, Yuan2019theoryofvariational}, where the depth of the circuit $N_{\mathrm{depth}} \propto N_{\mathrm{P}}$.
Thus, more compact ansatzes (circuits) are an important research direction for the success of simulating impurity models on quantum devices.

In this study, we develop compact ansatzes using a combination of two different strategies.
First, we employ the $k$-unitary coupled Jastrow ($k$-uCJ) ansatz originally proposed for quantum chemistry, where $\NP$ scales only as $O(N_{\mathrm{SO}}^2)$~\cite{matsuzawa2020jastrow}.
Second, we drop largely redundant variational parameters in both the UCCGSD and the $k$-uCJ ansatz based on physical intuition.
This exploits structures in the Hamiltonian which are specific to quantum impurity models with a star-like bath geometry where the bath sites are connected via the Hamiltonian only through the impurity (see Fig.~\ref{fig:sparse_ansatzes}).
In particular, we eliminate part of the two-particle excitations associated with direct excitations between bath sites, which does not change the scaling of $\NP$ but reduces the coefficient for a large number of bath sites.
The scalings of the proposed ansatzes are summarized in Table~\ref{tb:parameters}.
We numerically demonstrate that the compact ansatzes describe ground-state energies and dynamic quantities, especially imaginary-time Green's functions, without compromising accuracy for typical quantum impurity models with/without shot noise, validating their potential in quantum impurity models.

The following outlines the contents of each section.
Section~\ref{sec:green functions and VQA} provides an overview of Green's functions and variational quantum algorithms for computing ground-state energy and dynamic quantities.
This section also introduces the physics-inspired ansatzes used in this study.
Section~\ref{sec:sparse ansatzes} introduces compact quantum circuits for quantum impurity models and compares the scaling of their variational parameters to those of the original ansatzes.
Section~\ref{sec:state_vector_simulation} compares the accuracies of ground-state energy and dynamic quantities such as spectral functions and imaginary-time Green's functions among ansatzes for typical quantum impurity models.
Section~\ref{sec:shotnoise} explores the effect of finite shot noise within the single-site impurity model.
Section~\ref{sec:discussion} reviews our results, compares them to existing methods, and highlights areas for future research. 

\section{Review of Green's functions and variational quantum algorithms} \label{sec:green functions and VQA}

\subsection{Green's function}
We study a fermionic system in the grand-canonical ensemble, represented by the Hamiltonian $\mathcal{H}$, with

\begin{align}
   \label{eq:hamiltonian}
   \mathcal{ H} &= \sum_{ij}^N t_{ij} \hat{c}^\dagger_i \hat{c}_j + \frac{1}{2} \sum_{ijkl} U_{ijkl}\hat{c}_i^\dagger \hat{c}_k^\dagger \hat{c}_l \hat{c}_j - \mu \sum_i \hat{c}^\dagger_i \hat{c}_i,
\end{align}
where $c_i$/$c^\dagger_i$ are annihilation/creation operators for spin-orbital $i$, and $N$ represents the total number of spin-orbitals.
The hopping matrix, Coulomb interaction tensor, and chemical potential are denoted by $t_{ij}$, $U_{ijkl}$, and $\mu$, respectively. 
The retarded (fermionic) Green's function is defined as 
\begin{align}
	G^\mathrm{R}_{ab}(t) &= -\ii \theta(t)\expval{\hat c_a(t)\hat c^\dagger_b(0) + \hat c^\dagger_b(0) \hat c_a(t)},
\end{align}
where $\hat c_a(t) = e^{\ii \mathcal{ H} t} \hat c_a e^{-\ii \mathcal{ H} t}$ and $\hat c^\dagger_b(t) = e^{\ii \mathcal{ H} t}\hat c^\dagger_b e^{-\ii \mathcal{ H} t}$ represent the annihilation and creation operators for the spin-orbitals $a$ and $b$, respectively, in the Heisenberg representation.
The $\theta(t)$ denotes the Heaviside step function.
In this paper, we use $\hbar=k_\mathrm{B}=1$.
The thermal expectation, symbolized by $\expval{\cdots}$, is evaluated in the grand canonical ensemble. 

The retarded Green's function can be continued to the real frequency axis as
\begin{align}
G^\mathrm{R}_{ab}(\omega) &= \int_{-\infty}^\infty \dd t~e^{\ii\omega t}G^\mathrm{R}_{ab}(t),
\end{align}
where $\omega$ is a real frequency, while the imaginary-time Green's function is defined as
\begin{align}
    G_{ab}(\tau) &= -\theta(\tau)\expval{\hat c_a(\tau)\hat c^\dagger_b(0)}+
    \theta(-\tau)\expval{\hat c^\dagger_b(0)\hat c_a(\tau)},\label{eq:gtau}
\end{align}
where $\hat c_a(\tau) = e^{\tau \mathcal{H}} \hat c_a e^{-\tau \mathcal{H}}$.
Note that the imaginary-time Green's function is anti-periodic as $G_{ab}(\tau+\beta) = -G_{ab}(\tau)$.
The Fourier transform of the imaginary-time Green's function, known as the Matsubara Green's function, is given by
\begin{align}
     G_{ab}(\ii\omega) &= \int_0^\beta \dd \tau e^{i\omega \tau}G_{ab}(\tau),
\end{align}
where $\omega = (2n+1)\pi/\beta$ and $n\in \mathbb{N}$ and $\beta = 1/T$.

The Matsubara Green's function $G(i\omega)$ can be analytically continued from the imaginary axis to the full complex plane as $G_{ab}(z)$.
The analytically continued $G_{ab}(z)$ has the spectral representation
\begin{align}
    &G_{ab}(z)=
   \int_{-\infty}^\infty \dd \omega \frac{A_{ab}(\omega)}{z - \omega},
\end{align}
with
\begin{align}
    &A_{ab}(\omega) \equiv 
    \sum_{mn}(e^{-\beta E_n} + e^{-\beta E_m})\times \nonumber \\
     &\mel{n}{\hat c_a}{m}\mel{m}{\hat c^\dagger_b}{n}
     \delta(\omega - (E_m-E_n)),\label{eq:matsubara-lehmann}
\end{align}
where $z$ is a complex number and
$n$, $m$ runs over all eigenstates of the system with $E_m$ and $E_n$ being corresponding eigenvalues of $\mathcal{H}$.
On the real axis, these eigenvalues define individual poles for a finite system, or combine to form a branch cut for an infinite system.
The retarded and advanced Green's functions are given by the value of $G_{ab}(z)$ just above/below the real axis.
\begin{align}
   G^\mathrm{R}_{ab}(\omega) &= G_{ab}(\omega+\ii 0^+), \\
   G^\mathrm{A}_{ab}(\omega) &= G_{ab}(\omega+\ii 0^-).
\end{align}
Due to the branch cut on the real axis, $G^\mathrm{R}_{ab}(\omega) \neq G^\mathrm{A}_{ab}(\omega)$ in general.
The following relationship holds between the spectral function and the retarded and advanced Green's functions:
\begin{align}
    A_{ab}(\omega) = - \frac{1}{ 2 \pi \ii} (G^\mathrm{R}_{ab}(\omega) - G^\mathrm{A}_{ab}(\omega) ),
    \label{eq:spectral_greenfunc}
\end{align}
where we used the formula $1 /(x+  \mathrm{i}0^+)=\mathcal{P}(1 / x)-\ii \pi \delta(x)$, and $\mathcal{P}$ stands for the principal value.

We now consider the limit of $T\rightarrow 0$, where the ensemble average is restricted to the ground state(s) $\Psi_\mathrm{G}$.
At sufficiently low temperatures, Eq.~\eqref{eq:gtau} can be rewritten as
\begin{align}
\label{eq:gtau-zero-T}
    G_{ab}(\tau) &\underset{T\rightarrow 0}{=} \mp \mel{\Psi_\mathrm{G}}{ \hat A_\pm  e^{\mp (\mathcal{ H}-E_\mathrm{G})\tau}  \hat B_{\pm}}{\Psi_\mathrm{G}},
\end{align}
where ${A}_+ = \hat c_a$ and ${B}_+ = \hat c^\dagger_b$ for $0 < \tau < \beta/2$, and
${A}_- = \hat c^\dagger_b$ and ${B}_- = \hat c_a$ for $\beta/2 < \tau < 0$.
The signs $\mp$ are for $\tau>0$ and  $\tau<0$, respectively, and $E_{\mathrm{G}}= \mel{\Psi_{\mathrm{G}}}{\mathcal{ H}}{\Psi_{\mathrm{G}}}$. 
In the presence of degenerate ground states, Eq.~\eqref{eq:gtau-zero-T} should be averaged over all such states.
In general, \(|G_{ab}(\tau)|\) decays exponentially in an insulating system, while algebraic in a metallic system. 
To ensure that \(G_{ab}(\tau)\) is sufficiently small at the boundary, we need to increase \(\beta\), which determines the upper limit of time evolution.

\subsection{Variational quantum algorithms}
In quantum computing, it is necessary to convert fermionic operators into qubit representations. 
There are several methods for this, such as the Jordan-Wigner transformation~\cite{1928ZPhy...47..631J}, and the Bravyi-Kitaev transformation~\cite{bravyi2002fermionic, seeley2012bravyi}. 
In this study, we use the Jordan-Wigner transformation given by
\begin{align}
    \label{eq:jw_1}
    &\hat c_{j}^{\dagger} \rightarrow \frac{1}{2}\left(X_{j}-i Y_{j}\right) Z_{1} Z_{2} \cdots Z_{j-1}, \\
    \label{eq:jw_2}
    &\hat c_{j} \rightarrow \frac{1}{2}\left(X_{j}+i Y_{j}\right) Z_{1} Z_{2} \cdots Z_{j-1}.
\end{align}

\subsubsection{\rm Ground-state calculation using VQE}
We use variational quantum eigensolver (VQE)~\cite{peruzzo2014variational, TILLY20221}.
It begins by preparing an initial state $\ket{\Psi_\text{init}}$ on a quantum computer. 
Then, a unitary operator described by a parameterized circuit with variational parameters $\boldsymbol{\theta}$, denoted as $U(\boldsymbol{\theta})$, is applied to the initial state, producing a quantum state, $\ket{\Psi(\boldsymbol{\theta})}$.
Subsequently, the expectation value of each term in the Hamiltonian is measured using the quantum computer. 
This measured data is accumulated to compute the total expectation value of the Hamiltonian, $\expval{\mathcal{H}}$, on a classical computer.
The variational parameters are updated on the classical computer to minimize $\expval{\mathcal{H}}$, and the process is iterated until the parameters are stably minimized.
Provided the ansatz has sufficiently high expressive power and the optimization is carried out well using an appropriate initial state, the variational quantum state $\ket{\Psi(\boldsymbol{\theta^{*}})}$ with optimized variational parameters $\boldsymbol{\theta^{*}}$ approximates the ground state $\ket{\Psi_\mathrm{G}}$ accurately. 
The success of the VQE therefore relies on finding an appropriate representation of the quantum state in terms of a sufficiently compact parameteric quantum circuit that can be optimized classically.

\subsubsection{\rm Recursive VQE for spectral moments}
We detail here an approach to extend the scope of VQE to optimize the dynamics of the single-particle excitation spectrum via a compact moment expansion. 
This expansion allows access to a causal imaginary-time Green's function directly at zero temperature and in a fashion that allows for efficient quantum computation via a modified VQE~\cite{PhysRevB.103.085131, Backhouse_2022, backhouse2023dynamics}.
In a recent paper, direct measurements of the moment expansion expectation values via VQE have been proposed to compute the Green's functions~\cite{greenediniz2023quantum}. 
However, the proposed approach required measuring an increasing number of Pauli terms at higher-order moments and as systems increase in size, which we aim to mitigate via a recursive VQE approach to avoid this issue, as we will detail below.

The key physical quantities we aim to compute on the quantum device are the spectral moments of the Green's function. 
This quantity, which is classified as either hole or particle type at zero temperature, is defined in each case at the order $m$ as follows:
\begin{align}
    \label{eq:mom_h}
    M_{ r s}^{\mathrm{h},(m)}=\bra{\Psi_{\mathrm{G}}}\hat{c}^{\dagger}_{r} [\mathcal{H}_\mathcal{N}]^{m} \hat{c}_{s}\ket{\Psi_{\mathrm{G}}}, \\
    \label{eq:mom_p}
    M_{r s}^{\mathrm{p},(m)}=\bra{\Psi_{\mathrm{G}}}\hat{c}_{r} [\mathcal{H}_\mathcal{N}]^{m} \hat{c}^{\dagger}_{s}\ket{\Psi_{\mathrm{G}}},
\end{align}
where $\mathcal{H}_\mathcal{N}=\mathcal{H}-E_{\mathrm{G}}$.

These can be related to the matrix-valued spectral function, $A(\omega)_{rs}$ defined in Eq.~\eqref{eq:spectral_greenfunc}, as:
\begin{align}
M_{r s}^{\mathrm{h},(m)} & =\int_{-\infty}^0 A_{r s}(\omega) \omega^m \dd \omega , \label{eq:spec_mom_1} \\
M_{r s}^{\mathrm{p},(m)} & =\int_{0}^{\infty} A_{r s}(\omega) \omega^m \dd \omega. \label{eq:spec_mom_2}
\end{align}

The spectral moments defined in Eqs.~(\ref{eq:mom_h}) and (\ref{eq:mom_p}) correspond to the Taylor expansions of the imaginary-time Green's function at the discontinuity points $\tau=0^{-}$ and $\tau=0^{+}$, respectively. 
By increasing the number of moments, the imaginary-time Green's function can be systematically approximated over longer times $\tau$.

Once the spectral moments for the particle and hole sectors are determined up to a maximum order $N_{\mathrm{mom}}$, we can appeal to the block Lanczos algorithm~\cite{10.1063/1.471429} to constructively build an effective single-particle Hamiltonian from these moments.
This single-particle Hamiltonian spans the physical system and couples to it an auxiliary system whose dimensionality grows linearly with the number of system degrees of freedom and $N_{\mathrm{mom}}$. 
This auxiliary system acts as a zero-temperature dynamical self-energy, allowing correlation-driven changes to the original spectrum.
These changes result from the projection of the eigenstates of this effective Hamiltonian back into the physical system.
This auxiliary space is built in such a way that the resulting spectrum is causal, obeys required sum rules, and exactly preserves the initially provided moments, according to Eqs.~\eqref{eq:spec_mom_1} and \eqref{eq:spec_mom_2}. 
The resulting Green's function can be obtained directly in the Lehmann representation from the diagonalization of this effective Hamiltonian, providing the residues and energies of all the poles and allowing the Green's function to be easily transformed into any domain, including imaginary time. 
For more details of this procedure, see Refs.~\onlinecite{PhysRevB.103.085131,Backhouse_2022, backhouse2023dynamics}, while similar approaches has also recently been applied in classical perturbative electronic structure methods to expand the self-energy~\cite{10.1063/5.0143291,doi:10.1021/acs.jpclett.1c02383}.

We describe the procedure for calculating the moments defined by Eqs.~(\ref{eq:mom_h}) and (\ref{eq:mom_p}) using a hybrid quantum--classical optimization algorithm, similar to VQE approach for the ground state.
We assume that approximated $\ket{\Psi_\mathrm{G}}$ and $E_\mathrm{G}$ are already computed using VQE.
To simplify the exposition, we describe the construction of the particle sector moments, with the hole moments computed analogously.

First, we prepare a variational quantum state for the single-particle excited state ${\hat c_s}^{\dagger} \ket{\Psi_\mathrm{G}}$.
Because the operator is not unitary, we represent the resultant state as the action of a unitary multiplied by a scalar as
\begin{align}
    \label{eq:ex_state}
   {\hat c}^{\dagger}_s \ket{\Psi_\mathrm{G}} &\simeq d_0 \ket{\phi^0_\mathrm{EX}(\boldsymbol{\theta}^0_\mathrm{EX})},
\end{align}
where $d_0$ is a coefficient and the parametrized quantum state $\ket{\phi_\mathrm{EX}(\boldsymbol{\theta}^0_\mathrm{EX})}$ is defined by
\begin{align}
\label{eq:cdag_intermediate}
\ket{\phi^0_\mathrm{EX}(\boldsymbol{\theta}^0_\mathrm{EX})} &= U(\boldsymbol{\theta}^0_\mathrm{EX}) \ket{\phi^0_\mathrm{EX}}.
\end{align}
We choose to construct this state by defining an initial state $\ket{\phi^0_\mathrm{EX}}$ with $N+1$ electrons and ensure that our parameterization for $U(\boldsymbol{\theta}^0_\mathrm{EX})$ conserves the electron number of the state.

The variational parameters $\boldsymbol{\theta}^0_\mathrm{EX}$ and coefficient $d_0$ can be computed as follows:
After transforming ${\hat c}^{\dagger}_s$ into the qubit representation, we measure the cost function $C=-|\bra{\phi^0_{\mathrm{EX}}(\boldsymbol{\theta}^0_{\mathrm{EX}})}\hat c^{\dagger}_s\ket{\Psi_\mathrm{G }}|^{2}$ on the quantum computer via a circuit similar to a Hadamard test~\cite{PhysRevA.104.032405, ibe2022calculating} (see Appendix~\ref{sec:circuit_transition_am}).
The variational parameters are optimized to minimize the cost function $C$ until convergence is achieved.
After this optimization, the scaling coefficient $d_0=\mel{\phi^0_{\mathrm{EX}}(\boldsymbol{\theta}^{0*}_{\mathrm{EX}})}{\hat c^{\dagger}_s}{\Psi_\mathrm{G}}$ is measured on the quantum device.
Finally, the zeroth order moment can be computed via the sampling of $M_{r s}^{\mathrm{p},(0)} =\bra{\Psi_{\mathrm{G}}}\hat{c}_{r} \hat{c}^{\dagger}_{s}\ket{\Psi_{\mathrm{G}}} \approx d_0 \bra{\Psi_{\mathrm{G}}}\hat{c}_{r} \ket{\phi^0_{\mathrm{EX}}(\boldsymbol{\theta}^{0*}_{\mathrm{EX}})}$

We can then subsequently compute the higher order moments up to $N_{\mathrm{mom}}$ with ($1 \leq m \leq N_{\mathrm{mom}}$) via a recursive approach, avoiding the need to measure over increasingly large numbers of Pauli strings for higher-order moments, as considered in Ref.~\onlinecite{greenediniz2023quantum}.
Using $\ket{\phi^{(m-1)}_{\mathrm{EX}}(\boldsymbol{\theta}^{(m-1)*}_{\mathrm{EX}})}$ computed in the previous step, we approximate $\mathcal{H}_\mathcal{N}\ket{\phi^{(m-1)}_{\mathrm{EX}}(\boldsymbol{\theta}^{
(m-1)*}_{\mathrm{EX}})}$ as 
\begin{align}
    \label{eq:h_ex_state}
  \mathcal{H}_\mathcal{N} \ket{\phi^{(m-1)}_{\mathrm{EX}}(\boldsymbol{\theta}^{(m-1)*}_{\mathrm{EX}})} 
  &\simeq d_{m} \ket{\phi^{m}_\mathrm{EX}(\boldsymbol{\theta}^{m}_\mathrm{EX})}.
\end{align}
The variational parameters $\boldsymbol{\theta}^{m}_\mathrm{EX}$ and constant coefficient $d_{m}$ are determined by minimizing the cost function,
$C=-\left| \expval{\phi^{m}_{\mathrm{EX}}(\boldsymbol{\theta}^{m}_{\mathrm{EX}}) |
\mathcal{H}_\mathcal{N}|  \phi^{(m-1)}_{\mathrm{EX}}(\boldsymbol{\theta}^{(m-1)}_\mathrm{EX})}\right|^{2}$.
By performing $m$ VQE steps optimizing these states, we can calculate the moments of order $m$ as
\begin{align}
    \label{eq:moments_fitting}
    M_{r s}^{\mathrm{p},(m)} &=\bra{\Psi}\hat{c}_{r} [\mathcal{H}_\mathcal{N}]^{m} \hat{c}^{\dagger}_{s}\ket{\Psi} \nonumber \\
    &= d_0 d_1 \cdots d_{m}
    \bra{\Psi_{\mathrm{G}}}\hat{c}_{r} \ket{\phi^{m}_{\mathrm{EX}}(\boldsymbol{\theta}^{m*}_{\mathrm{EX}})}.
\end{align}
Similar ideas of hybrid quantum--classical variational optimization of alternative functionals for computing other (e.g. dynamical) properties have also been considered in other works~\cite{PhysRevA.104.032405, PhysRevA.99.062304, Higgott2019variationalquantum, mcardle2019variational,PhysRevResearch.2.043140,PhysRevB.104.075159} 

As the ansatz used in optimizing all $m$ states $|\phi_{\mathrm{EX}}^m(\boldsymbol{\theta}^{m}_{\mathrm{EX}}) \rangle$ becomes complete, it should enable the computation of the {\em exact} moments up to order $m$ using the described approach.
However, this optimization is also subject to various types of noises, including finite sampling errors of expectation values in a physical device, as well as optimization bottlenecks.
This can result in numerical errors, which would likely accumulate exponentially at high orders of $m$. 
Nevertheless, as the magnitude of the moment also increases exponentially with respect to its order, we find that the numerical relative error in these moments compared to their exact benchmarks remains almost constant (see Appendix~\ref{appendix:moment}). 
Finally, we note that while this approach has been presented for the computation of single-site Green's functions and moments, off-diagonal elements corresponding to matrix-valued Green's functions are possible, analogously to the approaches in Refs.~\onlinecite{PhysRevA.104.032405, greenediniz2023quantum}.

\subsection{Ansatzes}

We use two physics-inspired ansatzes: UCCGSD~\cite{lee2018generalized, nooijen2000can} and the $k$-uCJ~\cite{Matsuzawa_2020}, which we describe below.

\subsubsection{\rm  UCCGSD}

The UCCGSD is a generalization of a unitary coupled cluster (UCC) \cite{kutzelniggQuantumChemistryFock1982,kutzelniggQuantumChemistryFock1983,kutzelniggQuantumChemistryFock1985,bartlettAlternativeCoupledclusterAnsatze1989,kutzelniggErrorAnalysisImprovements1991,taubeNewPerspectivesUnitary2006} written as the exponential of an antisymmetric sum of excitation operators. The UCCGSD is formulated as follows:
\begin{align}
    \label{eq:uccgsd}
   \ket{\Psi_{\mathrm{UCCGSD}}} &=e^{(\hat{T_{2}}-\hat{T_{2}}^{\dagger}) + 
   (\hat{T_{1}}-\hat{T_{1}}^{\dagger})
   }\ket{\Psi_{\mathrm{init}}},
\end{align}
where $\ket{\Psi_{\mathrm{init}}}$ represents a product state, while $\hat T_{n}$ $(n=1,2)$ and their respective conjugates $\hat T_{n}^{\dagger}$ are excitation operators.
The excitation operators $\hat{T_n}$ are 
\begin{align}
    \hat{T_{1}} &=\sum_{pq, \alpha \beta} t_{pq}^{\alpha \beta} 
    \hat c_{p\alpha}^{\dagger} \hat c_{q\beta}, \\
    \hat{T_{2}} &= \frac{1}{4} \sum_{pqrs, \alpha \beta \gamma \zeta} 
    t_{p q r s}^{\alpha \beta \gamma \zeta} \hat c_{p\alpha}^{\dagger} \hat c_{q\beta}^{\dagger} \hat c_{r\gamma} \hat c_{s\zeta},
\end{align}
where $\hat{T_1}$ is a single-particle excitation operator, and $\hat{T_2}$ is a two-particle excitation operator. 
The indices $p, q, r, s$ represent spatial orbitals, and $\alpha, \beta, \gamma, \zeta$ represent spin. 
The composite indices $p\alpha, q\beta, r\gamma, s\zeta$ span all spin-orbitals $N_{\mathrm{SO}}$.
In this study, we removed 
one-particle and two-particle excitations that change total $S_z$.
The $t_{pq}^{\alpha \beta}$ and $t_{p q r s}^{\alpha \beta \gamma \zeta} $ are complex-number variational parameters.
The number of variational parameters $N_{\mathrm{P}}$ scales as $O(N_{\mathrm{SO}}^4)=O((N_{\mathrm{imp}} + N_{\mathrm{bath}})^{4})$, where $N_{\mathrm{imp}}$ represents the number of spin-orbitals of the impurity and $N_{\mathrm{bath}}$ the number in the bath.

Computing $\bra{\Psi_{\mathrm{UCCGSD}}} \mathcal{ H} \ket{\Psi_{\mathrm{UCCGSD}}}$ is exponentially expensive on classical computers because it results in a non-truncating Baker--Campbell--Hausdorff expansion. 
In contrast, quantum computers can compute this expectation value directly.
We use a Trotter decomposition to implement Eq.~(\ref{eq:uccgsd}) on a quantum computer. 
Classical optimization of variational quantum algorithms can partially mitigate the Trotterization error~\cite{o2016scalable,barkoutsos2018quantum}, but does result in a dependence of the final state on the ordering of the individual excitation operators. 
As commonly done, we set the Trotter step to 1, resulting in
\begin{align}
   \label{eq:UCCGSD_trotter}
   &\ket{\Psi_{\mathrm{UCCGSD}}}\nonumber \\
   &\simeq 
   e^{(\hat{T_{2}}-\hat{T_{2}}^{\dagger})} 
   e^{(\hat{T_{1}}-\hat{T_{1}}^{\dagger})
   }
   \ket{\Psi_{\mathrm{init}}} \nonumber \\ 
    &=\prod^{N_\mathrm{SO}}_{p\alpha, q\beta, r\gamma, s\zeta} 
   \{e^{t^{pqrs}_{\alpha\beta\gamma\zeta} \hat c^\dagger_{p\alpha} \hat c^\dagger_{q\beta} c_{r\gamma} c_{s\zeta} - 
   t^{pqrs*}_{\alpha\beta\gamma\zeta} \hat c^\dagger_{s\zeta} \hat c^\dagger_{r\gamma} \hat c_{ q\beta} \hat c_{p\alpha}}\}\nonumber\\
   &\hspace{2em}\times \prod^{N_{\mathrm{SO}}}_{p\alpha, q\beta} 
   \{e^{t^{\alpha \beta}_{pq} \hat c^\dagger_{p\alpha} \hat c_{q\beta} 
   -t^{\alpha \beta*}_{pq} \hat c^\dagger_{q\beta} \hat c_{p\alpha} }\}
   \ket{\Psi_{\mathrm{init}}}\nonumber\\
    &=\prod^{N_\mathrm{SO}}_{p\alpha, q\beta, r\gamma, s\zeta} 
   \{e^{t^{pqrs}_{\alpha\beta\gamma\zeta} \hat c^\dagger_{p\alpha} \hat c^\dagger_{q\beta} c_{r\gamma} c_{s\zeta} - 
   t^{pqrs*}_{\alpha\beta\gamma\zeta} \hat c^\dagger_{s\zeta} \hat c^\dagger_{r\gamma} \hat c_{ q\beta} \hat c_{p\alpha}}\}
   \ket{\Psi_{\mathrm{orb}}} ,
\end{align}
where $\ket{\Psi_{\mathrm{orb}}} \equiv \prod^{N_{\mathrm{SO}}}_{p\alpha, q\beta} 
   \{e^{t^{\alpha \beta}_{pq} \hat c^\dagger_{p\alpha} \hat c_{q\beta} 
   -t^{\alpha \beta*}_{pq} \hat c^\dagger_{q\beta} \hat c_{p\alpha} }\} \ket{\Psi_{\mathrm{init}}}$, demonstrating that the UCCGSD ansatz incorporates single-particle basis rotations into its definition~\cite{mizukami2020orbital}.

\subsubsection{k \rm  - uCJ}
Let us first define the unitary cluster Jastrow (uCJ) ansatz and then the $k$-uCJ ansatz~\cite{Matsuzawa_2020}.
The uCJ ansatz is defined as follows: 
\begin{align}
 \label{eq:ucj}
   \ket{\Psi_{\mathrm{uCJ}}} &=e^{\hat{K}} e^{\hat{J}} e^{-\hat{K}}\ket{\Psi_{\mathrm{orb}}},
\end{align}
where 
\begin{align}
    \hat{K} &=\sum_{pq,\alpha} \mathcal{K}_{pq}
    \hat c_{q\alpha}^{\dagger} \hat c_{p\alpha}, \\
    \label{eq:k-UCJ_J}
    \hat{J} &= \sum_{pq , \alpha \beta} \mathcal{J}_{p q}^{\alpha \beta} \hat c_{p\alpha}^{\dagger} \hat c_{p\alpha} \hat c_{q\beta}^{\dagger} \hat c_{q\beta}.
\end{align}
The matrix $\mathcal{K}$ is complex and anti-Hermitian.
The matrix $\mathcal{J}$ is symmetric, and its elements are purely imaginary.
The $\ket{\Psi_{\mathrm{orb}}}$ is the single-particle basis rotated state defined in Eq.~(\ref{eq:UCCGSD_trotter}). 
This ansatz preserves the particle's number and total $S_{z}$.
The scaling with $N_{\mathrm{P}}$ is $O(N_{\mathrm{SO}}^2)=O((N_{\mathrm{imp}} + N_{\mathrm{bath}})^{2})$.

The uCJ ansatz is motivated via a tensor decomposition process that compresses the generalized two-particle excitation operators in the coupled cluster method.
This compression results in a set of operators with only two indices.
Similar approaches based on tensor decomposition have been proposed in Refs.~\onlinecite{rubin2022compressing, PRXQuantum.2.040352, peng2017highly, motta2021low, PRXQuantum.2.030305}.
Equation~\eqref{eq:k-UCJ_J} can be implemented without Trotterization, as it involves only commuting number operators. By performing the Jordan-Wigner transformation on the equation, this term $\hat c_{p\alpha}^{\dagger} \hat c_{p\alpha} \hat c_{q\beta}^{\dagger} \hat c_{q\beta}$ can be simplified to $\frac{1}{4}(1-Z_{p\alpha})(1 - Z_{q\beta})$. 
The $k$-uCJ ansatz differs from the uCJ ansatz in that the operators $\mathcal{J}$ and $\mathcal{K}$ are applied multiple times, resulting in the $k$-uCJ ansatz,
\begin{align}
   \ket{\Psi_{k \text{-}\mathrm{uCJ}}} &=\sum_{i=1}^{k}e^{\hat{K_{\mathrm{i}}}} e^{\hat{J_{i}}} e^{-\hat{K_{i}}}\ket{\Psi_{\mathrm{orb}}},\label{eq:kuCJ}
\end{align}
where variational parameters for different $i$ are independently optimized.

\section{Sparse ansatzes}\label{sec:sparse ansatzes}
\begin{figure}[h]
    \centering
    \vspace{-2mm} 
\includegraphics[width=1.00\linewidth]{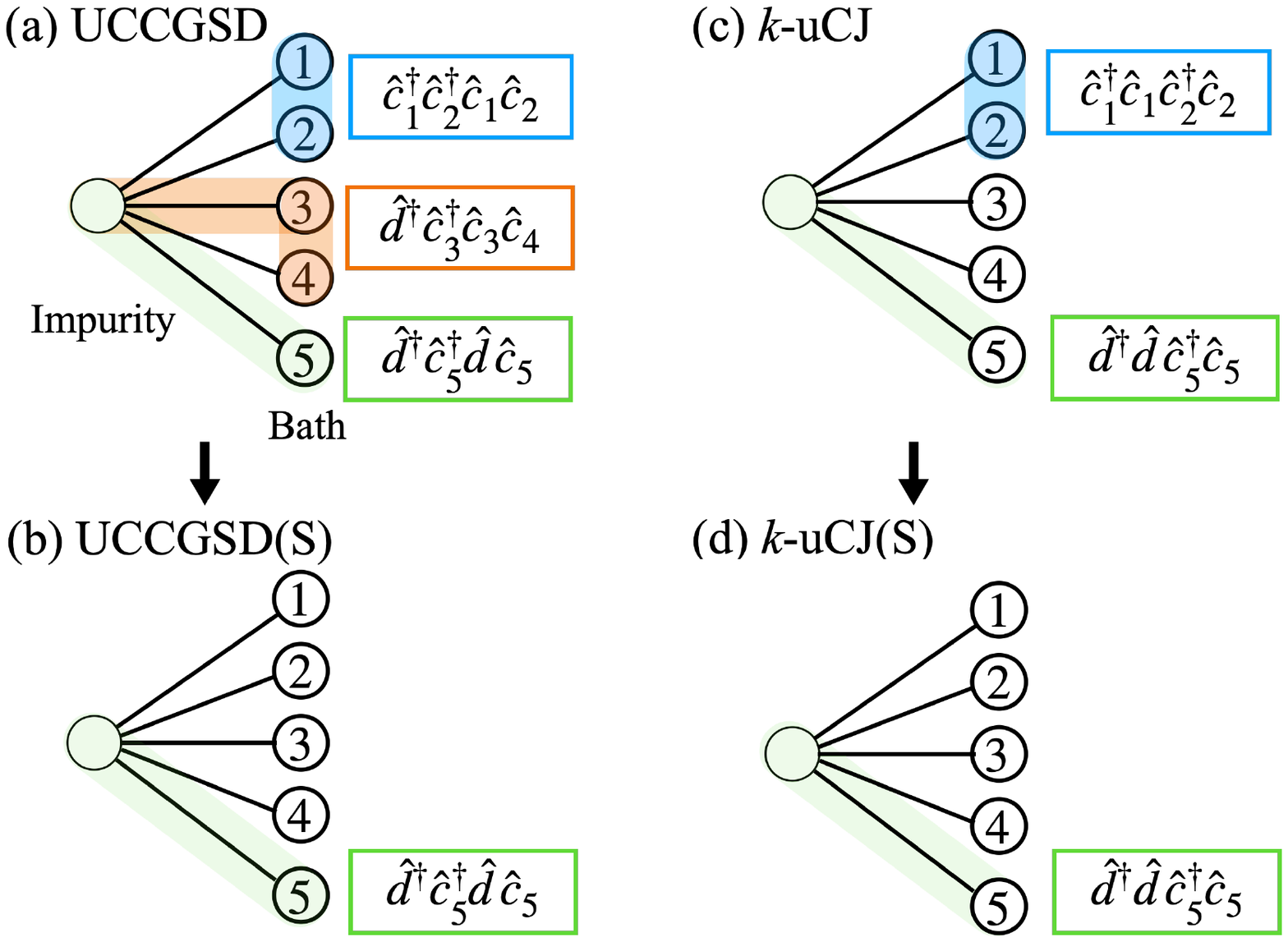}
 \vspace{-1mm} 
    \caption{
    Schematic illustrations for the construction of sparse ansatzes.
    Panels (a) and (b) show the eliminated operators that involve more than three bath orbitals when constructing the UCCGSD(S) from the UCCGSD.
    Panels (c) and (d) show the eliminated operators acting between the different bath sites when constructing the $k$-uCJ(S) from the $k$-uCJ.
    }
    \label{fig:sparse_ansatzes}
\end{figure}

In a quantum embedding calculation, a continuous hybridization can be discretized with a finite number of bath sites.
In particular, for a star-like geometry, the bath sites are connected only through the impurity.
The number of bath sites, $N_{\mathrm{bath}}$, required for an accurate discretization scales linearly with $N_\mathrm{imp}$, albeit with a significant prefactor (on the order of ten~\cite{sparsebathfitting2021}).
Given the significant number of variational parameters associated with the bath sites, reducing the number of these parameters is critical for efficient quantum simulation of impurity models.

We propose compact ansatzes for quantum impurity models with a star-like bath geometries.
We assume that two-particle excitation operators associated with two-body coupling between bath sites are not critical in the description of the ground states and spectral moments, given that two-particle interaction terms in the Hamiltonian are localized to the impurity space, and no Hamiltonian terms directly couple the bath sites.
The ansatzes incorporating this assumption are referred to as ``sparse ansatzes''.
In the present study, we construct sparse ansatzes based on the UCCGSD and the $k$-uCJ. 
We call them sparse UCCGSD and sparse $k$-uCJ, denoted UCCGSD(S) and $k$-uCJ(S), respectively.

For the UCCGSD, we remove two-particle excitation operators that involve more than three bath orbitals.
Examples of such operators that involve three or four bath orbitals are $\hat{c}_{1}^{\dagger} \hat{c}_{1}^{\dagger} \hat{c}_{2} \hat{c}_{2}$ and $\hat{d}_{}^{\dagger} \hat{c}_{3}^{\dagger} \hat{c}_{3} \hat{c}_{4}$, where $\hat{d}^{\dagger}$ ($\hat{c}^{\dagger}$) are 
fermionic creation operators for
the impurity (bath) degrees of freedom, respectively.
We illustrate these operators in Figs.~\ref{fig:sparse_ansatzes}(a) and (b). 
For the UCCGSD, this reduces the number of variational parameters $N_{\mathrm{P}}$ from $O\left((N_{\mathrm{imp}} + N_{\mathrm{bath}})^{4}\right)$ to 
$O\left(N_{\mathrm{imp}}^4 + N_{\mathrm{bath}}^{2}\right) \simeq O\left(N_{\mathrm{imp}}^{4}\right)$ for the sparse variant
(refer to Table~\ref{tb:parameters}). Although $N_{\mathrm{bath}}$ is proportional to $N_{\mathrm{imp}}$~\cite{sparsebathfitting2021}, ensuring that the scaling with respect to impurity size remains the same, the significant computational savings still result since $N_\mathrm{bath} \gg N_\mathrm{imp}$.

For the $k$-uCJ, we apply a similar motivation to remove the operators acting between different bath sites while keeping the two-particle excitation operators between the impurity and the bath. 
For example, $\hat{c}_{1}^{\dagger}\hat{c}_{1}\hat{c}_{2}^{\dagger}\hat{c}_{2}$ is dropped, as illustrated in Figs.~\ref{fig:sparse_ansatzes}(c) and (d). 
As summarized in Table~\ref{tb:parameters}, the scaling of $N_{\mathrm{P}}$ in the $k$-uCJ ansatz scales as $O\left((N_{\mathrm{imp}}+ N_{\mathrm{bath}})^{2}\right)$, while $N_{\mathrm{P}}$ in the corresponding $k$-uCJ(S) sparse ansatz scales as $O\left(N_{\mathrm{imp}}^2\right)$.
Again, the prefactor is substantially reduced when $N_\mathrm{bath} \gg N_\mathrm{imp}$.

\begin{table}[hbtp]
  \centering
\begin{tabular}{ll } \hline
   Ansatz & Number of variational parameters $N_\mathrm{P}$ \\ \hline
   UCCGSD & $O(N_{\mathrm{SO}}^4)$ = $O((N_{\mathrm{imp}} + N_{\mathrm{bath}})^4)$ \\
   UCCGSD(S) & $O(N_{\mathrm{imp}}^4)$ \\ 
   $k$-uCJ &$O(N_{\mathrm{SO}}^2)$ = $O((N_{\mathrm{imp}} + N_{\mathrm{bath}})^2)$   \\ 
    $k$-uCJ(S) & $O(N_{\mathrm{imp}}^2)$ \\\hline
 \end{tabular} 
  \caption{Number of variational parameters for the UCCGSD, UCCGSD(S), $k$-uCJ, and $k$-uCJ(S).
  }
  \label{tb:parameters}
\end{table}

\section{State vector simulation}\label{sec:state_vector_simulation}
In this section, we benchmark the $k$-uCJ and the proposed sparse ansaztes for typical quantum impurity models.
We consider both single-site and two-site impurity models with $N_{\mathrm{bath}}=3$ and $N_{\mathrm{bath}}=6$, respectively.
All calculations in this section are based on state vector simulations of quantum circuits. 

\subsection{Numerical details} \label{sec:state_vector_sim}
The calculations were performed using the following libraries:
\texttt{QCMaterialNew}~\cite{QCMaterialNew} was used as a quantum circuit simulator, which is a Julia wrapper of \texttt{Qulacs}~\cite{suzuki2021qulacs}.
We used \texttt{Openfermion}~\cite{mcclean2020openfermion} for the Jordan-Wigner transformation and to calculate the exact eigenvalues of Hamiltonians.
We performed DMFT calculations using \texttt{DCore}~\cite{shinaoka2021dcore} to generate the single-site impurity models.
We used \texttt{dyson}~\cite{dyson} library, in order to compute the Green's functions poles and residues from the spectral moments, as well as benchmark exact spectral moments via exact diagonalization.

For optimizing the variational parameters, we used the BFGS algorithm. 
We initialized variational parameters with random numbers.
We observed that setting the initial guess to zero could lead the optimization to converge to a metastable solution.
For ground-state calculations with VQE using the $k$-uCJ, we increased the number of terms $k$ in the ansatz one by one, reusing the optimized variational parameters.
In practice, at the beginning of the VQE calculations with $k$ terms,
we randomized the variational parameters in $\hat{K}_1$ and $\hat{J}_1$ but set those in $\hat{K}_i$ and $\hat{J}_i$ ($2 \leq i \leq k$) to the optimized variational parameters obtained in the previous calculation with $k-1$ terms.
This procedure ensures that the optimized energy decreases or remains nearly stable with an increasing number of terms in the $k$-uCJ.

It is worth noting that the initial parameters significantly influence the accuracy of the optimized ground state and spectral moments.
For ground-state calculations, we conducted VQE multiple times, each with a different set of initial parameters,
to find the best variational state for the ground state.
We used this best variational state for computing spectral moments.

Simulations were executed using an MPI parallelized program on a workstation with an AMD EPYC 7702P 64-core processor. 
Solving the largest model with 16 qubits and about 750 variational parameters in the $k$-uCJ took about five days on 55 cores using VQE and the recursive approach.

\subsection{Single-site impurity model}\label{sec:four-site}

\begin{figure}
\centering
\includegraphics[width=1.0\linewidth]{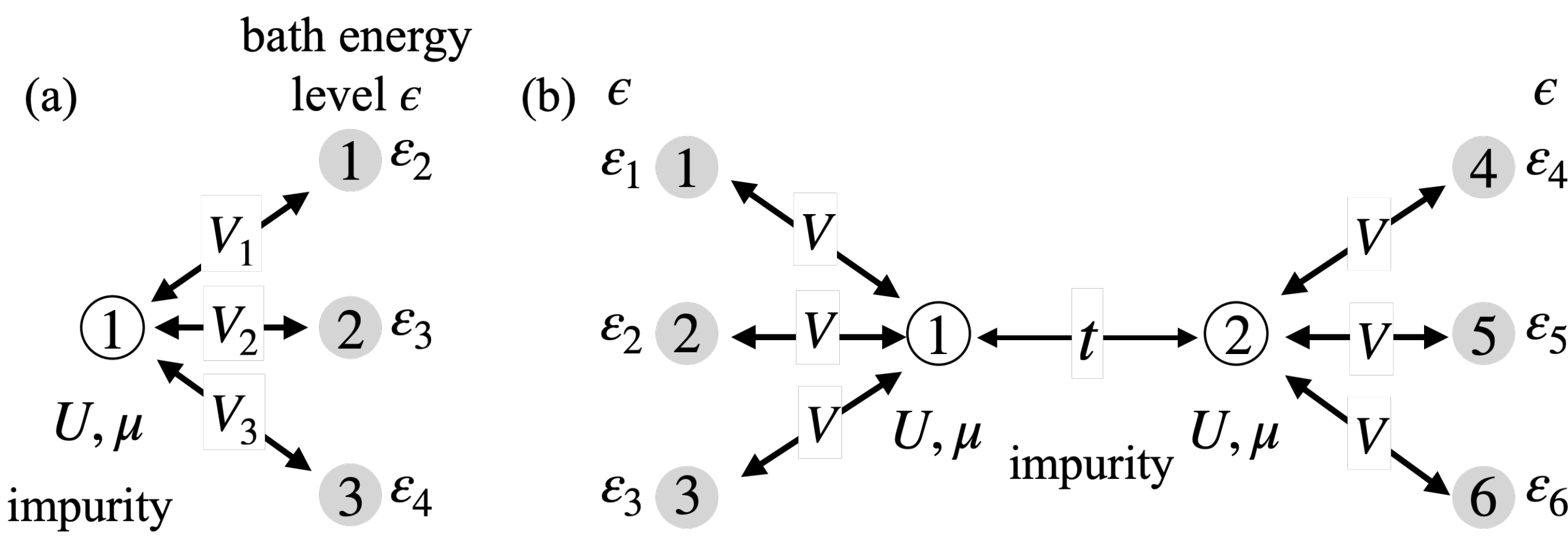}
\vspace{-2mm} 
\caption{Two quantum impurity models used in this study. Panels (a) and (b) show the single-site impurity model with $N_{\mathrm{bath}}=3$, and the two-site impurity model with $N_{\mathrm{bath}}=6$, respectively.
}
\label{fig:impurity_models}
\end{figure}

We consider the single-site impurity model with particle-hole symmetry and $N_\mathrm{bath}=3$ illustrated in Fig.~\ref{fig:impurity_models}(a).
The Hamiltonian is given by
\begin{align}
     \mathcal{ H} &= U 
     \hat{d}^{\dagger}_{1 \uparrow} 
     \hat{d}_{1 \uparrow}
     \hat{d}^{\dagger}_{1 \downarrow}
     \hat{d}_{1 \downarrow}-\mu\sum_{\sigma=\uparrow, \downarrow}
     \hat{d}^{\dagger}_{1 \sigma}
     \hat{d}_{1 \sigma}
    \nonumber \\&
     -\sum_{k=1}^{3}\sum_{\sigma=\uparrow, \downarrow}
     V_{k}\left(\hat{d}_{1 \sigma}^{\dagger} \hat{c}_{k \sigma}+\hat{c}_{k \sigma}^{\dagger} \hat{d}_{1 \sigma}\right)+
     \sum_{k=1}^{3}\sum_{\sigma=\uparrow, \downarrow}
     \epsilon_{k}\hat{c}^{\dagger}_{k \sigma}
     \hat{c}_{k \sigma},
\end{align}
where $\hat{d}^{\dagger}_{1 \sigma}$ ($\hat{c}_{k \sigma}^{\dagger}$) are the impurity (bath) degrees of freedom of the fermionic creation operator with $\sigma=\uparrow, \downarrow$, and $k$ is an index for bath sites. 
The $U$ represents the on-site Coulomb repulsion, $V_k$ is the hybridization term, $\mu~(=U/2)$ is the chemical potential, and $\epsilon_{k}$ denotes the bath energy.

We obtained the bath parameters using self-consistent DMFT calculations on a square lattice at zero temperature for $U=4$ (metallic phase) and $U=9$ (insulating phase).
The nearest neighbor hopping parameter was set to 1.
For $U=4$, we obtained
$V_k = \{-1.26264, 0.07702, -1.26264\}$ and  $\epsilon_{k} = \{1.11919, 0.0, -1.11919\}$.
For $U=9$, we obtained $V_k = \{1.31098, 0.07658, -1.38519\}$ and $\epsilon_{k} = \{-3.26141, 0.0, 3.26141\}$.

\subsubsection{\rm  Ground-state calculation}
\begin{figure}
    \centering
    \vspace{-1mm} 
    \includegraphics[width=0.85\linewidth]{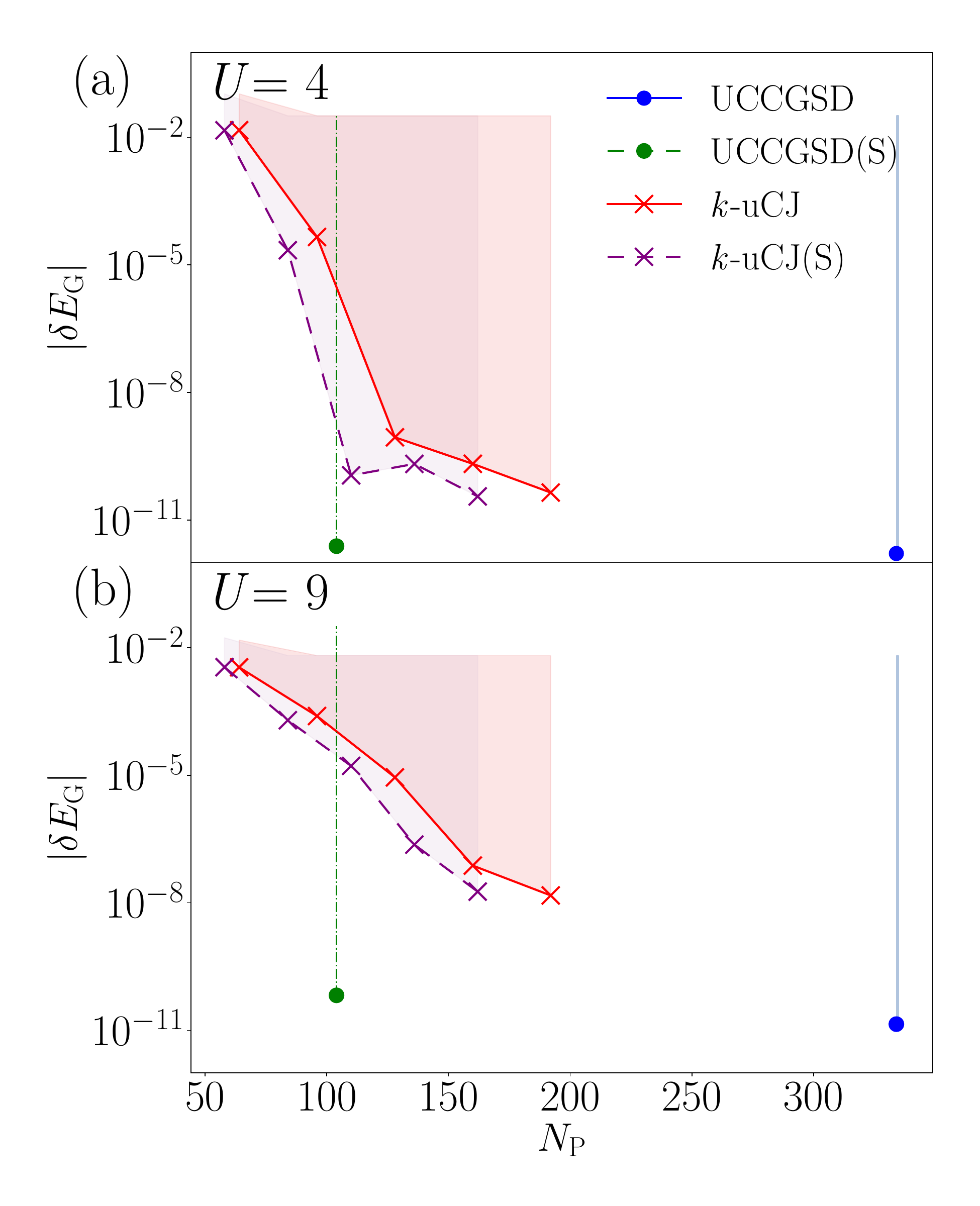}
    \vspace{-3mm}
    \vspace{-1mm} 
    \caption{
    Computed $|\delta E_\mathrm{G}|$ for the single-site impurity model.    
    Panels (a) and (b) show the results for $U=4$ and $U=9$, respectively.
    In the $k$-uCJ and the $k$-uCJ(S), $k$ increases from 1 to 5.
    The markers represent the smallest errors when the initial parameters are changed 50 times. 
    The lightly shaded areas in the figure illustrate the dependency of the absolute errors on the initial parameters.
    }
    \label{fig:4site_gs_all}
\end{figure}

Figures~\ref{fig:4site_gs_all}(a) and (b) show the absolute errors in ground-state energies ($|\delta E_{\mathrm{G}}|$) for $U=4$ and $U=9$, respectively, compared to exact diagonalization results.
For the $k$-uCJ and the $k$-uCJ(S), we varied $k$ from 1 to 5 to check convergence.
The markers represent the best results obtained by varying the initial parameters 50 times for each ansatz.
The lightly shaded areas indicate the variation in converged results depending on the choice of initial parameters for each ansatz.
In all four ansatzes, the best ground-state energies are well reproduced. 
We also confirmed that the $k$-uCJ reproduces the ground-state energy with smaller $N_{\mathrm{P}}$ than the UCCGSD.
Also, the results for the sparse ansatzes in Figs.~\ref{fig:4site_gs_all}(a) and (b) show that reducing the variational parameters associated with bath sites does not compromise the accuracy of the ground-state energies.
It should be noted that the sparse ansatzes are efficient even for the metal-like system ($U=4$), where the electronic structure is very much delocalized across the bath sites.

The following summarizes the reduction in $N_{\mathrm{P}}$ for each ansatz by using the sparseness.
In the UCCGSD, $N_{\mathrm{P}}$ is reduced from 334 to 104.
In the $k$-uCJ, $N_{\mathrm{P}}$ is reduced from $\{64, 96, 128, 160, 192\}$ to $\{58, 84, 110, 136, 162\}$ for $k=1,2,\cdots, 5$.
The $k$-uCJ(S) has a small reduction in the number of parameters for this system, but this reduction becomes more significant with increasing system size and complexity (see Sec.~\ref{sec:two-site_imp_gs}).

\subsubsection{\rm  Spectral functions}
\begin{figure}
    \centering
    \vspace{-1mm} 
    \includegraphics[width=0.90\linewidth]{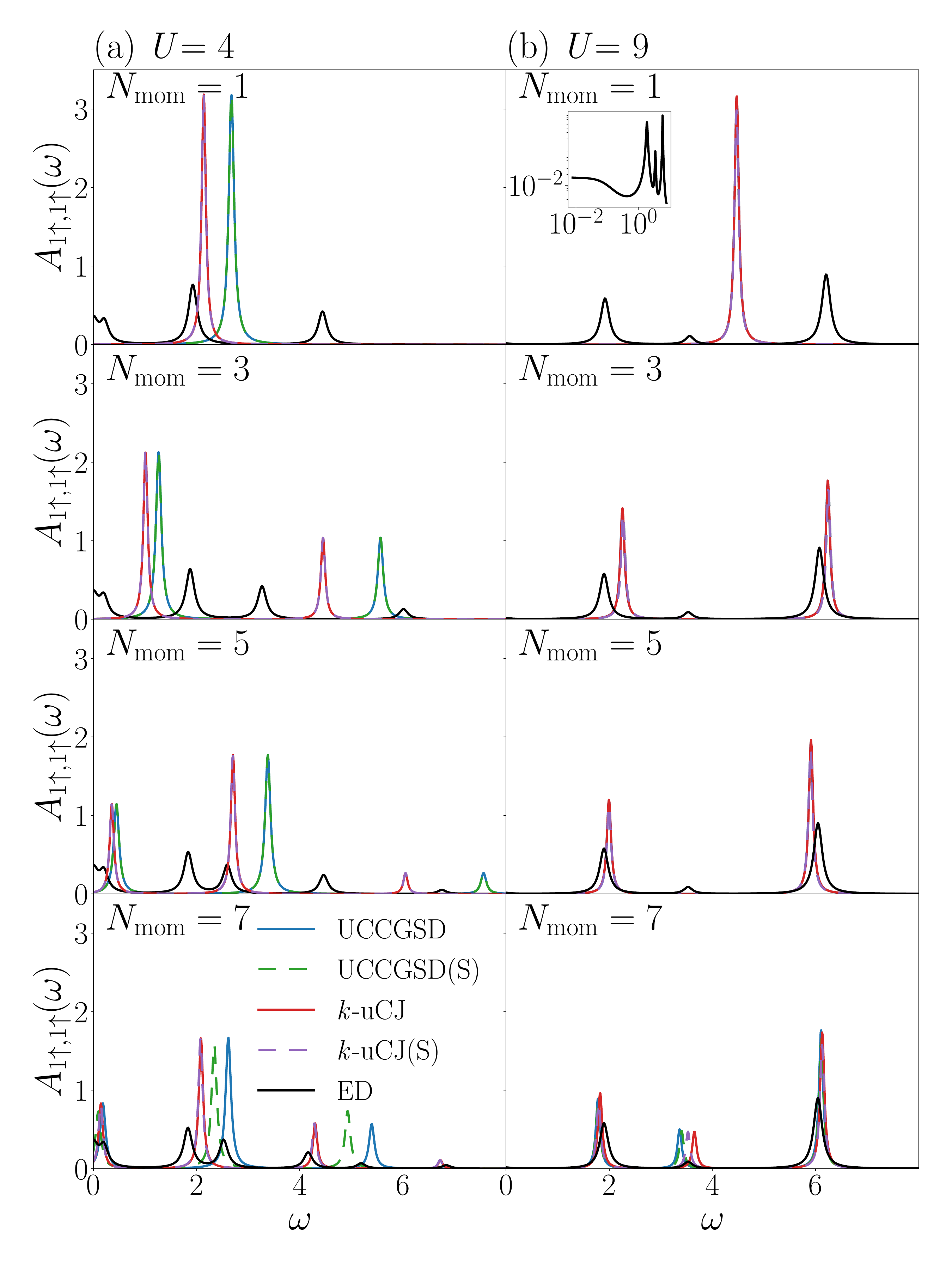}
    \vspace{-3mm}
    \vspace{-1mm} 
    \caption{
    Computed $A_{\mathrm{1 \uparrow, 1 \uparrow}}(\omega)$ for each $N_{\mathrm{mom}}$.
    Panels (a) and (b) show the results for $U=4$ and $U=9$, respectively.
    In the $k$-uCJ and the $k$-uCJ(S), we set $k=5$.
    ED refers to the spectral functions constructed from exact moments using exact diagonalization.
    The spectrum of $V=0.1$ around $\omega =0$ has a tiny magnitude of $10^{-2}$, as shown in the inset.
    }
    \label{fig:4site_spectrum}
\end{figure}

Figures~\ref{fig:4site_spectrum}(a) and (b) show the reconstructed spectral functions using the moment expansion for $U=4$ and $U=9$, respectively.
For the $k$-uCJ, we set $k=5$.
We computed the exact values of the moments using exact diagonalization (ED). As shown in Fig.~\ref{fig:4site_spectrum}(a), for $U=4$, all the ansatzes can reproduce the peaks around $\omega=0$.
However, the quality of reproduction drops for $\omega\ge 2$.
These discrepancies primarily arise from numerical errors during the moment computations via recursive VQE due to the limited representational ability of the ansatzes and the optimization issues.
This \textit{fitting error} in the recursive approach grows exponentially with $N_{\mathrm{mom}}$, which prevents systematic improvement of reconstructed spectral functions with increasing $N_\mathrm{mom}$.
Indeed, we observed no improvement for $N_\mathrm{mom} > 7$, although knowledge of the exact moments up to order $N_\mathrm{mom}=5$ is largely sufficient to converge the spectral function over all frequencies.

As shown in Fig.~\ref{fig:4site_spectrum}(b), for $U=9$, by increasing $N_{\mathrm{mom}}$ up to 7, all the ansatzes accurately reproduced
the positions of peaks for $\omega \lesssim 6$.
In general, an insulating system has fewer spectral peaks than metallic cases, allowing the moment expansion by the recursive approach to reproduce the peak positions more accurately.
Still, there is some variation among the ansatzes, likely due to the fitting error, especially around the small peak near $\omega =4$.
The spectral function shows a tiny peak near $\omega=0$ as shown in the inset of Fig.~\ref{fig:4site_spectrum}(b). 
This is due to the $k=3$ bath site nearly decoupled from the impurity, being physically irrelevant.

\begin{figure}
    \centering
    \vspace{-1mm} 
    \includegraphics[width=0.80\linewidth]{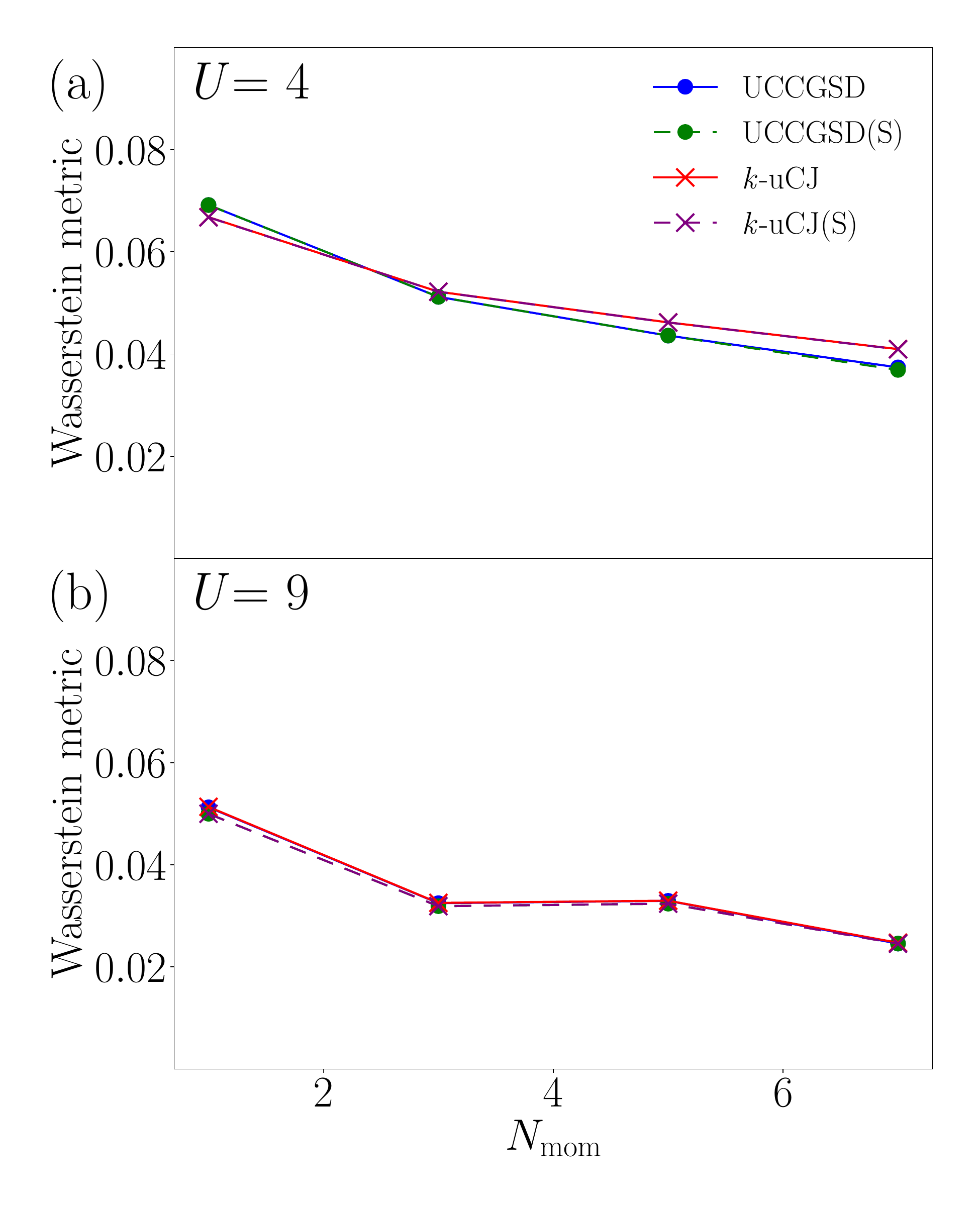}
    \vspace{-3mm}
    \vspace{-1mm} 
    \caption{
    Computed Wasserstein metric for each $N_{\mathrm{mom}}$.
    Panels (a) and (b) show the results for $U=4$ and 9, respectively.
    In the $k$-uCJ and the $k$-uCJ(S), we set $k=5$.
    }
    \label{fig:4site_metrix}
\end{figure}

Here, we aim to quantify the difference between the spectral functions reconstructed from the exact moments and those calculated via the recursive approach.
To this end, we utilize the Wasserstein metric, quantifying a difference between two distributions~\cite{10.1287/mnsc.6.4.366, ctx51223882530003681}.
Figures~\ref{fig:4site_metrix}(a) and (b) show the computed Wasserstein metric between the spectral functions from the exact moments at $N_{\mathrm{mom}}=7$ and those using ansatzes at each $N_{\mathrm{mom}}$ for $U=4$ and $U=9$, respectively.
As $N_{\mathrm{mom}}$  increases, the distance between the two distributions decreases, consistently with the enhanced reproducibility of the spectrum at large $N_\mathrm{mom}$.

\subsubsection{\rm  Imaginary-time Green's functions}

\begin{figure}
    \centering
    \vspace{-1mm} 
    \includegraphics[width=0.85\linewidth]{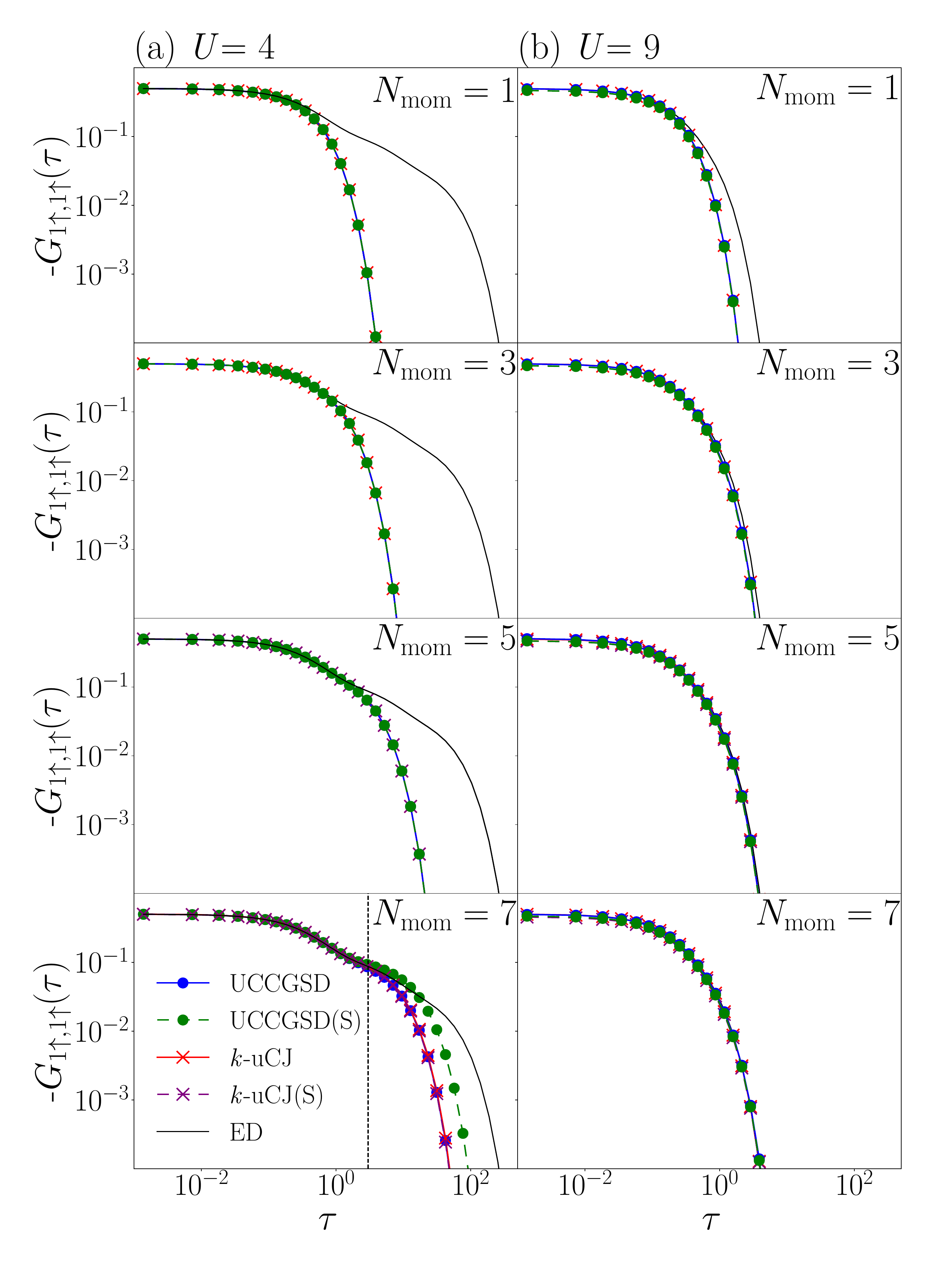}
    \vspace{-3mm}
    \caption{
    Computed $G_{1\uparrow,1\uparrow}(\tau)$ for each $N_{\mathrm{mom}}$.
    Panels (a) and (b) show the results for $U=4$ and $U=9$, respectively.
    In the $k$-uCJ and the $k$-uCJ(S), we set $k=5$.
    ED refers to the spectral functions constructed from exact moments using exact diagonalization.
    The black vertical lines in panel (a) for $N_{\mathrm{mom}}$ = 7 show where the deviation of the reconstructed spectral functions from the reference data starts.
    }
    \label{fig:4site_G}
\end{figure}

Figures~\ref{fig:4site_G}(a) and (b) show the imaginary-time Green's functions computed from the reconstructed spectral function by the moment expansion for $U=4$ and $U=9$, respectively.
We use the reconstructed spectral function from the exact moments for each $N_\mathrm{mom}$ as reference.
In the $k$-uCJ, we set $k=5$.
In computing the reference data, we filtered out peaks below $\omega \leq 10^{-2}$ that are physically irrelevant.

In Figs.~\ref{fig:4site_G}, for both $U=4$ and $U=9$, 
the differences among the ansatzes become less pronounced in the imaginary-time Green's functions 
compared to the differences in the spectral function.
In Fig.~\ref{fig:4site_G}(a), for $U=4$, imaginary-time Green's functions exhibit a power-law decay.
This necessitates a higher $N_\mathrm{mom}$ in the moment expansion.
However, for $\tau > 5$, we observed that increasing $N_\mathrm{mom}$ did not improve the accuracy due to the exponential growth in the fitting error with $N_\mathrm{mom}$ in the recursive approach. 
Only the UCCGSD(S) result seems to diverge from the rest. 
Nonetheless, its deviation starting at $\tau=5$ aligns with the trends observed in other ansatzes, displaying a comparable pattern.
In Fig.~\ref{fig:4site_G}(b), for $U=9$, imaginary-time Green's functions exhibit an exponential decay.
The Green's functions computed by the recursive approach, even at $N_\mathrm{mom}=5$, match the reference data, suggesting a smaller $N_\mathrm{mom}$ achieves convergence compared to the metallic case.

\subsection{Two-site impurity model}\label{section:two-site impurity}
We consider the two-site impurity model with particle-hole symmetry and $N_{\mathrm{bath}}$ = 6, shown in Fig.~\ref{fig:impurity_models}(b).
The Hamiltonian is given by 
\begin{align}
    \mathcal{H} &=U\sum_{i=1}^{2} \hat{d}^{\dagger}_{i \uparrow}  \hat{d}_{i \uparrow} \hat{d}^{\dagger}_{i \downarrow} \hat{d}_{i \downarrow}
    -\mu\sum_{i=1, 2}
    \sum_{\sigma=\uparrow, \downarrow}
    \hat{d}^{\dagger}_{i\sigma}\hat{d}_{i\sigma} \nonumber \\&
    -
    t_{}\sum_{\sigma=\uparrow, \downarrow} \left(\hat{d}_{1 \sigma}^{\dagger} \hat{d}_{2 \sigma}+\hat{d}_{2 \sigma}^{\dagger} 
    \hat{d}_{1 \sigma}\right) 
    \nonumber \\ 
    &-\sum_{j=1}^{2}\sum_{k_{1}={1}}^{3} \sum_{k_{2}=4}^{6}\sum_{\sigma=\uparrow, \downarrow} 
    V_{}\left(\hat{d}_{j \sigma}^{\dagger} \hat{c}_{k_{j} \sigma}+\hat{c}_{k_{j} \sigma}^{\dagger} \hat{d}_{j \sigma}\right) \nonumber \\
    &+
    \sum_{k=1}^{6} \sum_{\sigma=\uparrow, \downarrow} \epsilon_{k} \hat{c}^{\dagger}_{k \sigma} \hat{c}^{}_{k \sigma},
\end{align}
where $t$ represents the hopping between the two impurities.
For $V=0.5$ and $V=0.1$, we use common bath parameters: $U=4$, $\mu=U/2$, $t=1$, and $\epsilon_{k}= \{1, 0, -1, 1, 0, -1\}$.
The case of $V=0.5$ is expected to be more \textit{metallic} than $V=0.1$.

\subsubsection{\rm Ground-state calculation}\label{sec:two-site_imp_gs}
\begin{figure}
    \centering
    \vspace{-1mm} 
    \includegraphics[width=0.85\linewidth]{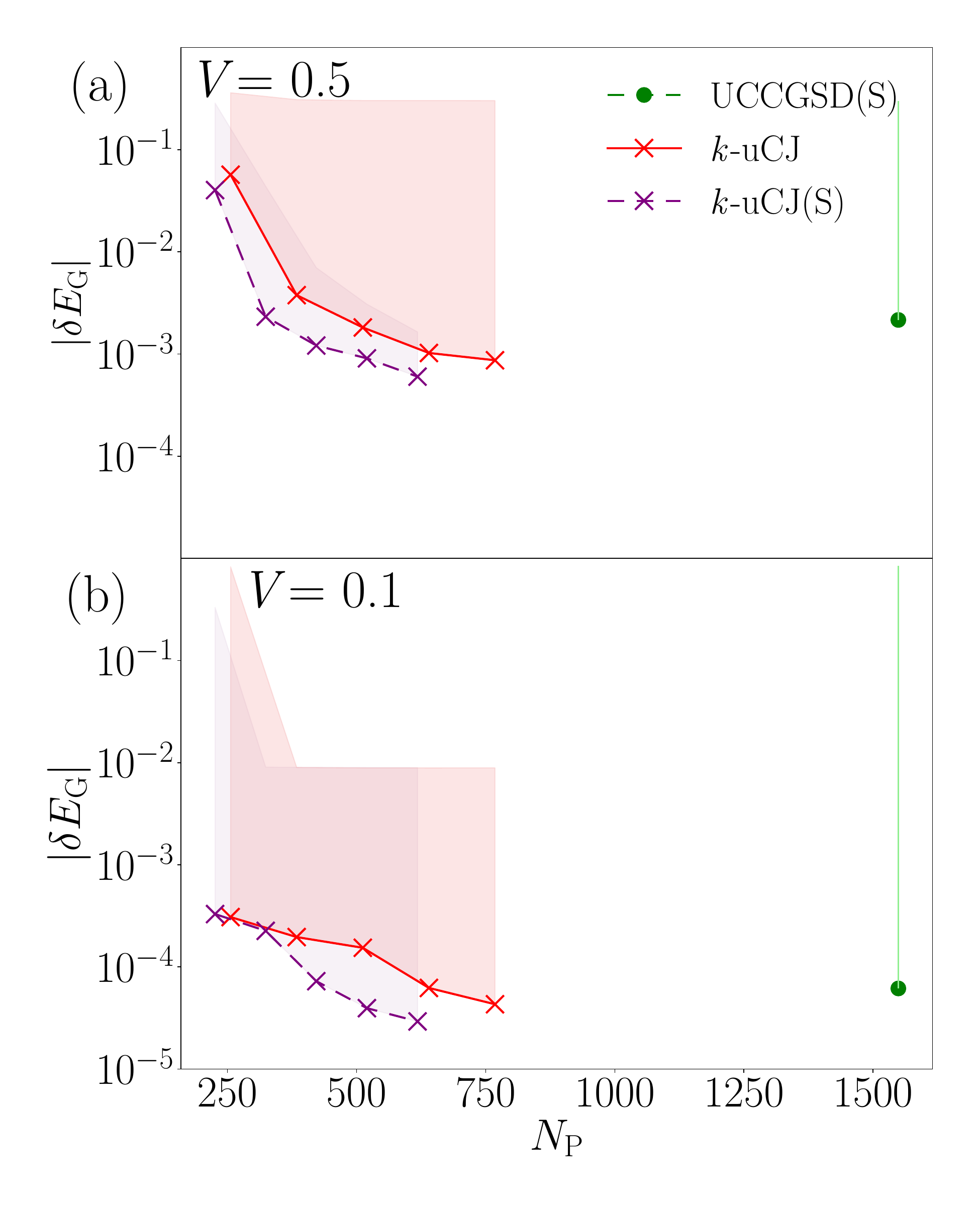}
    \vspace{-3mm}
    \vspace{-1mm} 
    \caption{
    Computed $|\delta E_\mathrm{G}|$ for the two-site impurity model.    
    The markers represent the smallest errors when the initial parameters are changed 20 times. 
    In the $k$-uCJ and the $k$-uCJ(S), $k$ increases from 1 to 5.
    The lightly shaded areas in the figure illustrate the dependency of the absolute errors on the initial parameters.
    Panels (a) and (b) show the results for $U=4$ and $U=9$, respectively.
    }
    \label{fig:8site_gs_all}
\end{figure}

Figures~\ref{fig:8site_gs_all}(a) and (b) show $|\delta E_{\mathrm{G}}|$ for $V=0.5$ and $V=0.1$, respectively. 
For the $k$-uCJ and the $k$-uCJ(S), $k$ was varied from 1 to 5 to check convergence.
We omitted the VQE calculation with the UCCGSD because of its prohibitively large number of variational parameters. 
As before, the markers represent the optimal results obtained from 20 variations of the initial variational parameters for each ansatz. 
The lightly shaded areas highlight the dependency of each ansatz on initial guesses.

In the three ansatzes, the ground-state energies are reproduced with comparable accuracy. 
Considering $N_{\mathrm{P}}$, both $k$-uCJ and $k$-uCJ(S) are more efficient than UCCGSD(S). 
The results for the sparse ansatzes in Figs.~\ref{fig:8site_gs_all}(a) and (b) show that we can reduce the number of the variational parameters associated with bath sites without sacrificing ground-state accuracy in the cluster impurity model. 
The sparse ansatzes are also applicable for the case of $V=0.5$, which exhibits more metallic characteristics.
For the $k$-uCJ, $N_\mathrm{P}$ is reduced from [256, 384, 512, 640, 768] to [226, 324, 422, 520, 618] for $k=1,2,\cdots, 5$.

\subsubsection{\rm Spectral functions}

\begin{figure}
    \centering
    \vspace{-1mm} 
    \includegraphics[width=0.85\linewidth]{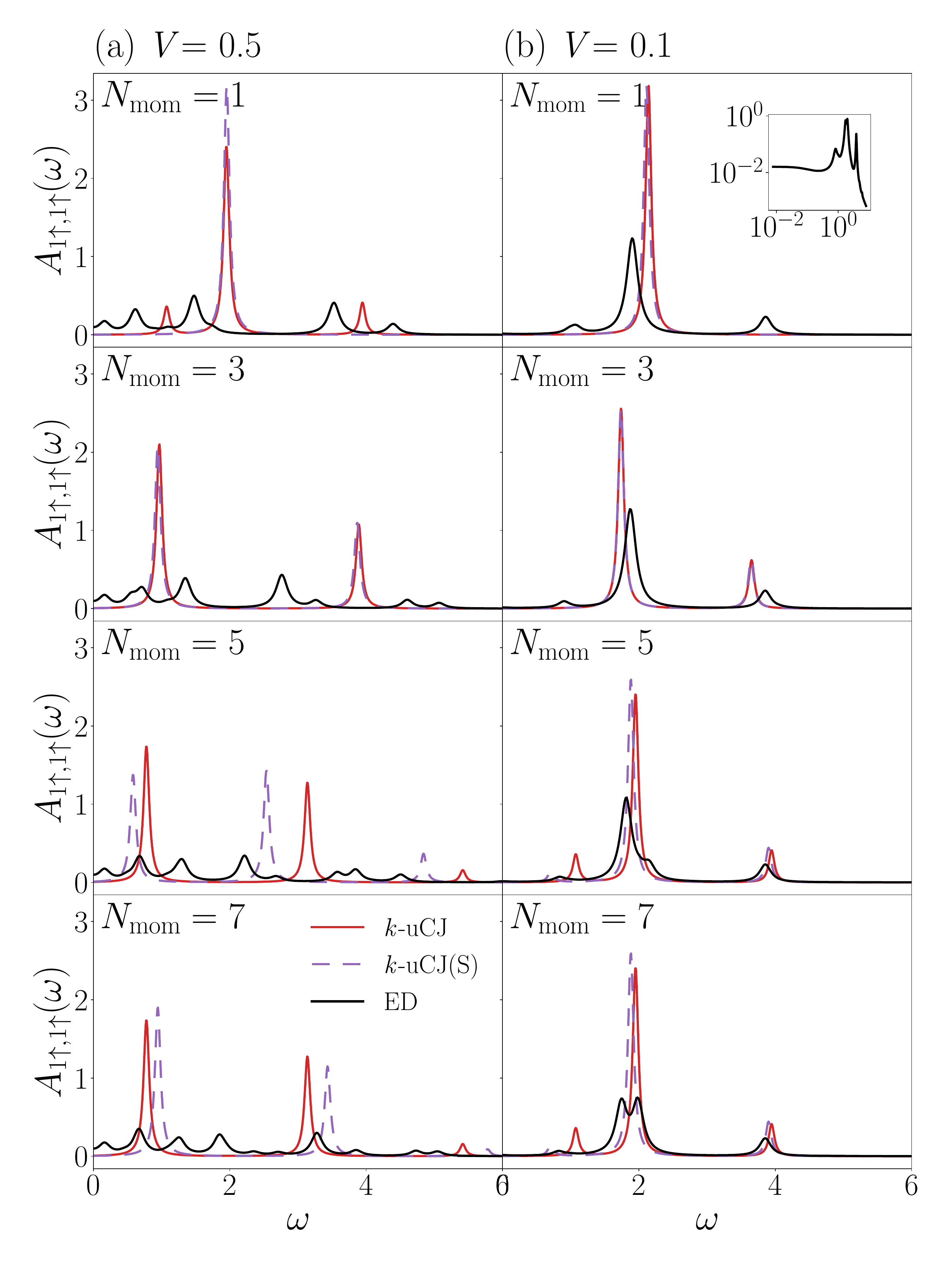}
    \vspace{-3mm}
    \vspace{-1mm} 
    \caption{
    Computed $A_{\mathrm{1 \uparrow, 1 \uparrow}}(\omega)$ for each $N_{\mathrm{mom}}$.
    Panels (a) and (b) show the results for $V=0.5$ and $V=0.1$, respectively.
    In the $k$-uCJ and the $k$-uCJ(S), we set $k=5$.
    ED refers to the spectral functions constructed from exact moments using exact diagonalization.
    The spectrum for $V=0.1$ has a tiny peak around $\omega =0$, as shown in the inset.}
    \label{fig:two_site_imp_sp}
\end{figure}

Figures~\ref{fig:two_site_imp_sp}(a) and (b) show the reconstructed spectral functions using the moment expansion for $V=0.5$ and $V=0.1$, respectively.
We computed the reference data from the exact moments for each $N_\mathrm{mom}$ using exact diagonalization.
In the $k$-uCJ, we set $k=5$.
Similar to the previous subsubsection, the UCCGSD and UCCGSD(S) calculations were omitted due to the prohibitive number of variational parameters.

In Fig.~\ref{fig:two_site_imp_sp}(a), for $V=0.5$, increasing $N_\mathrm{mom}$ tends to enhance the representation of several spectral peaks.
Yet, it remains challenging to comprehensively capture the entire structure, mainly due to the fitting error, observing no improvement beyond $N_\mathrm{mom} = 7$.
In Fig.~\ref{fig:two_site_imp_sp}(b), for $V=0.1$, by increasing $N_{\mathrm{mom}}$ up to 5, all the ansatzes accurately reproduced the positions of several peaks for $\omega \lesssim 4$. 
This indicates that an insulating system with fewer spectral peaks offers the advantage of accurately determining peak positions.
However, variations around the small peak near $\omega =1$ among the ansatzes likely result from the fitting error.
The spectral function shows a small peak around $\omega=0$ as shown in Fig.~\ref{fig:two_site_imp_sp}(b). 
This originates from bath sites weakly coupled with the impurity, being physically insignificant.

\begin{figure}
    \centering
    \vspace{-1mm} 
    \includegraphics[width=0.85\linewidth]{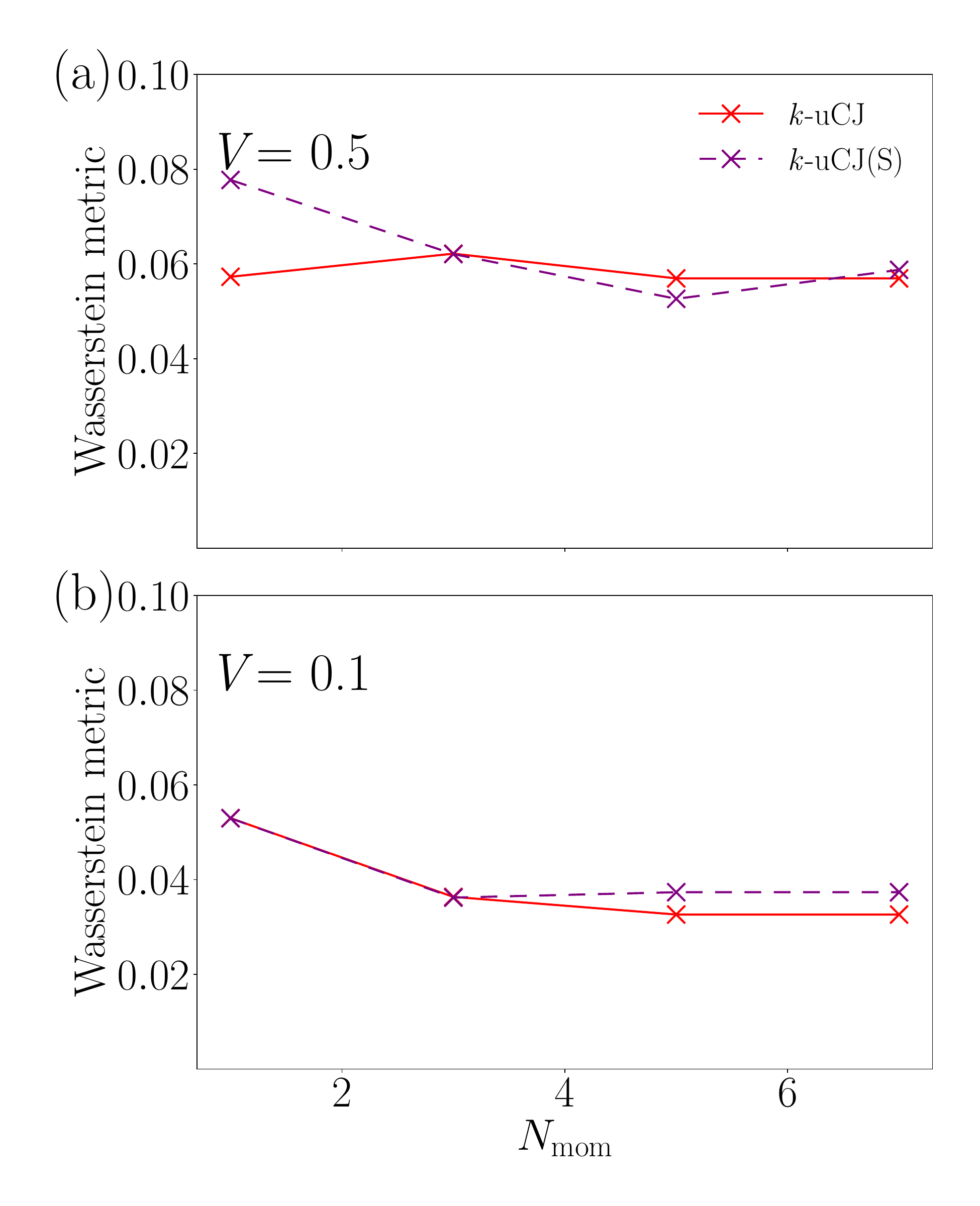}
    \vspace{-3mm}
    \vspace{-1mm} 
    \caption{
    Computed Wasserstein metric between the spectral functions constructed from exact moments and those computed using the ansatzes for the two-site impurity model for each $N_{\mathrm{mom}}$. 
    Panels (a) and (b) show the results for $V=0.5$ and 0.1, respectively.
    In the $k$-uCJ and the $k$-uCJ(S), we set $k=5$.}
    \label{fig:two-site_metric}
\end{figure}

Figures~\ref{fig:two-site_metric}(a) and (b) show the computed Wasserstein metrics between the spectral functions reconstructed from the exact moments at 
$N_{\mathrm{mom}}=7$ and those computed using the ansatzes at each $N_{\mathrm{mom}}$ for $V=0.5$ and $V=0.1$, respectively. 
Due to the influence of noise, the distances, especially for $N_{\mathrm{mom}} \geqq 5$, stay at higher values than those without shot noise.
Still, the Wasserstein metric tends to decrease as $N_{\mathrm{mom}}$ increases, which is consistent with the improved reproducibility of the spectral functions reconstructed by the moment expansion.

\subsubsection{\rm Imaginary-time Green's functions}
\begin{figure}
    \centering
    \vspace{-1mm} 
    \includegraphics[width=1.0\linewidth]{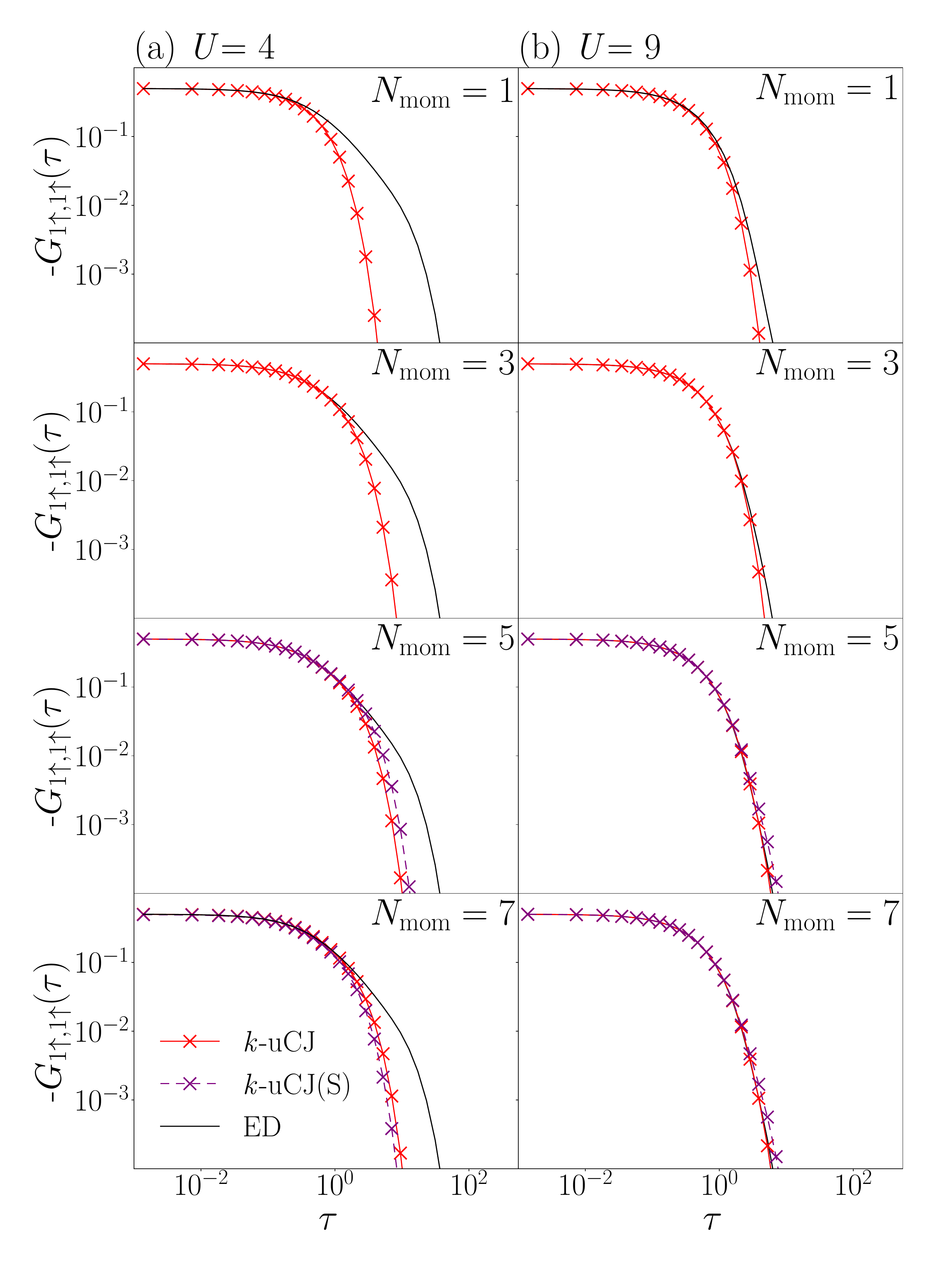}
    \vspace{-3mm}
    \vspace{-1mm} 
    \caption{
    Computed $G_{1\uparrow,1\uparrow}(\tau)$  for each $N_{\mathrm{mom}}$.
    Panels (a) and (b) show the results for $V=0.5$ and $V=0.1$, respectively.
    In the $k$-uCJ and the $k$-uCJ(S), we set $k=5$.
    ED refers to the spectral functions constructed from exact moments computed by exact diagonalization.
    The black vertical lines in panel (a) for $N_{\mathrm{mom}} = 7$ show where the deviation of the reconstructed spectral functions from the reference data starts.
    }
    \label{fig:8site_G_all}
\end{figure}

Figures~\ref{fig:8site_G_all}(a) and (b) show the imaginary-time Green's functions computed from the moment expansion for $V=0.5$ and $V=0.1$, respectively, with the $k$-uCJ ansatz with $k=5$. 
We computed the reference data from the reconstructed spectral function using exact moments for each $N_\mathrm{mom}$.
In this computation, we removed the physically irrelevant peaks below $\omega = 10^{-2}$ in the spectrum (see the inset of \ref{fig:two_site_imp_sp}).
In Fig.~\ref{fig:8site_G_all}, for both  $V=0.5$ and $ V=0.1$, the differences among the ansatzes become less pronounced in imaginary-time Green's functions compared to the cases of spectral functions. 
In Fig.~\ref{fig:8site_G_all}(a), for $V=0.5$, the imaginary-time Green's functions exhibit a power-law decay. 
We observed no improvement by increasing $N_{\mathrm{P}}$, likely due to the fitting error in computing spectral moments.
In Fig.~\ref{fig:8site_G_all}(b), for $V=0.1$, imaginary-time Green's functions exhibit an exponential decay.
The results with $N_\mathrm{mom} = 5$ agree with the reference data.

\section{Finite shot simulations}\label{sec:shotnoise}\label{section:noise}

This section investigates the effects of shot noise for the single-site impurity model with $N_{\mathrm{bath}}=3$.
We first optimize variational parameters for the ground state and the intermediate states in the computation of the spectral moments [Eqs.~\eqref{eq:ex_state}, \eqref{eq:h_ex_state}] using state vector simulations as detailed in Sec.~\ref{sec:state_vector_simulation}.
Then, we measure the expectation values of the Hamiltonian and the transition amplitude \eqref{eq:moments_fitting} for each order of the moment $m$ with a finite number of measurements.
It should be noted that the effect of the shot noise was not considered during the optimization steps.
This noise affects the measured scalar values, the ground-state energy $E_{\mathrm{G}}$ and coefficients $d_{0}, d_{1}, \cdots, d_{\mathrm{mom}}$ in the recursive approach [Eq.~\eqref{eq:moments_fitting}].
We set the number of measurements to 30,000.

\subsection{Ground-state calculation}
\begin{figure}
    \centering
    \vspace{-1mm} 
    \includegraphics[width=0.85\linewidth]{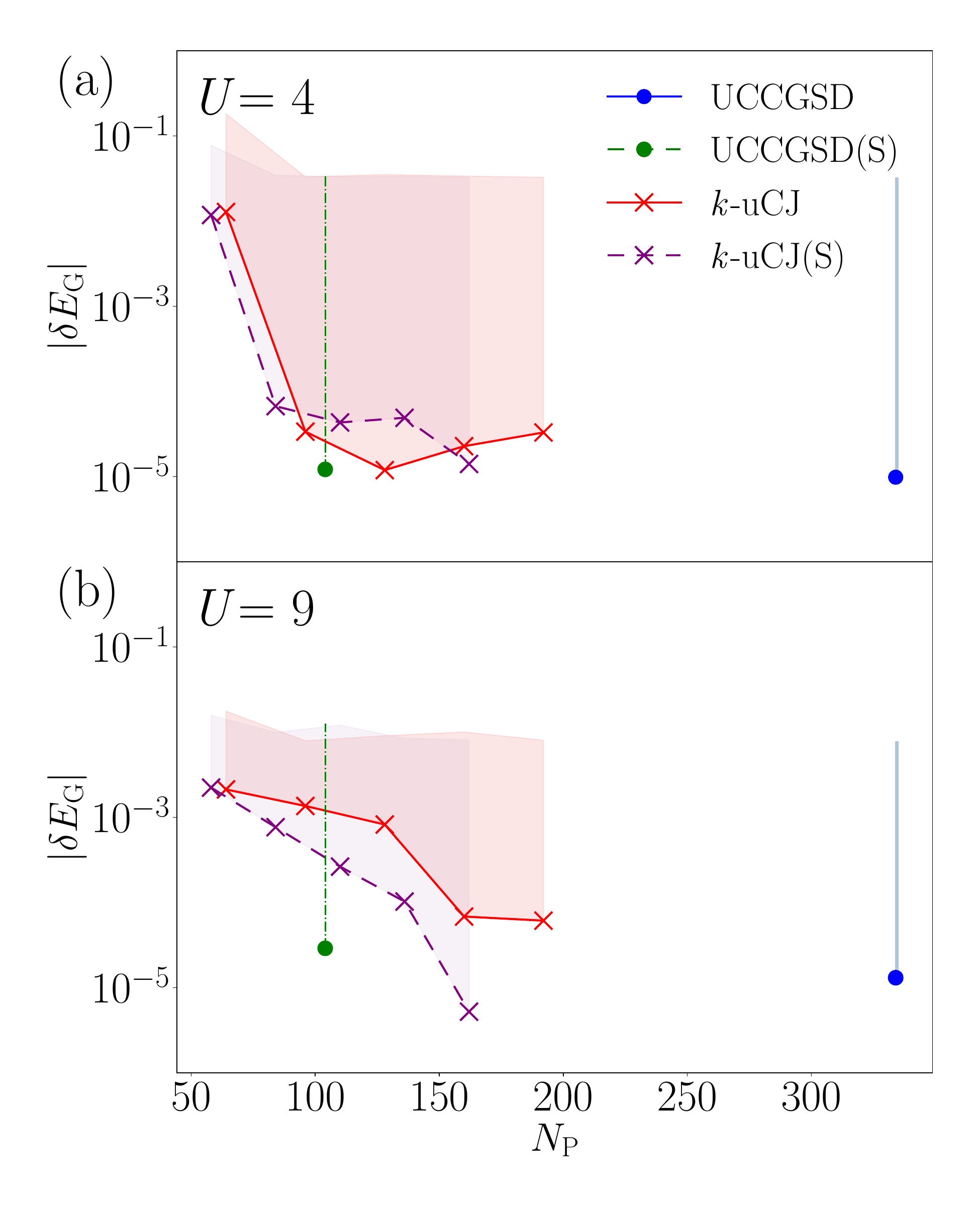}
    \vspace{-3mm}
    \vspace{-1mm} 
    \caption{Computed $|\delta E_\mathrm{G}|$ with 30000 measurements for the single-site impurity model.    
    In the $k$-uCJ and the $k$-uCJ(S), $k$ was varied from 1 to 5.
    The markers represent the best result obtained by varying the initial parameters 50 times. 
    The lightly shaded areas in the figure illustrate the dependency of the absolute errors on the initial parameters.
    Panels (a) and (b) show the results for $U=4$ and $U=9$, respectively.
    }
    \label{fig:4site_gs_all_noise}
\end{figure}

Figures~\ref{fig:4site_gs_all_noise}(a) and (b) show $|\delta E_{\mathrm{G}}|$ computed with shot noise for $U=4$ and $U=9$, respectively. 
The markers represent the best results obtained by varying the initial
parameters 50 times for each ansatz.
As indicated by the shaded area, the issue of initial parameter dependency remains significant in the presence of shot noise.

In all four ansatzes, statistical errors with a finite number of measurements reduce the overall accuracy compared to the results without shot noise (see Fig.~\ref{fig:4site_gs_all}).
Still, the ground-state energies can be reproduced with comparative accuracy among ansatzes.
The results for the sparse ansatzes in Figs.~\ref{fig:4site_gs_all_noise}(a) and (b) show that reducing the variational parameters associated with bath sites does not compromise the accuracy of $E_\mathrm{G}$.
The accuracy of the $k$-uCJ (S) is lower than that of the $k$-uCJ for the \textit{metallic} system ($U=4$), which may be attributed to statistical error.

\subsection{\rm Spectral functions}
\begin{figure}
    \centering
    \vspace{-1mm} 
    \includegraphics[width=0.85\linewidth]{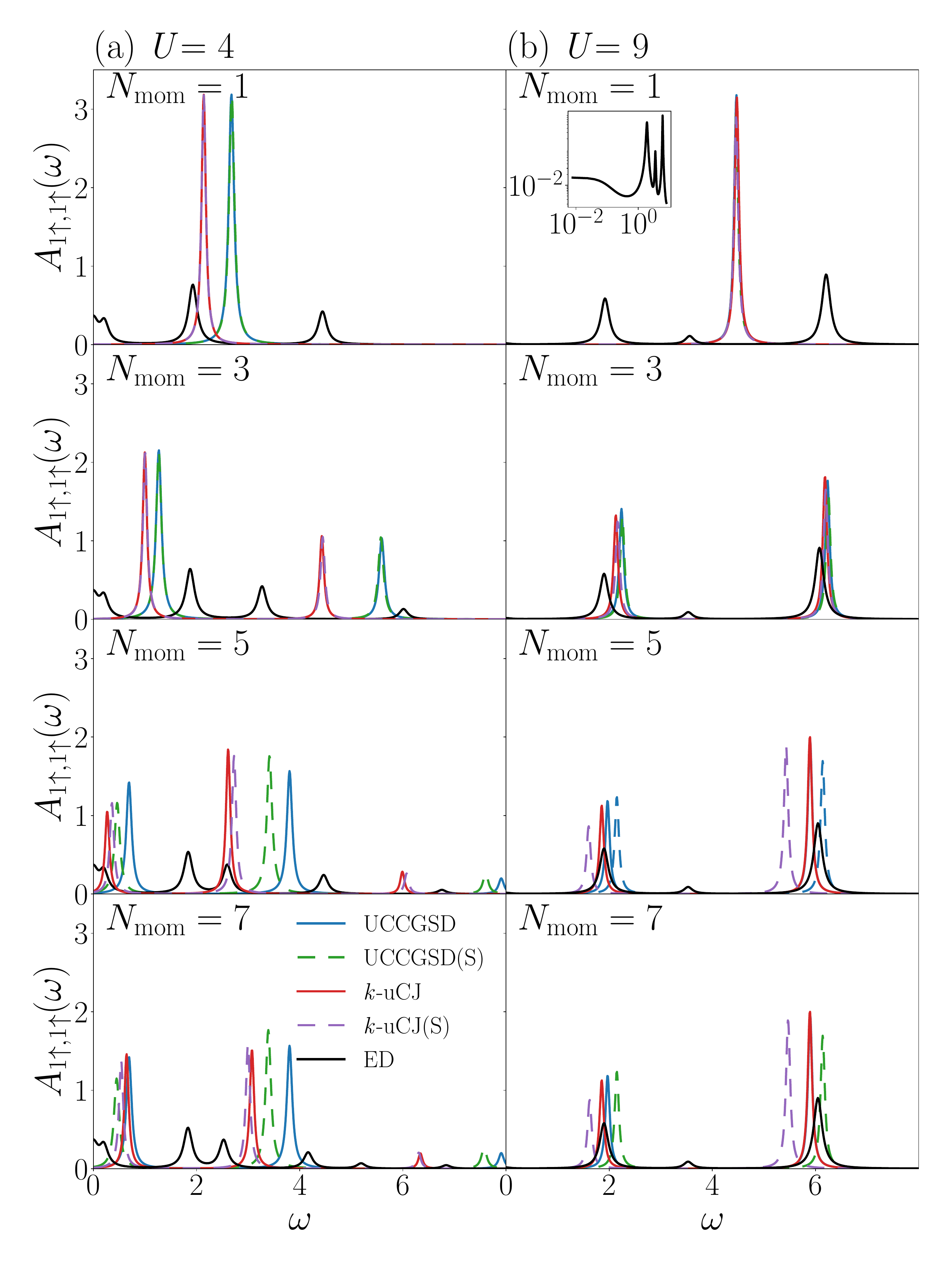}
    \vspace{-3mm}
    \vspace{-1mm} 
    \caption{
    Computed $A_{\mathrm{1 \uparrow, 1 \uparrow}}(\omega)$ with 30000 measurements for each $N_{\mathrm{mom}}$.
    In the $k$-uCJ and the $k$-uCJ(S), we set $k=5$.
    ED refers to the spectral functions constructed from exact moments using exact diagonalization.
    Panels (a) and (b) show the results for $U=4$ and $U=9$, respectively.}
    \label{4sitedmft_sp_noise}
\end{figure}

Figures~\ref{4sitedmft_sp_noise}(a) and (b) show the reconstructed spectral functions using the spectral moment computed with shot noise for $U=4$ and $U=9$, respectively.
We set $k=5$ in the $k$-uCJ. 
In Fig.~\ref{4sitedmft_sp_noise}(a), for $U=4$, none of the ansatzes reconstruct the spectral peaks.
These discrepancies primarily stem from numerical errors in the moment calculations. 
It should be noted that reconstructing a spectral function from the moments is not a well-conditioned problem (although more robust than traditional numerical analytic continuation from imaginary time due to the analytic procedure). 
Specifically, in the shot noise simulation, such errors are attributed to statistical error, the limited representational capability of the ansatzes, and optimization issues. 
The effect of statistical noise is dominant when comparing the calculation results to the case without shot noise Fig.~\ref{fig:4site_spectrum}.
In Fig.~\ref{4sitedmft_sp_noise}(b), for $U=9$, the shot noise induces small shifts in the positions of several peaks for $\omega \lesssim 6$ compared to the results computed without the shot noise. 
There are some variations among the ansatzes, likely due to the fitting error, but generally, the agreement is much improved compared to the more metallic $U=4$ results.

\subsection{\rm Imaginary-time Green's functions}
\begin{figure}
    \centering
    \vspace{-1mm} 
    \includegraphics[width=0.85\linewidth]{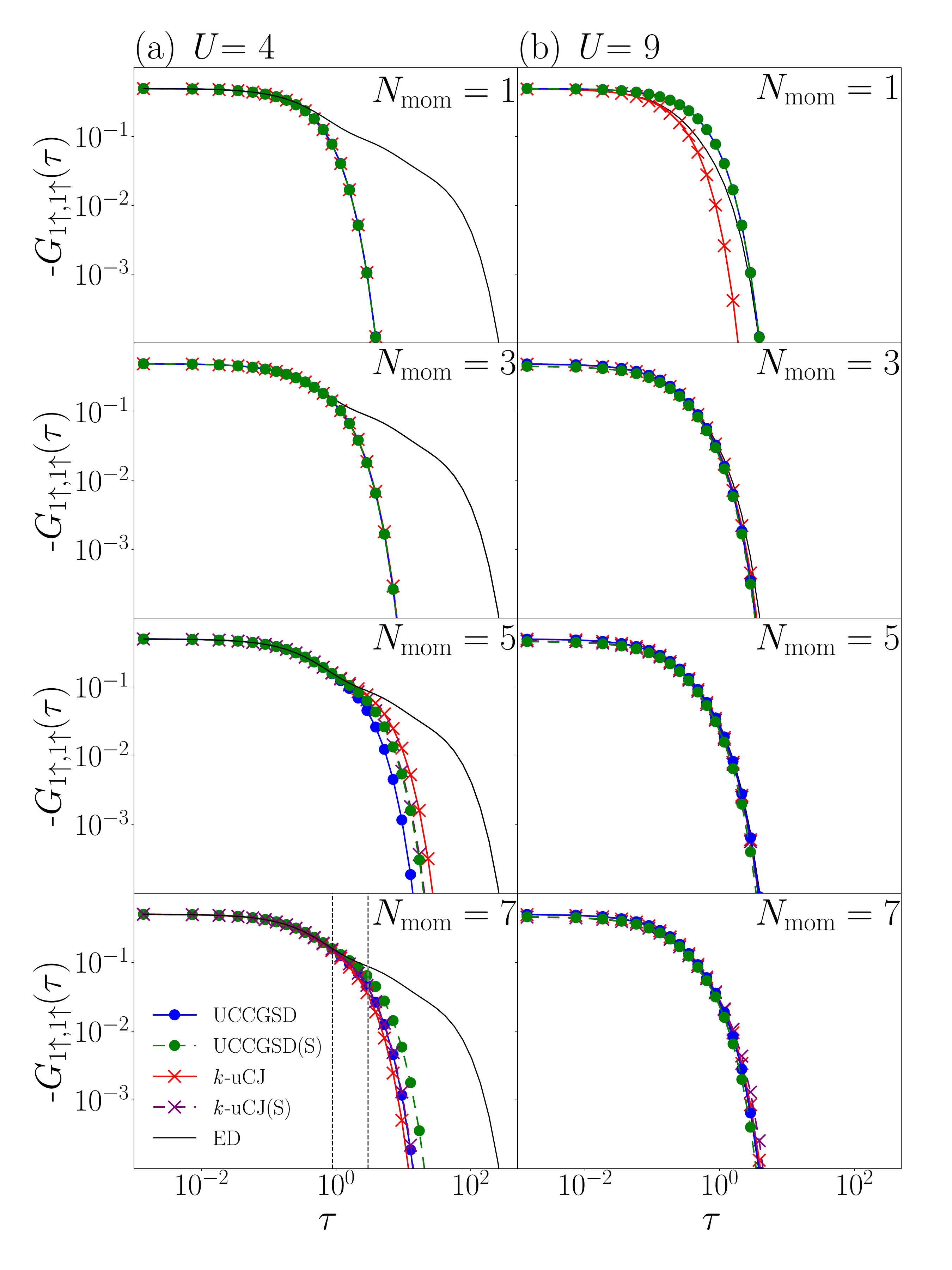}
    \vspace{-3mm}
    \vspace{-1mm} 
    \caption{
    Computed $G_{1\uparrow,1\uparrow}(\tau)$ with 30000 measurementsfor each $N_{\mathrm{mom}}$.
    Panels (a) and (b) show the results for $U=4$ and $U=9$, respectively.
    In the $k$-uCJ and the $k$-uCJ(S), we set $k=5$.
    ED refers to the spectral functions constructed from exact moments using exact diagonalization.
    The black vertical lines in panel (a) for $N_{\mathrm{mom}}$ = 7 indicate where the reconstructed spectral functions with shot noise begin to differ from those derived from exact moments.
    The gray line indicates the case without shot noise (see Fig.~\ref{fig:4site_G}).
    }
    \label{4sitedmft_G_all_noise}
\end{figure}

We now compute the imaginary-time Green's functions from the reconstructed spectral functions by the moment expansion with the shot noise.
Figures~\ref{4sitedmft_G_all_noise}(a) and (b) show the results for $U=4$ and 9, respectively.
In the $k$-uCJ, we set $k=5$.

For both $U=4$ and $U=9$, despite the large deviations in the spectral functions due to the fitting error, these variations are suppressed in the reconstructed imaginary-time Green's functions.
The results from all the ansatzes are consistent up to $\tau \approx 1$; then they start to deviate.
This is because the imaginary-time Green's function is insensitive to changes in the associated spectral function.
In Fig.~\ref{4sitedmft_G_all_noise}(a), for $U=4$ with $N_{\mathrm{mom}}=7$, due to the shot noise, the black vertical line at $\tau=1$ marks the earlier start of deviation, while the gray vertical line at $\tau=5$ indicates the start without shot noise (see Fig.~\ref{fig:4site_G}).
In Fig.~\ref{4sitedmft_G_all_noise}(b), for $U=4$ with $N_{\mathrm{mom}}=5$, the results with shot noise are in good agreement with the reference data.
These results indicate the moment expansion can successfully calculate the imaginary-time Green's functions under the influence of shot noise.
The imaginary-time Green's function, as calculated in this way, is sufficient for performing self-consistent calculations of DMFT. 
After convergence, some quantities computed from the imaginary-time Green's function (e.g., electron occupancy) are expected to be less sensitive to noise than real-frequency spectral functions.

\section{Summary and Discussion}\label{sec:discussion}
In this paper, we proposed compact quantum circuits for quantum impurity models with a star-like bath geometry by sparsifying the UCCGSD and $k$-uCJ ansatz. 
These forms have a significant parameter scaling of $N^{4}_{\mathrm{SO}}$ and $N^{2}_{\mathrm{SO}}$ respectively, which are reduced by removing insignificant variational parameters associated with two-body coupling between bath sites. 
This results in a reduced number of variational parameters scaling as $O(N^{4}_{\mathrm{imp}})$ and $O(N^{2}_{\mathrm{imp}})$ for the UCCGSD(S) and $k$-uCJ(S) ansatz respectively.
We numerically demonstrated that the compact ansatzes can accurately reproduce the ground-state energies for typical quantum impurity models, with and without shot noise. 
In the moment calculations for dynamic quantities, to avoid measuring more Pauli-operator terms at higher orders, we proposed a recursive method similar to VQE.
We also demonstrated that, when combined with the suggested ansatzes, the moment expansion effectively computes the imaginary-time Green's function, even in the presence of shot noise.

Before concluding this paper, we consider the proposed ansatzes and the spectral moments in the context of other approaches. 
A previous study utilized an adaptive variational quantum eigensolver (ADAPT-VQE) for impurity models~\cite{gomes2021adaptive, mukherjee2022comparative, Mukherjee_2023}. 
While ADAPT-VQE can provide near-exact solutions with a deep circuit, it demands more measurements for gradient computation than traditional VQE. 
Also, its success depends on the selected operator pool, which makes it hard to compare it to other approaches.
In addition, it is instructive to compare the moment expansion to alternative approaches such as the VQS approach~\cite{PhysRevResearch.4.023219}, with which the method bears many similarities. 
The moment expansion preserves the causal nature of the spectral functions; however, it encounters growing fitting errors in the recursive approach, most significantly in metallic systems. 
The VQS method might handle these systems more effectively via time evolution over a longer time span. Still, it could be costly since it requires computing all variational parameters at every time step.
A more detailed comparison is left in future studies.

Finally, we discuss the potential future research directions.
The initial parameter selection plays a crucial role in the accuracy of ground-state energies and moments.
Specifically, the accuracy of the moment is closely tied to that of the ground state.
The selection of optimal initial parameters to avoid local minima should be a critical area of future studies.
Moreover, minimizing the number of measurements in VQE and the recursive approach is essential in the context of utilizing near-term quantum devices. 
One viable solution is the efficient grouping of observables for simultaneous measurement~\cite{Gokhale2019MinimizingSP}. 
It is also important to investigate how the noise in the measured Green's function and discretization errors of the bath propagate during self-consistent calculations in DMFT and affect quantities of interest, e.g., momentum-resolved spectrum.
Developing methods for suppressing such errors is crucial.
When optimizing in VQE under the presence of noise, it is crucial to employ optimization methods resilient to noise~\cite{kingma2017adam, PhysRevResearch.2.043158}.
At the same time, noise mitigation techniques~\cite{rogers2021error, endo2018practical} are essential. 
Furthermore, the potential applicability of sparse ansatzes to other impurity models with star-like geometry, such as multi-orbital systems,  requires further investigation.
Lastly, incorporating the concept of sparsity into classical variational algorithmic approaches, such as machine learning wave functions~\cite{PMID:28183973, RevModPhys.91.045002}, may improve computational efficiency.

\begin{acknowledgments}
H.S. was supported by JSPS KAKENHI Grants No.18H01158, No. 21H01041, No. 21H01003, and No. 23H03817, Japan.
R.S. and H.S. were supported by JST PRESTO Grant No. JPMJPR2012, Japan.
W.M. was supported by funding from the MEXT Quantum Leap Flagship Program (MEXTQLEAP) through Grant No. JPMXS0120319794, and the JST COI-NEXT Program through Grant No. JPMJPF2014. 
W.M. also acknowledges support from the JSPS KAKENHI Grant Nos. JP23H03819 and JP21K18933. 
G.H.B. and O.J.B. were supported by funding from European Union's Horizon 2020 research and innovation programme under grant agreement No. 759063, as well as EPSRC grants EP/Y005244/1 (VSL-Q) and EP/W032635/1 (RoaRQ).
We thank the Supercomputer Center, the Institute for Solid State Physics, the University of Tokyo for the use of the facilities.
\end{acknowledgments}

\bibliography{ref}

\begin{thebibliography}{86}%
\makeatletter
\providecommand \@ifxundefined [1]{%
 \@ifx{#1\undefined}
}%
\providecommand \@ifnum [1]{%
 \ifnum #1\expandafter \@firstoftwo
 \else \expandafter \@secondoftwo
 \fi
}%
\providecommand \@ifx [1]{%
 \ifx #1\expandafter \@firstoftwo
 \else \expandafter \@secondoftwo
 \fi
}%
\providecommand \natexlab [1]{#1}%
\providecommand \enquote  [1]{``#1''}%
\providecommand \bibnamefont  [1]{#1}%
\providecommand \bibfnamefont [1]{#1}%
\providecommand \citenamefont [1]{#1}%
\providecommand \href@noop [0]{\@secondoftwo}%
\providecommand \href [0]{\begingroup \@sanitize@url \@href}%
\providecommand \@href[1]{\@@startlink{#1}\@@href}%
\providecommand \@@href[1]{\endgroup#1\@@endlink}%
\providecommand \@sanitize@url [0]{\catcode `\\12\catcode `\$12\catcode
  `\&12\catcode `\#12\catcode `\^12\catcode `\_12\catcode `\%12\relax}%
\providecommand \@@startlink[1]{}%
\providecommand \@@endlink[0]{}%
\providecommand \url  [0]{\begingroup\@sanitize@url \@url }%
\providecommand \@url [1]{\endgroup\@href {#1}{\urlprefix }}%
\providecommand \urlprefix  [0]{URL }%
\providecommand \Eprint [0]{\href }%
\providecommand \doibase [0]{https://doi.org/}%
\providecommand \selectlanguage [0]{\@gobble}%
\providecommand \bibinfo  [0]{\@secondoftwo}%
\providecommand \bibfield  [0]{\@secondoftwo}%
\providecommand \translation [1]{[#1]}%
\providecommand \BibitemOpen [0]{}%
\providecommand \bibitemStop [0]{}%
\providecommand \bibitemNoStop [0]{.\EOS\space}%
\providecommand \EOS [0]{\spacefactor3000\relax}%
\providecommand \BibitemShut  [1]{\csname bibitem#1\endcsname}%
\let\auto@bib@innerbib\@empty
\bibitem [{\citenamefont {{Leggett}}(2006)}]{2006NatPh...2..134L}%
  \BibitemOpen
  \bibfield  {author} {\bibinfo {author} {\bibfnamefont {A.~J.}\ \bibnamefont
  {{Leggett}}},\ }\bibfield  {title} {\bibinfo {title} {{What DO we know about
  high T$_{c}$?}},\ }\href {https://doi.org/10.1038/nphys254} {\bibfield
  {journal} {\bibinfo  {journal} {Nature Physics}\ }\textbf {\bibinfo {volume}
  {2}},\ \bibinfo {pages} {134} (\bibinfo {year} {2006})}\BibitemShut {NoStop}%
\bibitem [{\citenamefont {Keimer}\ \emph {et~al.}(2015)\citenamefont {Keimer},
  \citenamefont {Kivelson}, \citenamefont {Norman}, \citenamefont {Uchida},\
  and\ \citenamefont
  {Zaanen}}]{RePEc:nat:nature:v:518:y:2015:i:7538:d:10.1038_nature14165}%
  \BibitemOpen
  \bibfield  {author} {\bibinfo {author} {\bibfnamefont {B.}~\bibnamefont
  {Keimer}}, \bibinfo {author} {\bibfnamefont {S.~A.}\ \bibnamefont
  {Kivelson}}, \bibinfo {author} {\bibfnamefont {M.~R.}\ \bibnamefont
  {Norman}}, \bibinfo {author} {\bibfnamefont {S.}~\bibnamefont {Uchida}},\
  and\ \bibinfo {author} {\bibfnamefont {J.}~\bibnamefont {Zaanen}},\
  }\bibfield  {title} {\bibinfo {title} {From quantum matter to
  high-temperature superconductivity in copper oxides},\ }\href
  {https://EconPapers.repec.org/RePEc:nat:nature:v:518:y:2015:i:7538:d:10.1038_nature14165}
  {\bibfield  {journal} {\bibinfo  {journal} {Nature}\ }\textbf {\bibinfo
  {volume} {518}},\ \bibinfo {pages} {179} (\bibinfo {year}
  {2015})}\BibitemShut {NoStop}%
\bibitem [{\citenamefont {Georges}\ \emph {et~al.}(1996)\citenamefont
  {Georges}, \citenamefont {Kotliar}, \citenamefont {Krauth},\ and\
  \citenamefont {Rozenberg}}]{georges1996dynamical}%
  \BibitemOpen
  \bibfield  {author} {\bibinfo {author} {\bibfnamefont {A.}~\bibnamefont
  {Georges}}, \bibinfo {author} {\bibfnamefont {G.}~\bibnamefont {Kotliar}},
  \bibinfo {author} {\bibfnamefont {W.}~\bibnamefont {Krauth}},\ and\ \bibinfo
  {author} {\bibfnamefont {M.~J.}\ \bibnamefont {Rozenberg}},\ }\bibfield
  {title} {\bibinfo {title} {Dynamical mean-field theory of strongly correlated
  fermion systems and the limit of infinite dimensions},\ }\href@noop {}
  {\bibfield  {journal} {\bibinfo  {journal} {Reviews of Modern Physics}\
  }\textbf {\bibinfo {volume} {68}},\ \bibinfo {pages} {13} (\bibinfo {year}
  {1996})}\BibitemShut {NoStop}%
\bibitem [{\citenamefont {Kotliar}\ \emph {et~al.}(2006)\citenamefont
  {Kotliar}, \citenamefont {Savrasov}, \citenamefont {Haule}, \citenamefont
  {Oudovenko}, \citenamefont {Parcollet},\ and\ \citenamefont
  {Marianetti}}]{kotliar2006electronic}%
  \BibitemOpen
  \bibfield  {author} {\bibinfo {author} {\bibfnamefont {G.}~\bibnamefont
  {Kotliar}}, \bibinfo {author} {\bibfnamefont {S.~Y.}\ \bibnamefont
  {Savrasov}}, \bibinfo {author} {\bibfnamefont {K.}~\bibnamefont {Haule}},
  \bibinfo {author} {\bibfnamefont {V.~S.}\ \bibnamefont {Oudovenko}}, \bibinfo
  {author} {\bibfnamefont {O.}~\bibnamefont {Parcollet}},\ and\ \bibinfo
  {author} {\bibfnamefont {C.}~\bibnamefont {Marianetti}},\ }\bibfield  {title}
  {\bibinfo {title} {Electronic structure calculations with dynamical
  mean-field theory},\ }\href@noop {} {\bibfield  {journal} {\bibinfo
  {journal} {Reviews of Modern Physics}\ }\textbf {\bibinfo {volume} {78}},\
  \bibinfo {pages} {865} (\bibinfo {year} {2006})}\BibitemShut {NoStop}%
\bibitem [{\citenamefont {Knizia}\ and\ \citenamefont
  {Chan}(2012)}]{PhysRevLett.109.186404}%
  \BibitemOpen
  \bibfield  {author} {\bibinfo {author} {\bibfnamefont {G.}~\bibnamefont
  {Knizia}}\ and\ \bibinfo {author} {\bibfnamefont {G.~K.-L.}\ \bibnamefont
  {Chan}},\ }\bibfield  {title} {\bibinfo {title} {Density matrix embedding: A
  simple alternative to dynamical mean-field theory},\ }\href
  {https://doi.org/10.1103/PhysRevLett.109.186404} {\bibfield  {journal}
  {\bibinfo  {journal} {Phys. Rev. Lett.}\ }\textbf {\bibinfo {volume} {109}},\
  \bibinfo {pages} {186404} (\bibinfo {year} {2012})}\BibitemShut {NoStop}%
\bibitem [{\citenamefont {Scott}\ and\ \citenamefont
  {Booth}(2021)}]{PhysRevB.104.245114}%
  \BibitemOpen
  \bibfield  {author} {\bibinfo {author} {\bibfnamefont {C.~J.~C.}\
  \bibnamefont {Scott}}\ and\ \bibinfo {author} {\bibfnamefont {G.~H.}\
  \bibnamefont {Booth}},\ }\bibfield  {title} {\bibinfo {title} {Extending
  density matrix embedding: A static two-particle theory},\ }\href
  {https://doi.org/10.1103/PhysRevB.104.245114} {\bibfield  {journal} {\bibinfo
   {journal} {Phys. Rev. B}\ }\textbf {\bibinfo {volume} {104}},\ \bibinfo
  {pages} {245114} (\bibinfo {year} {2021})}\BibitemShut {NoStop}%
\bibitem [{\citenamefont {Linden}\ \emph {et~al.}(2020)\citenamefont {Linden},
  \citenamefont {Zingl}, \citenamefont {Hubig}, \citenamefont {Parcollet},\
  and\ \citenamefont {Schollw\"ock}}]{PhysRevB.101.041101}%
  \BibitemOpen
  \bibfield  {author} {\bibinfo {author} {\bibfnamefont {N.-O.}\ \bibnamefont
  {Linden}}, \bibinfo {author} {\bibfnamefont {M.}~\bibnamefont {Zingl}},
  \bibinfo {author} {\bibfnamefont {C.}~\bibnamefont {Hubig}}, \bibinfo
  {author} {\bibfnamefont {O.}~\bibnamefont {Parcollet}},\ and\ \bibinfo
  {author} {\bibfnamefont {U.}~\bibnamefont {Schollw\"ock}},\ }\bibfield
  {title} {\bibinfo {title} {Imaginary-time matrix product state impurity
  solver in a real material calculation: Spin-orbit coupling in
  $\mathrm{Sr}{}_{2}\mathrm{RuO}{}_{4}$},\ }\href
  {https://doi.org/10.1103/PhysRevB.101.041101} {\bibfield  {journal} {\bibinfo
   {journal} {Phys. Rev. B}\ }\textbf {\bibinfo {volume} {101}},\ \bibinfo
  {pages} {041101} (\bibinfo {year} {2020})}\BibitemShut {NoStop}%
\bibitem [{\citenamefont {Wolf}\ \emph {et~al.}(2015)\citenamefont {Wolf},
  \citenamefont {Go}, \citenamefont {McCulloch}, \citenamefont {Millis},\ and\
  \citenamefont {Schollw{\"o}ck}}]{wolf2015imaginary}%
  \BibitemOpen
  \bibfield  {author} {\bibinfo {author} {\bibfnamefont {F.~A.}\ \bibnamefont
  {Wolf}}, \bibinfo {author} {\bibfnamefont {A.}~\bibnamefont {Go}}, \bibinfo
  {author} {\bibfnamefont {I.~P.}\ \bibnamefont {McCulloch}}, \bibinfo {author}
  {\bibfnamefont {A.~J.}\ \bibnamefont {Millis}},\ and\ \bibinfo {author}
  {\bibfnamefont {U.}~\bibnamefont {Schollw{\"o}ck}},\ }\bibfield  {title}
  {\bibinfo {title} {Imaginary-time matrix product state impurity solver for
  dynamical mean-field theory},\ }\href@noop {} {\bibfield  {journal} {\bibinfo
   {journal} {Physical Review X}\ }\textbf {\bibinfo {volume} {5}},\ \bibinfo
  {pages} {041032} (\bibinfo {year} {2015})}\BibitemShut {NoStop}%
\bibitem [{\citenamefont {Bauernfeind}\ \emph {et~al.}(2017)\citenamefont
  {Bauernfeind}, \citenamefont {Zingl}, \citenamefont {Triebl}, \citenamefont
  {Aichhorn},\ and\ \citenamefont {Evertz}}]{bauernfeind2017fork}%
  \BibitemOpen
  \bibfield  {author} {\bibinfo {author} {\bibfnamefont {D.}~\bibnamefont
  {Bauernfeind}}, \bibinfo {author} {\bibfnamefont {M.}~\bibnamefont {Zingl}},
  \bibinfo {author} {\bibfnamefont {R.}~\bibnamefont {Triebl}}, \bibinfo
  {author} {\bibfnamefont {M.}~\bibnamefont {Aichhorn}},\ and\ \bibinfo
  {author} {\bibfnamefont {H.~G.}\ \bibnamefont {Evertz}},\ }\bibfield  {title}
  {\bibinfo {title} {Fork tensor-product states: Efficient multiorbital
  real-time dmft solver},\ }\href@noop {} {\bibfield  {journal} {\bibinfo
  {journal} {Physical Review X}\ }\textbf {\bibinfo {volume} {7}},\ \bibinfo
  {pages} {031013} (\bibinfo {year} {2017})}\BibitemShut {NoStop}%
\bibitem [{\citenamefont {Gull}\ \emph {et~al.}(2011)\citenamefont {Gull},
  \citenamefont {Millis}, \citenamefont {Lichtenstein}, \citenamefont
  {Rubtsov}, \citenamefont {Troyer},\ and\ \citenamefont
  {Werner}}]{RevModPhys.83.349}%
  \BibitemOpen
  \bibfield  {author} {\bibinfo {author} {\bibfnamefont {E.}~\bibnamefont
  {Gull}}, \bibinfo {author} {\bibfnamefont {A.~J.}\ \bibnamefont {Millis}},
  \bibinfo {author} {\bibfnamefont {A.~I.}\ \bibnamefont {Lichtenstein}},
  \bibinfo {author} {\bibfnamefont {A.~N.}\ \bibnamefont {Rubtsov}}, \bibinfo
  {author} {\bibfnamefont {M.}~\bibnamefont {Troyer}},\ and\ \bibinfo {author}
  {\bibfnamefont {P.}~\bibnamefont {Werner}},\ }\bibfield  {title} {\bibinfo
  {title} {Continuous-time monte carlo methods for quantum impurity models},\
  }\href {https://doi.org/10.1103/RevModPhys.83.349} {\bibfield  {journal}
  {\bibinfo  {journal} {Rev. Mod. Phys.}\ }\textbf {\bibinfo {volume} {83}},\
  \bibinfo {pages} {349} (\bibinfo {year} {2011})}\BibitemShut {NoStop}%
\bibitem [{\citenamefont {Kitaev}(1995)}]{kitaev1995quantum}%
  \BibitemOpen
  \bibfield  {author} {\bibinfo {author} {\bibfnamefont {A.~Y.}\ \bibnamefont
  {Kitaev}},\ }\href@noop {} {\bibinfo {title} {Quantum measurements and the
  abelian stabilizer problem}} (\bibinfo {year} {1995}),\ \Eprint
  {https://arxiv.org/abs/quant-ph/9511026} {arXiv:quant-ph/9511026 [quant-ph]}
  \BibitemShut {NoStop}%
\bibitem [{\citenamefont {Aspuru-Guzik}\ \emph {et~al.}(2005)\citenamefont
  {Aspuru-Guzik}, \citenamefont {Dutoi}, \citenamefont {Love},\ and\
  \citenamefont {Head-Gordon}}]{doi:10.1126/science.1113479}%
  \BibitemOpen
  \bibfield  {author} {\bibinfo {author} {\bibfnamefont {A.}~\bibnamefont
  {Aspuru-Guzik}}, \bibinfo {author} {\bibfnamefont {A.~D.}\ \bibnamefont
  {Dutoi}}, \bibinfo {author} {\bibfnamefont {P.~J.}\ \bibnamefont {Love}},\
  and\ \bibinfo {author} {\bibfnamefont {M.}~\bibnamefont {Head-Gordon}},\
  }\bibfield  {title} {\bibinfo {title} {Simulated quantum computation of
  molecular energies},\ }\href {https://doi.org/10.1126/science.1113479}
  {\bibfield  {journal} {\bibinfo  {journal} {Science}\ }\textbf {\bibinfo
  {volume} {309}},\ \bibinfo {pages} {1704} (\bibinfo {year} {2005})},\ \Eprint
  {https://arxiv.org/abs/https://www.science.org/doi/pdf/10.1126/science.1113479}
  {https://www.science.org/doi/pdf/10.1126/science.1113479} \BibitemShut
  {NoStop}%
\bibitem [{\citenamefont {Farhi}\ \emph {et~al.}(2000)\citenamefont {Farhi},
  \citenamefont {Goldstone}, \citenamefont {Gutmann},\ and\ \citenamefont
  {Sipser}}]{farhi2000quantum}%
  \BibitemOpen
  \bibfield  {author} {\bibinfo {author} {\bibfnamefont {E.}~\bibnamefont
  {Farhi}}, \bibinfo {author} {\bibfnamefont {J.}~\bibnamefont {Goldstone}},
  \bibinfo {author} {\bibfnamefont {S.}~\bibnamefont {Gutmann}},\ and\ \bibinfo
  {author} {\bibfnamefont {M.}~\bibnamefont {Sipser}},\ }\href@noop {}
  {\bibinfo {title} {Quantum computation by adiabatic evolution}} (\bibinfo
  {year} {2000}),\ \Eprint {https://arxiv.org/abs/quant-ph/0001106}
  {arXiv:quant-ph/0001106 [quant-ph]} \BibitemShut {NoStop}%
\bibitem [{\citenamefont {Albash}\ and\ \citenamefont
  {Lidar}(2018)}]{RevModPhys.90.015002}%
  \BibitemOpen
  \bibfield  {author} {\bibinfo {author} {\bibfnamefont {T.}~\bibnamefont
  {Albash}}\ and\ \bibinfo {author} {\bibfnamefont {D.~A.}\ \bibnamefont
  {Lidar}},\ }\bibfield  {title} {\bibinfo {title} {Adiabatic quantum
  computation},\ }\href {https://doi.org/10.1103/RevModPhys.90.015002}
  {\bibfield  {journal} {\bibinfo  {journal} {Rev. Mod. Phys.}\ }\textbf
  {\bibinfo {volume} {90}},\ \bibinfo {pages} {015002} (\bibinfo {year}
  {2018})}\BibitemShut {NoStop}%
\bibitem [{\citenamefont {Bauer}\ \emph {et~al.}(2016)\citenamefont {Bauer},
  \citenamefont {Wecker}, \citenamefont {Millis}, \citenamefont {Hastings},\
  and\ \citenamefont {Troyer}}]{bauer2016hybrid}%
  \BibitemOpen
  \bibfield  {author} {\bibinfo {author} {\bibfnamefont {B.}~\bibnamefont
  {Bauer}}, \bibinfo {author} {\bibfnamefont {D.}~\bibnamefont {Wecker}},
  \bibinfo {author} {\bibfnamefont {A.~J.}\ \bibnamefont {Millis}}, \bibinfo
  {author} {\bibfnamefont {M.~B.}\ \bibnamefont {Hastings}},\ and\ \bibinfo
  {author} {\bibfnamefont {M.}~\bibnamefont {Troyer}},\ }\bibfield  {title}
  {\bibinfo {title} {Hybrid quantum-classical approach to correlated
  materials},\ }\href@noop {} {\bibfield  {journal} {\bibinfo  {journal}
  {Physical Review X}\ }\textbf {\bibinfo {volume} {6}},\ \bibinfo {pages}
  {031045} (\bibinfo {year} {2016})}\BibitemShut {NoStop}%
\bibitem [{\citenamefont {Peruzzo}\ \emph {et~al.}(2014)\citenamefont
  {Peruzzo}, \citenamefont {McClean}, \citenamefont {Shadbolt}, \citenamefont
  {Yung}, \citenamefont {Zhou}, \citenamefont {Love}, \citenamefont
  {Aspuru-Guzik},\ and\ \citenamefont {O’brien}}]{peruzzo2014variational}%
  \BibitemOpen
  \bibfield  {author} {\bibinfo {author} {\bibfnamefont {A.}~\bibnamefont
  {Peruzzo}}, \bibinfo {author} {\bibfnamefont {J.}~\bibnamefont {McClean}},
  \bibinfo {author} {\bibfnamefont {P.}~\bibnamefont {Shadbolt}}, \bibinfo
  {author} {\bibfnamefont {M.-H.}\ \bibnamefont {Yung}}, \bibinfo {author}
  {\bibfnamefont {X.-Q.}\ \bibnamefont {Zhou}}, \bibinfo {author}
  {\bibfnamefont {P.~J.}\ \bibnamefont {Love}}, \bibinfo {author}
  {\bibfnamefont {A.}~\bibnamefont {Aspuru-Guzik}},\ and\ \bibinfo {author}
  {\bibfnamefont {J.~L.}\ \bibnamefont {O’brien}},\ }\bibfield  {title}
  {\bibinfo {title} {A variational eigenvalue solver on a photonic quantum
  processor},\ }\href@noop {} {\bibfield  {journal} {\bibinfo  {journal}
  {Nature communications}\ }\textbf {\bibinfo {volume} {5}},\ \bibinfo {pages}
  {4213} (\bibinfo {year} {2014})}\BibitemShut {NoStop}%
\bibitem [{\citenamefont {McArdle}\ \emph {et~al.}(2019)\citenamefont
  {McArdle}, \citenamefont {Jones}, \citenamefont {Endo}, \citenamefont {Li},
  \citenamefont {Benjamin},\ and\ \citenamefont
  {Yuan}}]{mcardle2019variational}%
  \BibitemOpen
  \bibfield  {author} {\bibinfo {author} {\bibfnamefont {S.}~\bibnamefont
  {McArdle}}, \bibinfo {author} {\bibfnamefont {T.}~\bibnamefont {Jones}},
  \bibinfo {author} {\bibfnamefont {S.}~\bibnamefont {Endo}}, \bibinfo {author}
  {\bibfnamefont {Y.}~\bibnamefont {Li}}, \bibinfo {author} {\bibfnamefont
  {S.~C.}\ \bibnamefont {Benjamin}},\ and\ \bibinfo {author} {\bibfnamefont
  {X.}~\bibnamefont {Yuan}},\ }\bibfield  {title} {\bibinfo {title}
  {Variational ansatz-based quantum simulation of imaginary time evolution},\
  }\href@noop {} {\bibfield  {journal} {\bibinfo  {journal} {npj Quantum
  Information}\ }\textbf {\bibinfo {volume} {5}},\ \bibinfo {pages} {75}
  (\bibinfo {year} {2019})}\BibitemShut {NoStop}%
\bibitem [{\citenamefont {Preskill}(2018)}]{preskill2018quantum}%
  \BibitemOpen
  \bibfield  {author} {\bibinfo {author} {\bibfnamefont {J.}~\bibnamefont
  {Preskill}},\ }\bibfield  {title} {\bibinfo {title} {Quantum computing in the
  nisq era and beyond},\ }\href@noop {} {\bibfield  {journal} {\bibinfo
  {journal} {Quantum}\ }\textbf {\bibinfo {volume} {2}},\ \bibinfo {pages} {79}
  (\bibinfo {year} {2018})}\BibitemShut {NoStop}%
\bibitem [{\citenamefont {Besserve}\ and\ \citenamefont
  {Ayral}(2022)}]{PhysRevB.105.115108}%
  \BibitemOpen
  \bibfield  {author} {\bibinfo {author} {\bibfnamefont {P.}~\bibnamefont
  {Besserve}}\ and\ \bibinfo {author} {\bibfnamefont {T.}~\bibnamefont
  {Ayral}},\ }\bibfield  {title} {\bibinfo {title} {Unraveling correlated
  material properties with noisy quantum computers: Natural orbitalized
  variational quantum eigensolving of extended impurity models within a
  slave-boson approach},\ }\href {https://doi.org/10.1103/PhysRevB.105.115108}
  {\bibfield  {journal} {\bibinfo  {journal} {Phys. Rev. B}\ }\textbf {\bibinfo
  {volume} {105}},\ \bibinfo {pages} {115108} (\bibinfo {year}
  {2022})}\BibitemShut {NoStop}%
\bibitem [{\citenamefont {Backes}\ \emph {et~al.}(2023)\citenamefont {Backes},
  \citenamefont {Murakami}, \citenamefont {Sakai},\ and\ \citenamefont
  {Arita}}]{PhysRevB.107.165155}%
  \BibitemOpen
  \bibfield  {author} {\bibinfo {author} {\bibfnamefont {S.}~\bibnamefont
  {Backes}}, \bibinfo {author} {\bibfnamefont {Y.}~\bibnamefont {Murakami}},
  \bibinfo {author} {\bibfnamefont {S.}~\bibnamefont {Sakai}},\ and\ \bibinfo
  {author} {\bibfnamefont {R.}~\bibnamefont {Arita}},\ }\bibfield  {title}
  {\bibinfo {title} {Dynamical mean-field theory for the hubbard-holstein model
  on a quantum device},\ }\href {https://doi.org/10.1103/PhysRevB.107.165155}
  {\bibfield  {journal} {\bibinfo  {journal} {Phys. Rev. B}\ }\textbf {\bibinfo
  {volume} {107}},\ \bibinfo {pages} {165155} (\bibinfo {year}
  {2023})}\BibitemShut {NoStop}%
\bibitem [{\citenamefont {Rungger}\ \emph {et~al.}(2020)\citenamefont
  {Rungger}, \citenamefont {Fitzpatrick}, \citenamefont {Chen}, \citenamefont
  {Alderete}, \citenamefont {Apel}, \citenamefont {Cowtan}, \citenamefont
  {Patterson}, \citenamefont {Ramo}, \citenamefont {Zhu}, \citenamefont
  {Nguyen}, \citenamefont {Grant}, \citenamefont {Chretien}, \citenamefont
  {Wossnig}, \citenamefont {Linke},\ and\ \citenamefont
  {Duncan}}]{rungger2020dynamical}%
  \BibitemOpen
  \bibfield  {author} {\bibinfo {author} {\bibfnamefont {I.}~\bibnamefont
  {Rungger}}, \bibinfo {author} {\bibfnamefont {N.}~\bibnamefont
  {Fitzpatrick}}, \bibinfo {author} {\bibfnamefont {H.}~\bibnamefont {Chen}},
  \bibinfo {author} {\bibfnamefont {C.~H.}\ \bibnamefont {Alderete}}, \bibinfo
  {author} {\bibfnamefont {H.}~\bibnamefont {Apel}}, \bibinfo {author}
  {\bibfnamefont {A.}~\bibnamefont {Cowtan}}, \bibinfo {author} {\bibfnamefont
  {A.}~\bibnamefont {Patterson}}, \bibinfo {author} {\bibfnamefont {D.~M.}\
  \bibnamefont {Ramo}}, \bibinfo {author} {\bibfnamefont {Y.}~\bibnamefont
  {Zhu}}, \bibinfo {author} {\bibfnamefont {N.~H.}\ \bibnamefont {Nguyen}},
  \bibinfo {author} {\bibfnamefont {E.}~\bibnamefont {Grant}}, \bibinfo
  {author} {\bibfnamefont {S.}~\bibnamefont {Chretien}}, \bibinfo {author}
  {\bibfnamefont {L.}~\bibnamefont {Wossnig}}, \bibinfo {author} {\bibfnamefont
  {N.~M.}\ \bibnamefont {Linke}},\ and\ \bibinfo {author} {\bibfnamefont
  {R.}~\bibnamefont {Duncan}},\ }\href@noop {} {\bibinfo {title} {Dynamical
  mean field theory algorithm and experiment on quantum computers}} (\bibinfo
  {year} {2020}),\ \Eprint {https://arxiv.org/abs/1910.04735} {arXiv:1910.04735
  [quant-ph]} \BibitemShut {NoStop}%
\bibitem [{\citenamefont {Greene-Diniz}\ \emph {et~al.}(2023)\citenamefont
  {Greene-Diniz}, \citenamefont {Manrique}, \citenamefont {Yamamoto},
  \citenamefont {Plekhanov}, \citenamefont {Fitzpatrick}, \citenamefont
  {Krompiec}, \citenamefont {Sakuma},\ and\ \citenamefont
  {Ramo}}]{greenediniz2023quantum}%
  \BibitemOpen
  \bibfield  {author} {\bibinfo {author} {\bibfnamefont {G.}~\bibnamefont
  {Greene-Diniz}}, \bibinfo {author} {\bibfnamefont {D.~Z.}\ \bibnamefont
  {Manrique}}, \bibinfo {author} {\bibfnamefont {K.}~\bibnamefont {Yamamoto}},
  \bibinfo {author} {\bibfnamefont {E.}~\bibnamefont {Plekhanov}}, \bibinfo
  {author} {\bibfnamefont {N.}~\bibnamefont {Fitzpatrick}}, \bibinfo {author}
  {\bibfnamefont {M.}~\bibnamefont {Krompiec}}, \bibinfo {author}
  {\bibfnamefont {R.}~\bibnamefont {Sakuma}},\ and\ \bibinfo {author}
  {\bibfnamefont {D.~M.}\ \bibnamefont {Ramo}},\ }\href@noop {} {\bibinfo
  {title} {Quantum computed green's functions using a cumulant expansion of the
  lanczos method}} (\bibinfo {year} {2023}),\ \Eprint
  {https://arxiv.org/abs/2309.09685} {arXiv:2309.09685 [cond-mat.str-el]}
  \BibitemShut {NoStop}%
\bibitem [{\citenamefont {Dhawan}\ \emph {et~al.}(2023)\citenamefont {Dhawan},
  \citenamefont {Zgid},\ and\ \citenamefont {Motta}}]{dhawan2023quantum}%
  \BibitemOpen
  \bibfield  {author} {\bibinfo {author} {\bibfnamefont {D.}~\bibnamefont
  {Dhawan}}, \bibinfo {author} {\bibfnamefont {D.}~\bibnamefont {Zgid}},\ and\
  \bibinfo {author} {\bibfnamefont {M.}~\bibnamefont {Motta}},\ }\href@noop {}
  {\bibinfo {title} {Quantum algorithm for imaginary-time green's functions}}
  (\bibinfo {year} {2023}),\ \Eprint {https://arxiv.org/abs/2309.09914}
  {arXiv:2309.09914 [quant-ph]} \BibitemShut {NoStop}%
\bibitem [{\citenamefont {Bravyi}\ and\ \citenamefont
  {Gosset}(2017)}]{Bravyi:2017cc}%
  \BibitemOpen
  \bibfield  {author} {\bibinfo {author} {\bibfnamefont {S.}~\bibnamefont
  {Bravyi}}\ and\ \bibinfo {author} {\bibfnamefont {D.}~\bibnamefont
  {Gosset}},\ }\bibfield  {title} {\bibinfo {title} {Complexity of quantum
  impurity problems},\ }\href {https://doi.org/10.1007/s00220-017-2976-9}
  {\bibfield  {journal} {\bibinfo  {journal} {Communications in Mathematical
  Physics}\ }\textbf {\bibinfo {volume} {356}},\ \bibinfo {pages} {451 500}
  (\bibinfo {year} {2017})}\BibitemShut {NoStop}%
\bibitem [{\citenamefont {Nusspickel}\ and\ \citenamefont
  {Booth}(2020)}]{Nusspickel2020}%
  \BibitemOpen
  \bibfield  {author} {\bibinfo {author} {\bibfnamefont {M.}~\bibnamefont
  {Nusspickel}}\ and\ \bibinfo {author} {\bibfnamefont {G.~H.}\ \bibnamefont
  {Booth}},\ }\bibfield  {title} {\bibinfo {title} {{Efficient compression of
  the environment of an open quantum system}},\ }\href
  {https://doi.org/10.1103/physrevb.102.165107} {\bibfield  {journal} {\bibinfo
   {journal} {Physical Review B}\ }\textbf {\bibinfo {volume} {102}},\ \bibinfo
  {pages} {165107} (\bibinfo {year} {2020})}\BibitemShut {NoStop}%
\bibitem [{\citenamefont {Shinaoka}\ and\ \citenamefont
  {Nagai}(2021)}]{sparsebathfitting2021}%
  \BibitemOpen
  \bibfield  {author} {\bibinfo {author} {\bibfnamefont {H.}~\bibnamefont
  {Shinaoka}}\ and\ \bibinfo {author} {\bibfnamefont {Y.}~\bibnamefont
  {Nagai}},\ }\bibfield  {title} {\bibinfo {title} {{Sparse modeling of
  large-scale quantum impurity models with low symmetries}},\ }\href
  {https://doi.org/10.1103/physrevb.103.045120} {\bibfield  {journal} {\bibinfo
   {journal} {Physical Review B}\ }\textbf {\bibinfo {volume} {103}},\ \bibinfo
  {pages} {045120} (\bibinfo {year} {2021})}\BibitemShut {NoStop}%
\bibitem [{\citenamefont {de~Vega}\ \emph {et~al.}(2015)\citenamefont
  {de~Vega}, \citenamefont {Schollw\"ock},\ and\ \citenamefont
  {Wolf}}]{PhysRevB.92.155126}%
  \BibitemOpen
  \bibfield  {author} {\bibinfo {author} {\bibfnamefont {I.}~\bibnamefont
  {de~Vega}}, \bibinfo {author} {\bibfnamefont {U.}~\bibnamefont
  {Schollw\"ock}},\ and\ \bibinfo {author} {\bibfnamefont {F.~A.}\ \bibnamefont
  {Wolf}},\ }\bibfield  {title} {\bibinfo {title} {How to discretize a quantum
  bath for real-time evolution},\ }\href
  {https://doi.org/10.1103/PhysRevB.92.155126} {\bibfield  {journal} {\bibinfo
  {journal} {Phys. Rev. B}\ }\textbf {\bibinfo {volume} {92}},\ \bibinfo
  {pages} {155126} (\bibinfo {year} {2015})}\BibitemShut {NoStop}%
\bibitem [{\citenamefont {Mejuto-Zaera}\ \emph {et~al.}(2020)\citenamefont
  {Mejuto-Zaera}, \citenamefont {Zepeda-N\'u\~nez}, \citenamefont {Lindsey},
  \citenamefont {Tubman}, \citenamefont {Whaley},\ and\ \citenamefont
  {Lin}}]{PhysRevB.101.035143}%
  \BibitemOpen
  \bibfield  {author} {\bibinfo {author} {\bibfnamefont {C.}~\bibnamefont
  {Mejuto-Zaera}}, \bibinfo {author} {\bibfnamefont {L.}~\bibnamefont
  {Zepeda-N\'u\~nez}}, \bibinfo {author} {\bibfnamefont {M.}~\bibnamefont
  {Lindsey}}, \bibinfo {author} {\bibfnamefont {N.}~\bibnamefont {Tubman}},
  \bibinfo {author} {\bibfnamefont {B.}~\bibnamefont {Whaley}},\ and\ \bibinfo
  {author} {\bibfnamefont {L.}~\bibnamefont {Lin}},\ }\bibfield  {title}
  {\bibinfo {title} {Efficient hybridization fitting for dynamical mean-field
  theory via semi-definite relaxation},\ }\href
  {https://doi.org/10.1103/PhysRevB.101.035143} {\bibfield  {journal} {\bibinfo
   {journal} {Phys. Rev. B}\ }\textbf {\bibinfo {volume} {101}},\ \bibinfo
  {pages} {035143} (\bibinfo {year} {2020})}\BibitemShut {NoStop}%
\bibitem [{\citenamefont {Bartlett}\ \emph
  {et~al.}(1989{\natexlab{a}})\citenamefont {Bartlett}, \citenamefont
  {Kucharski},\ and\ \citenamefont {Noga}}]{BARTLETT1989133}%
  \BibitemOpen
  \bibfield  {author} {\bibinfo {author} {\bibfnamefont {R.~J.}\ \bibnamefont
  {Bartlett}}, \bibinfo {author} {\bibfnamefont {S.~A.}\ \bibnamefont
  {Kucharski}},\ and\ \bibinfo {author} {\bibfnamefont {J.}~\bibnamefont
  {Noga}},\ }\bibfield  {title} {\bibinfo {title} {Alternative coupled-cluster
  ansätze ii. the unitary coupled-cluster method},\ }\href
  {https://doi.org/https://doi.org/10.1016/S0009-2614(89)87372-5} {\bibfield
  {journal} {\bibinfo  {journal} {Chemical Physics Letters}\ }\textbf {\bibinfo
  {volume} {155}},\ \bibinfo {pages} {133} (\bibinfo {year}
  {1989}{\natexlab{a}})}\BibitemShut {NoStop}%
\bibitem [{\citenamefont {Taube}\ and\ \citenamefont
  {Bartlett}(2006{\natexlab{a}})}]{https://doi.org/10.1002/qua.21198}%
  \BibitemOpen
  \bibfield  {author} {\bibinfo {author} {\bibfnamefont {A.~G.}\ \bibnamefont
  {Taube}}\ and\ \bibinfo {author} {\bibfnamefont {R.~J.}\ \bibnamefont
  {Bartlett}},\ }\bibfield  {title} {\bibinfo {title} {New perspectives on
  unitary coupled-cluster theory},\ }\href
  {https://doi.org/https://doi.org/10.1002/qua.21198} {\bibfield  {journal}
  {\bibinfo  {journal} {International Journal of Quantum Chemistry}\ }\textbf
  {\bibinfo {volume} {106}},\ \bibinfo {pages} {3393} (\bibinfo {year}
  {2006}{\natexlab{a}})},\ \Eprint
  {https://arxiv.org/abs/https://onlinelibrary.wiley.com/doi/pdf/10.1002/qua.21198}
  {https://onlinelibrary.wiley.com/doi/pdf/10.1002/qua.21198} \BibitemShut
  {NoStop}%
\bibitem [{\citenamefont {Magoulas}\ and\ \citenamefont
  {Evangelista}(2023)}]{doi:10.1021/acs.jpca.3c02781}%
  \BibitemOpen
  \bibfield  {author} {\bibinfo {author} {\bibfnamefont {I.}~\bibnamefont
  {Magoulas}}\ and\ \bibinfo {author} {\bibfnamefont {F.~A.}\ \bibnamefont
  {Evangelista}},\ }\bibfield  {title} {\bibinfo {title} {Unitary coupled
  cluster: Seizing the quantum moment},\ }\href
  {https://doi.org/10.1021/acs.jpca.3c02781} {\bibfield  {journal} {\bibinfo
  {journal} {The Journal of Physical Chemistry A}\ }\textbf {\bibinfo {volume}
  {127}},\ \bibinfo {pages} {6567} (\bibinfo {year} {2023})},\ \bibinfo {note}
  {pMID: 37523485},\ \Eprint
  {https://arxiv.org/abs/https://doi.org/10.1021/acs.jpca.3c02781}
  {https://doi.org/10.1021/acs.jpca.3c02781} \BibitemShut {NoStop}%
\bibitem [{\citenamefont {Yao}\ \emph {et~al.}(2021)\citenamefont {Yao},
  \citenamefont {Zhang}, \citenamefont {Wang}, \citenamefont {Ho},\ and\
  \citenamefont {Orth}}]{PhysRevResearch.3.013184}%
  \BibitemOpen
  \bibfield  {author} {\bibinfo {author} {\bibfnamefont {Y.}~\bibnamefont
  {Yao}}, \bibinfo {author} {\bibfnamefont {F.}~\bibnamefont {Zhang}}, \bibinfo
  {author} {\bibfnamefont {C.-Z.}\ \bibnamefont {Wang}}, \bibinfo {author}
  {\bibfnamefont {K.-M.}\ \bibnamefont {Ho}},\ and\ \bibinfo {author}
  {\bibfnamefont {P.~P.}\ \bibnamefont {Orth}},\ }\bibfield  {title} {\bibinfo
  {title} {Gutzwiller hybrid quantum-classical computing approach for
  correlated materials},\ }\href
  {https://doi.org/10.1103/PhysRevResearch.3.013184} {\bibfield  {journal}
  {\bibinfo  {journal} {Phys. Rev. Res.}\ }\textbf {\bibinfo {volume} {3}},\
  \bibinfo {pages} {013184} (\bibinfo {year} {2021})}\BibitemShut {NoStop}%
\bibitem [{\citenamefont {Mukherjee}\ \emph {et~al.}(2023)\citenamefont
  {Mukherjee}, \citenamefont {Berthusen}, \citenamefont {Getelina},
  \citenamefont {Orth},\ and\ \citenamefont {Yao}}]{Mukherjee_2023}%
  \BibitemOpen
  \bibfield  {author} {\bibinfo {author} {\bibfnamefont {A.}~\bibnamefont
  {Mukherjee}}, \bibinfo {author} {\bibfnamefont {N.~F.}\ \bibnamefont
  {Berthusen}}, \bibinfo {author} {\bibfnamefont {J.~C.}\ \bibnamefont
  {Getelina}}, \bibinfo {author} {\bibfnamefont {P.~P.}\ \bibnamefont {Orth}},\
  and\ \bibinfo {author} {\bibfnamefont {Y.-X.}\ \bibnamefont {Yao}},\
  }\bibfield  {title} {\bibinfo {title} {Comparative study of adaptive
  variational quantum eigensolvers for multi-orbital impurity models},\
  }\bibfield  {journal} {\bibinfo  {journal} {Communications Physics}\ }\textbf
  {\bibinfo {volume} {6}},\ \href {https://doi.org/10.1038/s42005-022-01089-6}
  {10.1038/s42005-022-01089-6} (\bibinfo {year} {2023})\BibitemShut {NoStop}%
\bibitem [{\citenamefont {Lee}\ \emph {et~al.}(2019)\citenamefont {Lee},
  \citenamefont {Huggins}, \citenamefont {Head-Gordon},\ and\ \citenamefont
  {Whaley}}]{lee2018generalized}%
  \BibitemOpen
  \bibfield  {author} {\bibinfo {author} {\bibfnamefont {J.}~\bibnamefont
  {Lee}}, \bibinfo {author} {\bibfnamefont {W.~J.}\ \bibnamefont {Huggins}},
  \bibinfo {author} {\bibfnamefont {M.}~\bibnamefont {Head-Gordon}},\ and\
  \bibinfo {author} {\bibfnamefont {K.~B.}\ \bibnamefont {Whaley}},\ }\bibfield
   {title} {\bibinfo {title} {Generalized unitary coupled cluster wave
  functions for quantum computation},\ }\href@noop {} {\bibfield  {journal}
  {\bibinfo  {journal} {Journal of chemical theory and computation}\ }\textbf
  {\bibinfo {volume} {15}},\ \bibinfo {pages} {311} (\bibinfo {year}
  {2019})}\BibitemShut {NoStop}%
\bibitem [{\citenamefont {Sakurai}\ \emph {et~al.}(2022)\citenamefont
  {Sakurai}, \citenamefont {Mizukami},\ and\ \citenamefont
  {Shinaoka}}]{PhysRevResearch.4.023219}%
  \BibitemOpen
  \bibfield  {author} {\bibinfo {author} {\bibfnamefont {R.}~\bibnamefont
  {Sakurai}}, \bibinfo {author} {\bibfnamefont {W.}~\bibnamefont {Mizukami}},\
  and\ \bibinfo {author} {\bibfnamefont {H.}~\bibnamefont {Shinaoka}},\
  }\bibfield  {title} {\bibinfo {title} {Hybrid quantum-classical algorithm for
  computing imaginary-time correlation functions},\ }\href
  {https://doi.org/10.1103/PhysRevResearch.4.023219} {\bibfield  {journal}
  {\bibinfo  {journal} {Phys. Rev. Res.}\ }\textbf {\bibinfo {volume} {4}},\
  \bibinfo {pages} {023219} (\bibinfo {year} {2022})}\BibitemShut {NoStop}%
\bibitem [{\citenamefont {Yuan}\ \emph {et~al.}(2019)\citenamefont {Yuan},
  \citenamefont {Endo}, \citenamefont {Zhao}, \citenamefont {Li},\ and\
  \citenamefont {Benjamin}}]{Yuan2019theoryofvariational}%
  \BibitemOpen
  \bibfield  {author} {\bibinfo {author} {\bibfnamefont {X.}~\bibnamefont
  {Yuan}}, \bibinfo {author} {\bibfnamefont {S.}~\bibnamefont {Endo}}, \bibinfo
  {author} {\bibfnamefont {Q.}~\bibnamefont {Zhao}}, \bibinfo {author}
  {\bibfnamefont {Y.}~\bibnamefont {Li}},\ and\ \bibinfo {author}
  {\bibfnamefont {S.~C.}\ \bibnamefont {Benjamin}},\ }\bibfield  {title}
  {\bibinfo {title} {Theory of variational quantum simulation},\ }\href
  {https://doi.org/10.22331/q-2019-10-07-191} {\bibfield  {journal} {\bibinfo
  {journal} {{Quantum}}\ }\textbf {\bibinfo {volume} {3}},\ \bibinfo {pages}
  {191} (\bibinfo {year} {2019})}\BibitemShut {NoStop}%
\bibitem [{\citenamefont {Matsuzawa}\ and\ \citenamefont
  {Kurashige}(2020{\natexlab{a}})}]{matsuzawa2020jastrow}%
  \BibitemOpen
  \bibfield  {author} {\bibinfo {author} {\bibfnamefont {Y.}~\bibnamefont
  {Matsuzawa}}\ and\ \bibinfo {author} {\bibfnamefont {Y.}~\bibnamefont
  {Kurashige}},\ }\bibfield  {title} {\bibinfo {title} {Jastrow-type
  decomposition in quantum chemistry for low-depth quantum circuits},\
  }\href@noop {} {\bibfield  {journal} {\bibinfo  {journal} {Journal of
  chemical theory and computation}\ }\textbf {\bibinfo {volume} {16}},\
  \bibinfo {pages} {944} (\bibinfo {year} {2020}{\natexlab{a}})}\BibitemShut
  {NoStop}%
\bibitem [{\citenamefont {{Jordan}}\ and\ \citenamefont
  {{Wigner}}(1928)}]{1928ZPhy...47..631J}%
  \BibitemOpen
  \bibfield  {author} {\bibinfo {author} {\bibfnamefont {P.}~\bibnamefont
  {{Jordan}}}\ and\ \bibinfo {author} {\bibfnamefont {E.}~\bibnamefont
  {{Wigner}}},\ }\bibfield  {title} {\bibinfo {title} {{{\"U}ber das Paulische
  {\"A}quivalenzverbot}},\ }\href {https://doi.org/10.1007/BF01331938}
  {\bibfield  {journal} {\bibinfo  {journal} {Zeitschrift fur Physik}\ }\textbf
  {\bibinfo {volume} {47}},\ \bibinfo {pages} {631} (\bibinfo {year}
  {1928})}\BibitemShut {NoStop}%
\bibitem [{\citenamefont {Bravyi}\ and\ \citenamefont
  {Kitaev}(2002)}]{bravyi2002fermionic}%
  \BibitemOpen
  \bibfield  {author} {\bibinfo {author} {\bibfnamefont {S.~B.}\ \bibnamefont
  {Bravyi}}\ and\ \bibinfo {author} {\bibfnamefont {A.~Y.}\ \bibnamefont
  {Kitaev}},\ }\bibfield  {title} {\bibinfo {title} {Fermionic quantum
  computation},\ }\href@noop {} {\bibfield  {journal} {\bibinfo  {journal}
  {Annals of Physics}\ }\textbf {\bibinfo {volume} {298}},\ \bibinfo {pages}
  {210} (\bibinfo {year} {2002})}\BibitemShut {NoStop}%
\bibitem [{\citenamefont {Seeley}\ \emph {et~al.}(2012)\citenamefont {Seeley},
  \citenamefont {Richard},\ and\ \citenamefont {Love}}]{seeley2012bravyi}%
  \BibitemOpen
  \bibfield  {author} {\bibinfo {author} {\bibfnamefont {J.~T.}\ \bibnamefont
  {Seeley}}, \bibinfo {author} {\bibfnamefont {M.~J.}\ \bibnamefont
  {Richard}},\ and\ \bibinfo {author} {\bibfnamefont {P.~J.}\ \bibnamefont
  {Love}},\ }\bibfield  {title} {\bibinfo {title} {The bravyi-kitaev
  transformation for quantum computation of electronic structure},\ }\href@noop
  {} {\bibfield  {journal} {\bibinfo  {journal} {The Journal of chemical
  physics}\ }\textbf {\bibinfo {volume} {137}},\ \bibinfo {pages} {224109}
  (\bibinfo {year} {2012})}\BibitemShut {NoStop}%
\bibitem [{\citenamefont {Tilly}\ \emph {et~al.}(2022)\citenamefont {Tilly},
  \citenamefont {Chen}, \citenamefont {Cao}, \citenamefont {Picozzi},
  \citenamefont {Setia}, \citenamefont {Li}, \citenamefont {Grant},
  \citenamefont {Wossnig}, \citenamefont {Rungger}, \citenamefont {Booth},\
  and\ \citenamefont {Tennyson}}]{TILLY20221}%
  \BibitemOpen
  \bibfield  {author} {\bibinfo {author} {\bibfnamefont {J.}~\bibnamefont
  {Tilly}}, \bibinfo {author} {\bibfnamefont {H.}~\bibnamefont {Chen}},
  \bibinfo {author} {\bibfnamefont {S.}~\bibnamefont {Cao}}, \bibinfo {author}
  {\bibfnamefont {D.}~\bibnamefont {Picozzi}}, \bibinfo {author} {\bibfnamefont
  {K.}~\bibnamefont {Setia}}, \bibinfo {author} {\bibfnamefont
  {Y.}~\bibnamefont {Li}}, \bibinfo {author} {\bibfnamefont {E.}~\bibnamefont
  {Grant}}, \bibinfo {author} {\bibfnamefont {L.}~\bibnamefont {Wossnig}},
  \bibinfo {author} {\bibfnamefont {I.}~\bibnamefont {Rungger}}, \bibinfo
  {author} {\bibfnamefont {G.~H.}\ \bibnamefont {Booth}},\ and\ \bibinfo
  {author} {\bibfnamefont {J.}~\bibnamefont {Tennyson}},\ }\bibfield  {title}
  {\bibinfo {title} {The variational quantum eigensolver: A review of methods
  and best practices},\ }\href
  {https://doi.org/https://doi.org/10.1016/j.physrep.2022.08.003} {\bibfield
  {journal} {\bibinfo  {journal} {Physics Reports}\ }\textbf {\bibinfo {volume}
  {986}},\ \bibinfo {pages} {1} (\bibinfo {year} {2022})},\ \bibinfo {note}
  {the Variational Quantum Eigensolver: a review of methods and best
  practices}\BibitemShut {NoStop}%
\bibitem [{\citenamefont {Sriluckshmy}\ \emph {et~al.}(2021)\citenamefont
  {Sriluckshmy}, \citenamefont {Nusspickel}, \citenamefont {Fertitta},\ and\
  \citenamefont {Booth}}]{PhysRevB.103.085131}%
  \BibitemOpen
  \bibfield  {author} {\bibinfo {author} {\bibfnamefont {P.~V.}\ \bibnamefont
  {Sriluckshmy}}, \bibinfo {author} {\bibfnamefont {M.}~\bibnamefont
  {Nusspickel}}, \bibinfo {author} {\bibfnamefont {E.}~\bibnamefont
  {Fertitta}},\ and\ \bibinfo {author} {\bibfnamefont {G.~H.}\ \bibnamefont
  {Booth}},\ }\bibfield  {title} {\bibinfo {title} {Fully algebraic and
  self-consistent effective dynamics in a static quantum embedding},\ }\href
  {https://doi.org/10.1103/PhysRevB.103.085131} {\bibfield  {journal} {\bibinfo
   {journal} {Phys. Rev. B}\ }\textbf {\bibinfo {volume} {103}},\ \bibinfo
  {pages} {085131} (\bibinfo {year} {2021})}\BibitemShut {NoStop}%
\bibitem [{\citenamefont {Backhouse}\ and\ \citenamefont
  {Booth}(2022)}]{Backhouse_2022}%
  \BibitemOpen
  \bibfield  {author} {\bibinfo {author} {\bibfnamefont {O.~J.}\ \bibnamefont
  {Backhouse}}\ and\ \bibinfo {author} {\bibfnamefont {G.~H.}\ \bibnamefont
  {Booth}},\ }\bibfield  {title} {\bibinfo {title} {Constructing
  {\textquotedblleft}full-frequency{\textquotedblright} spectra via moment
  constraints for coupled cluster green's functions},\ }\href
  {https://doi.org/10.1021/acs.jctc.2c00670} {\bibfield  {journal} {\bibinfo
  {journal} {Journal of Chemical Theory and Computation}\ }\textbf {\bibinfo
  {volume} {18}},\ \bibinfo {pages} {6622} (\bibinfo {year}
  {2022})}\BibitemShut {NoStop}%
\bibitem [{\citenamefont {Backhouse}\ and\ \citenamefont
  {Booth}(2023)}]{backhouse2023dynamics}%
  \BibitemOpen
  \bibfield  {author} {\bibinfo {author} {\bibfnamefont {O.~J.}\ \bibnamefont
  {Backhouse}}\ and\ \bibinfo {author} {\bibfnamefont {G.~H.}\ \bibnamefont
  {Booth}},\ }\emph {\bibinfo {title} {Dynamics from statics: A conceptual
  reformulation of Green’s}},\ \href@noop {} {Ph.D. thesis},\ \bibinfo
  {school} {King’s College London} (\bibinfo {year} {2023})\BibitemShut
  {NoStop}%
\bibitem [{\citenamefont {Weikert}\ \emph {et~al.}(1996)\citenamefont
  {Weikert}, \citenamefont {Meyer}, \citenamefont {Cederbaum},\ and\
  \citenamefont {Tarantelli}}]{10.1063/1.471429}%
  \BibitemOpen
  \bibfield  {author} {\bibinfo {author} {\bibfnamefont {H.}~\bibnamefont
  {Weikert}}, \bibinfo {author} {\bibfnamefont {H.}~\bibnamefont {Meyer}},
  \bibinfo {author} {\bibfnamefont {L.~S.}\ \bibnamefont {Cederbaum}},\ and\
  \bibinfo {author} {\bibfnamefont {F.}~\bibnamefont {Tarantelli}},\ }\bibfield
   {title} {\bibinfo {title} {{Block Lanczos and many‐body theory:
  Application to the one‐particle Green’s function}},\ }\href
  {https://doi.org/10.1063/1.471429} {\bibfield  {journal} {\bibinfo  {journal}
  {The Journal of Chemical Physics}\ }\textbf {\bibinfo {volume} {104}},\
  \bibinfo {pages} {7122} (\bibinfo {year} {1996})},\ \Eprint
  {https://arxiv.org/abs/https://pubs.aip.org/aip/jcp/article-pdf/104/18/7122/9437631/7122\_1\_online.pdf}
  {https://pubs.aip.org/aip/jcp/article-pdf/104/18/7122/9437631/7122\_1\_online.pdf}
  \BibitemShut {NoStop}%
\bibitem [{\citenamefont {Scott}\ \emph {et~al.}(2023)\citenamefont {Scott},
  \citenamefont {Backhouse},\ and\ \citenamefont {Booth}}]{10.1063/5.0143291}%
  \BibitemOpen
  \bibfield  {author} {\bibinfo {author} {\bibfnamefont {C.~J.~C.}\
  \bibnamefont {Scott}}, \bibinfo {author} {\bibfnamefont {O.~J.}\ \bibnamefont
  {Backhouse}},\ and\ \bibinfo {author} {\bibfnamefont {G.~H.}\ \bibnamefont
  {Booth}},\ }\bibfield  {title} {\bibinfo {title} {{A “moment-conserving”
  reformulation of GW theory}},\ }\href {https://doi.org/10.1063/5.0143291}
  {\bibfield  {journal} {\bibinfo  {journal} {The Journal of Chemical Physics}\
  }\textbf {\bibinfo {volume} {158}},\ \bibinfo {pages} {124102} (\bibinfo
  {year} {2023})},\ \Eprint
  {https://arxiv.org/abs/https://pubs.aip.org/aip/jcp/article-pdf/doi/10.1063/5.0143291/16791990/124102\_1\_online.pdf}
  {https://pubs.aip.org/aip/jcp/article-pdf/doi/10.1063/5.0143291/16791990/124102\_1\_online.pdf}
  \BibitemShut {NoStop}%
\bibitem [{\citenamefont {Backhouse}\ \emph {et~al.}(2021)\citenamefont
  {Backhouse}, \citenamefont {Santana-Bonilla},\ and\ \citenamefont
  {Booth}}]{doi:10.1021/acs.jpclett.1c02383}%
  \BibitemOpen
  \bibfield  {author} {\bibinfo {author} {\bibfnamefont {O.~J.}\ \bibnamefont
  {Backhouse}}, \bibinfo {author} {\bibfnamefont {A.}~\bibnamefont
  {Santana-Bonilla}},\ and\ \bibinfo {author} {\bibfnamefont {G.~H.}\
  \bibnamefont {Booth}},\ }\bibfield  {title} {\bibinfo {title} {Scalable and
  predictive spectra of correlated molecules with moment truncated iterated
  perturbation theory},\ }\href {https://doi.org/10.1021/acs.jpclett.1c02383}
  {\bibfield  {journal} {\bibinfo  {journal} {The Journal of Physical Chemistry
  Letters}\ }\textbf {\bibinfo {volume} {12}},\ \bibinfo {pages} {7650}
  (\bibinfo {year} {2021})}\BibitemShut {NoStop}%
\bibitem [{\citenamefont {Chen}\ \emph {et~al.}(2021)\citenamefont {Chen},
  \citenamefont {Nusspickel}, \citenamefont {Tilly},\ and\ \citenamefont
  {Booth}}]{PhysRevA.104.032405}%
  \BibitemOpen
  \bibfield  {author} {\bibinfo {author} {\bibfnamefont {H.}~\bibnamefont
  {Chen}}, \bibinfo {author} {\bibfnamefont {M.}~\bibnamefont {Nusspickel}},
  \bibinfo {author} {\bibfnamefont {J.}~\bibnamefont {Tilly}},\ and\ \bibinfo
  {author} {\bibfnamefont {G.~H.}\ \bibnamefont {Booth}},\ }\bibfield  {title}
  {\bibinfo {title} {Variational quantum eigensolver for dynamic correlation
  functions},\ }\href {https://doi.org/10.1103/PhysRevA.104.032405} {\bibfield
  {journal} {\bibinfo  {journal} {Phys. Rev. A}\ }\textbf {\bibinfo {volume}
  {104}},\ \bibinfo {pages} {032405} (\bibinfo {year} {2021})}\BibitemShut
  {NoStop}%
\bibitem [{\citenamefont {Ibe}\ \emph {et~al.}(2022)\citenamefont {Ibe},
  \citenamefont {Nakagawa}, \citenamefont {Earnest}, \citenamefont {Yamamoto},
  \citenamefont {Mitarai}, \citenamefont {Gao},\ and\ \citenamefont
  {Kobayashi}}]{ibe2022calculating}%
  \BibitemOpen
  \bibfield  {author} {\bibinfo {author} {\bibfnamefont {Y.}~\bibnamefont
  {Ibe}}, \bibinfo {author} {\bibfnamefont {Y.~O.}\ \bibnamefont {Nakagawa}},
  \bibinfo {author} {\bibfnamefont {N.}~\bibnamefont {Earnest}}, \bibinfo
  {author} {\bibfnamefont {T.}~\bibnamefont {Yamamoto}}, \bibinfo {author}
  {\bibfnamefont {K.}~\bibnamefont {Mitarai}}, \bibinfo {author} {\bibfnamefont
  {Q.}~\bibnamefont {Gao}},\ and\ \bibinfo {author} {\bibfnamefont
  {T.}~\bibnamefont {Kobayashi}},\ }\bibfield  {title} {\bibinfo {title}
  {Calculating transition amplitudes by variational quantum deflation},\
  }\href@noop {} {\bibfield  {journal} {\bibinfo  {journal} {Physical Review
  Research}\ }\textbf {\bibinfo {volume} {4}},\ \bibinfo {pages} {013173}
  (\bibinfo {year} {2022})}\BibitemShut {NoStop}%
\bibitem [{\citenamefont {Jones}\ \emph {et~al.}(2019)\citenamefont {Jones},
  \citenamefont {Endo}, \citenamefont {McArdle}, \citenamefont {Yuan},\ and\
  \citenamefont {Benjamin}}]{PhysRevA.99.062304}%
  \BibitemOpen
  \bibfield  {author} {\bibinfo {author} {\bibfnamefont {T.}~\bibnamefont
  {Jones}}, \bibinfo {author} {\bibfnamefont {S.}~\bibnamefont {Endo}},
  \bibinfo {author} {\bibfnamefont {S.}~\bibnamefont {McArdle}}, \bibinfo
  {author} {\bibfnamefont {X.}~\bibnamefont {Yuan}},\ and\ \bibinfo {author}
  {\bibfnamefont {S.~C.}\ \bibnamefont {Benjamin}},\ }\bibfield  {title}
  {\bibinfo {title} {Variational quantum algorithms for discovering hamiltonian
  spectra},\ }\href {https://doi.org/10.1103/PhysRevA.99.062304} {\bibfield
  {journal} {\bibinfo  {journal} {Phys. Rev. A}\ }\textbf {\bibinfo {volume}
  {99}},\ \bibinfo {pages} {062304} (\bibinfo {year} {2019})}\BibitemShut
  {NoStop}%
\bibitem [{\citenamefont {Higgott}\ \emph {et~al.}(2019)\citenamefont
  {Higgott}, \citenamefont {Wang},\ and\ \citenamefont
  {Brierley}}]{Higgott2019variationalquantum}%
  \BibitemOpen
  \bibfield  {author} {\bibinfo {author} {\bibfnamefont {O.}~\bibnamefont
  {Higgott}}, \bibinfo {author} {\bibfnamefont {D.}~\bibnamefont {Wang}},\ and\
  \bibinfo {author} {\bibfnamefont {S.}~\bibnamefont {Brierley}},\ }\bibfield
  {title} {\bibinfo {title} {Variational {Q}uantum {C}omputation of {E}xcited
  {S}tates},\ }\href {https://doi.org/10.22331/q-2019-07-01-156} {\bibfield
  {journal} {\bibinfo  {journal} {{Quantum}}\ }\textbf {\bibinfo {volume}
  {3}},\ \bibinfo {pages} {156} (\bibinfo {year} {2019})}\BibitemShut {NoStop}%
\bibitem [{\citenamefont {Ollitrault}\ \emph {et~al.}(2020)\citenamefont
  {Ollitrault}, \citenamefont {Kandala}, \citenamefont {Chen}, \citenamefont
  {Barkoutsos}, \citenamefont {Mezzacapo}, \citenamefont {Pistoia},
  \citenamefont {Sheldon}, \citenamefont {Woerner}, \citenamefont {Gambetta},\
  and\ \citenamefont {Tavernelli}}]{PhysRevResearch.2.043140}%
  \BibitemOpen
  \bibfield  {author} {\bibinfo {author} {\bibfnamefont {P.~J.}\ \bibnamefont
  {Ollitrault}}, \bibinfo {author} {\bibfnamefont {A.}~\bibnamefont {Kandala}},
  \bibinfo {author} {\bibfnamefont {C.-F.}\ \bibnamefont {Chen}}, \bibinfo
  {author} {\bibfnamefont {P.~K.}\ \bibnamefont {Barkoutsos}}, \bibinfo
  {author} {\bibfnamefont {A.}~\bibnamefont {Mezzacapo}}, \bibinfo {author}
  {\bibfnamefont {M.}~\bibnamefont {Pistoia}}, \bibinfo {author} {\bibfnamefont
  {S.}~\bibnamefont {Sheldon}}, \bibinfo {author} {\bibfnamefont
  {S.}~\bibnamefont {Woerner}}, \bibinfo {author} {\bibfnamefont {J.~M.}\
  \bibnamefont {Gambetta}},\ and\ \bibinfo {author} {\bibfnamefont
  {I.}~\bibnamefont {Tavernelli}},\ }\bibfield  {title} {\bibinfo {title}
  {Quantum equation of motion for computing molecular excitation energies on a
  noisy quantum processor},\ }\href
  {https://doi.org/10.1103/PhysRevResearch.2.043140} {\bibfield  {journal}
  {\bibinfo  {journal} {Phys. Rev. Res.}\ }\textbf {\bibinfo {volume} {2}},\
  \bibinfo {pages} {043140} (\bibinfo {year} {2020})}\BibitemShut {NoStop}%
\bibitem [{\citenamefont {Zhang}\ \emph {et~al.}(2021)\citenamefont {Zhang},
  \citenamefont {Gomes}, \citenamefont {Yao}, \citenamefont {Orth},\ and\
  \citenamefont {Iadecola}}]{PhysRevB.104.075159}%
  \BibitemOpen
  \bibfield  {author} {\bibinfo {author} {\bibfnamefont {F.}~\bibnamefont
  {Zhang}}, \bibinfo {author} {\bibfnamefont {N.}~\bibnamefont {Gomes}},
  \bibinfo {author} {\bibfnamefont {Y.}~\bibnamefont {Yao}}, \bibinfo {author}
  {\bibfnamefont {P.~P.}\ \bibnamefont {Orth}},\ and\ \bibinfo {author}
  {\bibfnamefont {T.}~\bibnamefont {Iadecola}},\ }\bibfield  {title} {\bibinfo
  {title} {Adaptive variational quantum eigensolvers for highly excited
  states},\ }\href {https://doi.org/10.1103/PhysRevB.104.075159} {\bibfield
  {journal} {\bibinfo  {journal} {Phys. Rev. B}\ }\textbf {\bibinfo {volume}
  {104}},\ \bibinfo {pages} {075159} (\bibinfo {year} {2021})}\BibitemShut
  {NoStop}%
\bibitem [{\citenamefont {Nooijen}(2000)}]{nooijen2000can}%
  \BibitemOpen
  \bibfield  {author} {\bibinfo {author} {\bibfnamefont {M.}~\bibnamefont
  {Nooijen}},\ }\bibfield  {title} {\bibinfo {title} {Can the eigenstates of a
  many-body hamiltonian be represented exactly using a general two-body cluster
  expansion?},\ }\href@noop {} {\bibfield  {journal} {\bibinfo  {journal}
  {Physical review letters}\ }\textbf {\bibinfo {volume} {84}},\ \bibinfo
  {pages} {2108} (\bibinfo {year} {2000})}\BibitemShut {NoStop}%
\bibitem [{\citenamefont {Matsuzawa}\ and\ \citenamefont
  {Kurashige}(2020{\natexlab{b}})}]{Matsuzawa_2020}%
  \BibitemOpen
  \bibfield  {author} {\bibinfo {author} {\bibfnamefont {Y.}~\bibnamefont
  {Matsuzawa}}\ and\ \bibinfo {author} {\bibfnamefont {Y.}~\bibnamefont
  {Kurashige}},\ }\bibfield  {title} {\bibinfo {title} {Jastrow-type
  decomposition in quantum chemistry for low-depth quantum circuits},\ }\href
  {https://doi.org/10.1021/acs.jctc.9b00963} {\bibfield  {journal} {\bibinfo
  {journal} {Journal of Chemical Theory and Computation}\ }\textbf {\bibinfo
  {volume} {16}},\ \bibinfo {pages} {944} (\bibinfo {year}
  {2020}{\natexlab{b}})}\BibitemShut {NoStop}%
\bibitem [{\citenamefont
  {Kutzelnigg}(1982)}]{kutzelniggQuantumChemistryFock1982}%
  \BibitemOpen
  \bibfield  {author} {\bibinfo {author} {\bibfnamefont {W.}~\bibnamefont
  {Kutzelnigg}},\ }\bibfield  {title} {\bibinfo {title} {Quantum chemistry in
  {{Fock}} space. {{I}}. {{The}} universal wave and energy operators},\ }\href
  {http://aip.scitation.org/doi/10.1063/1.444231} {\bibfield  {journal}
  {\bibinfo  {journal} {The Journal of Chemical Physics}\ }\textbf {\bibinfo
  {volume} {77}},\ \bibinfo {pages} {3081} (\bibinfo {year}
  {1982})}\BibitemShut {NoStop}%
\bibitem [{\citenamefont {Kutzelnigg}\ and\ \citenamefont
  {Koch}(1983)}]{kutzelniggQuantumChemistryFock1983}%
  \BibitemOpen
  \bibfield  {author} {\bibinfo {author} {\bibfnamefont {W.}~\bibnamefont
  {Kutzelnigg}}\ and\ \bibinfo {author} {\bibfnamefont {S.}~\bibnamefont
  {Koch}},\ }\bibfield  {title} {\bibinfo {title} {Quantum chemistry in
  {{Fock}} space. {{II}}. {{Effective Hamiltonians}} in {{Fock}} space},\
  }\href {https://aip.scitation.org/doi/10.1063/1.446313} {\bibfield  {journal}
  {\bibinfo  {journal} {J. Chem. Phys.}\ }\textbf {\bibinfo {volume} {79}},\
  \bibinfo {pages} {4315} (\bibinfo {year} {1983})}\BibitemShut {NoStop}%
\bibitem [{\citenamefont
  {Kutzelnigg}(1985)}]{kutzelniggQuantumChemistryFock1985}%
  \BibitemOpen
  \bibfield  {author} {\bibinfo {author} {\bibfnamefont {W.}~\bibnamefont
  {Kutzelnigg}},\ }\bibfield  {title} {\bibinfo {title} {Quantum chemistry in
  {{Fock}} space. {{IV}}. {{The}} treatment of permutational symmetry.
  {{Spin}}-free diagrams with symmetrized vertices},\ }\href
  {https://aip.scitation.org/doi/10.1063/1.448859} {\bibfield  {journal}
  {\bibinfo  {journal} {J. Chem. Phys.}\ }\textbf {\bibinfo {volume} {82}},\
  \bibinfo {pages} {4166} (\bibinfo {year} {1985})}\BibitemShut {NoStop}%
\bibitem [{\citenamefont {Bartlett}\ \emph
  {et~al.}(1989{\natexlab{b}})\citenamefont {Bartlett}, \citenamefont
  {Kucharski},\ and\ \citenamefont
  {Noga}}]{bartlettAlternativeCoupledclusterAnsatze1989}%
  \BibitemOpen
  \bibfield  {author} {\bibinfo {author} {\bibfnamefont {R.~J.}\ \bibnamefont
  {Bartlett}}, \bibinfo {author} {\bibfnamefont {S.~A.}\ \bibnamefont
  {Kucharski}},\ and\ \bibinfo {author} {\bibfnamefont {J.}~\bibnamefont
  {Noga}},\ }\bibfield  {title} {\bibinfo {title} {Alternative coupled-cluster
  ans\"atze {{II}}. {{The}} unitary coupled-cluster method},\ }\href
  {https://www.sciencedirect.com/science/article/pii/S0009261489873725}
  {\bibfield  {journal} {\bibinfo  {journal} {Chemical Physics Letters}\
  }\textbf {\bibinfo {volume} {155}},\ \bibinfo {pages} {133} (\bibinfo {year}
  {1989}{\natexlab{b}})}\BibitemShut {NoStop}%
\bibitem [{\citenamefont
  {Kutzelnigg}(1991)}]{kutzelniggErrorAnalysisImprovements1991}%
  \BibitemOpen
  \bibfield  {author} {\bibinfo {author} {\bibfnamefont {W.}~\bibnamefont
  {Kutzelnigg}},\ }\bibfield  {title} {\bibinfo {title} {Error analysis and
  improvements of coupled-cluster theory},\ }\href
  {https://doi.org/10.1007/BF01117418} {\bibfield  {journal} {\bibinfo
  {journal} {Theoret. Chim. Acta}\ }\textbf {\bibinfo {volume} {80}},\ \bibinfo
  {pages} {349} (\bibinfo {year} {1991})}\BibitemShut {NoStop}%
\bibitem [{\citenamefont {Taube}\ and\ \citenamefont
  {Bartlett}(2006{\natexlab{b}})}]{taubeNewPerspectivesUnitary2006}%
  \BibitemOpen
  \bibfield  {author} {\bibinfo {author} {\bibfnamefont {A.~G.}\ \bibnamefont
  {Taube}}\ and\ \bibinfo {author} {\bibfnamefont {R.~J.}\ \bibnamefont
  {Bartlett}},\ }\bibfield  {title} {\bibinfo {title} {New perspectives on
  unitary coupled-cluster theory},\ }\href
  {https://onlinelibrary.wiley.com/doi/abs/10.1002/qua.21198} {\bibfield
  {journal} {\bibinfo  {journal} {International Journal of Quantum Chemistry}\
  }\textbf {\bibinfo {volume} {106}},\ \bibinfo {pages} {3393} (\bibinfo {year}
  {2006}{\natexlab{b}})}\BibitemShut {NoStop}%
\bibitem [{\citenamefont {O’Malley}\ \emph {et~al.}(2016)\citenamefont
  {O’Malley}, \citenamefont {Babbush}, \citenamefont {Kivlichan},
  \citenamefont {Romero}, \citenamefont {McClean}, \citenamefont {Barends},
  \citenamefont {Kelly}, \citenamefont {Roushan}, \citenamefont {Tranter},
  \citenamefont {Ding} \emph {et~al.}}]{o2016scalable}%
  \BibitemOpen
  \bibfield  {author} {\bibinfo {author} {\bibfnamefont {P.~J.}\ \bibnamefont
  {O’Malley}}, \bibinfo {author} {\bibfnamefont {R.}~\bibnamefont {Babbush}},
  \bibinfo {author} {\bibfnamefont {I.~D.}\ \bibnamefont {Kivlichan}}, \bibinfo
  {author} {\bibfnamefont {J.}~\bibnamefont {Romero}}, \bibinfo {author}
  {\bibfnamefont {J.~R.}\ \bibnamefont {McClean}}, \bibinfo {author}
  {\bibfnamefont {R.}~\bibnamefont {Barends}}, \bibinfo {author} {\bibfnamefont
  {J.}~\bibnamefont {Kelly}}, \bibinfo {author} {\bibfnamefont
  {P.}~\bibnamefont {Roushan}}, \bibinfo {author} {\bibfnamefont
  {A.}~\bibnamefont {Tranter}}, \bibinfo {author} {\bibfnamefont
  {N.}~\bibnamefont {Ding}}, \emph {et~al.},\ }\bibfield  {title} {\bibinfo
  {title} {Scalable quantum simulation of molecular energies},\ }\href@noop {}
  {\bibfield  {journal} {\bibinfo  {journal} {Physical Review X}\ }\textbf
  {\bibinfo {volume} {6}},\ \bibinfo {pages} {031007} (\bibinfo {year}
  {2016})}\BibitemShut {NoStop}%
\bibitem [{\citenamefont {Barkoutsos}\ \emph {et~al.}(2018)\citenamefont
  {Barkoutsos}, \citenamefont {Gonthier}, \citenamefont {Sokolov},
  \citenamefont {Moll}, \citenamefont {Salis}, \citenamefont {Fuhrer},
  \citenamefont {Ganzhorn}, \citenamefont {Egger}, \citenamefont {Troyer},
  \citenamefont {Mezzacapo} \emph {et~al.}}]{barkoutsos2018quantum}%
  \BibitemOpen
  \bibfield  {author} {\bibinfo {author} {\bibfnamefont {P.~K.}\ \bibnamefont
  {Barkoutsos}}, \bibinfo {author} {\bibfnamefont {J.~F.}\ \bibnamefont
  {Gonthier}}, \bibinfo {author} {\bibfnamefont {I.}~\bibnamefont {Sokolov}},
  \bibinfo {author} {\bibfnamefont {N.}~\bibnamefont {Moll}}, \bibinfo {author}
  {\bibfnamefont {G.}~\bibnamefont {Salis}}, \bibinfo {author} {\bibfnamefont
  {A.}~\bibnamefont {Fuhrer}}, \bibinfo {author} {\bibfnamefont
  {M.}~\bibnamefont {Ganzhorn}}, \bibinfo {author} {\bibfnamefont {D.~J.}\
  \bibnamefont {Egger}}, \bibinfo {author} {\bibfnamefont {M.}~\bibnamefont
  {Troyer}}, \bibinfo {author} {\bibfnamefont {A.}~\bibnamefont {Mezzacapo}},
  \emph {et~al.},\ }\bibfield  {title} {\bibinfo {title} {Quantum algorithms
  for electronic structure calculations: Particle-hole hamiltonian and
  optimized wave-function expansions},\ }\href@noop {} {\bibfield  {journal}
  {\bibinfo  {journal} {Physical Review A}\ }\textbf {\bibinfo {volume} {98}},\
  \bibinfo {pages} {022322} (\bibinfo {year} {2018})}\BibitemShut {NoStop}%
\bibitem [{\citenamefont {Mizukami}\ \emph {et~al.}(2020)\citenamefont
  {Mizukami}, \citenamefont {Mitarai}, \citenamefont {Nakagawa}, \citenamefont
  {Yamamoto}, \citenamefont {Yan},\ and\ \citenamefont
  {Ohnishi}}]{mizukami2020orbital}%
  \BibitemOpen
  \bibfield  {author} {\bibinfo {author} {\bibfnamefont {W.}~\bibnamefont
  {Mizukami}}, \bibinfo {author} {\bibfnamefont {K.}~\bibnamefont {Mitarai}},
  \bibinfo {author} {\bibfnamefont {Y.~O.}\ \bibnamefont {Nakagawa}}, \bibinfo
  {author} {\bibfnamefont {T.}~\bibnamefont {Yamamoto}}, \bibinfo {author}
  {\bibfnamefont {T.}~\bibnamefont {Yan}},\ and\ \bibinfo {author}
  {\bibfnamefont {Y.-y.}\ \bibnamefont {Ohnishi}},\ }\bibfield  {title}
  {\bibinfo {title} {Orbital optimized unitary coupled cluster theory for
  quantum computer},\ }\href@noop {} {\bibfield  {journal} {\bibinfo  {journal}
  {Physical Review Research}\ }\textbf {\bibinfo {volume} {2}},\ \bibinfo
  {pages} {033421} (\bibinfo {year} {2020})}\BibitemShut {NoStop}%
\bibitem [{\citenamefont {Rubin}\ \emph {et~al.}(2022)\citenamefont {Rubin},
  \citenamefont {Lee},\ and\ \citenamefont {Babbush}}]{rubin2022compressing}%
  \BibitemOpen
  \bibfield  {author} {\bibinfo {author} {\bibfnamefont {N.~C.}\ \bibnamefont
  {Rubin}}, \bibinfo {author} {\bibfnamefont {J.}~\bibnamefont {Lee}},\ and\
  \bibinfo {author} {\bibfnamefont {R.}~\bibnamefont {Babbush}},\ }\bibfield
  {title} {\bibinfo {title} {Compressing many-body fermion operators under
  unitary constraints},\ }\href@noop {} {\bibfield  {journal} {\bibinfo
  {journal} {Journal of Chemical Theory and Computation}\ }\textbf {\bibinfo
  {volume} {18}},\ \bibinfo {pages} {1480} (\bibinfo {year}
  {2022})}\BibitemShut {NoStop}%
\bibitem [{\citenamefont {Cohn}\ \emph {et~al.}(2021)\citenamefont {Cohn},
  \citenamefont {Motta},\ and\ \citenamefont {Parrish}}]{PRXQuantum.2.040352}%
  \BibitemOpen
  \bibfield  {author} {\bibinfo {author} {\bibfnamefont {J.}~\bibnamefont
  {Cohn}}, \bibinfo {author} {\bibfnamefont {M.}~\bibnamefont {Motta}},\ and\
  \bibinfo {author} {\bibfnamefont {R.~M.}\ \bibnamefont {Parrish}},\
  }\bibfield  {title} {\bibinfo {title} {Quantum filter diagonalization with
  compressed double-factorized hamiltonians},\ }\href
  {https://doi.org/10.1103/PRXQuantum.2.040352} {\bibfield  {journal} {\bibinfo
   {journal} {PRX Quantum}\ }\textbf {\bibinfo {volume} {2}},\ \bibinfo {pages}
  {040352} (\bibinfo {year} {2021})}\BibitemShut {NoStop}%
\bibitem [{\citenamefont {Peng}\ and\ \citenamefont
  {Kowalski}(2017)}]{peng2017highly}%
  \BibitemOpen
  \bibfield  {author} {\bibinfo {author} {\bibfnamefont {B.}~\bibnamefont
  {Peng}}\ and\ \bibinfo {author} {\bibfnamefont {K.}~\bibnamefont
  {Kowalski}},\ }\bibfield  {title} {\bibinfo {title} {Highly efficient and
  scalable compound decomposition of two-electron integral tensor and its
  application in coupled cluster calculations},\ }\href@noop {} {\bibfield
  {journal} {\bibinfo  {journal} {Journal of chemical theory and computation}\
  }\textbf {\bibinfo {volume} {13}},\ \bibinfo {pages} {4179} (\bibinfo {year}
  {2017})}\BibitemShut {NoStop}%
\bibitem [{\citenamefont {Motta}\ \emph {et~al.}(2021)\citenamefont {Motta},
  \citenamefont {Ye}, \citenamefont {McClean}, \citenamefont {Li},
  \citenamefont {Minnich}, \citenamefont {Babbush},\ and\ \citenamefont
  {Chan}}]{motta2021low}%
  \BibitemOpen
  \bibfield  {author} {\bibinfo {author} {\bibfnamefont {M.}~\bibnamefont
  {Motta}}, \bibinfo {author} {\bibfnamefont {E.}~\bibnamefont {Ye}}, \bibinfo
  {author} {\bibfnamefont {J.~R.}\ \bibnamefont {McClean}}, \bibinfo {author}
  {\bibfnamefont {Z.}~\bibnamefont {Li}}, \bibinfo {author} {\bibfnamefont
  {A.~J.}\ \bibnamefont {Minnich}}, \bibinfo {author} {\bibfnamefont
  {R.}~\bibnamefont {Babbush}},\ and\ \bibinfo {author} {\bibfnamefont {G.~K.}\
  \bibnamefont {Chan}},\ }\bibfield  {title} {\bibinfo {title} {Low rank
  representations for quantum simulation of electronic structure},\ }\href@noop
  {} {\bibfield  {journal} {\bibinfo  {journal} {npj Quantum Information}\
  }\textbf {\bibinfo {volume} {7}},\ \bibinfo {pages} {1} (\bibinfo {year}
  {2021})}\BibitemShut {NoStop}%
\bibitem [{\citenamefont {Lee}\ \emph {et~al.}(2021)\citenamefont {Lee},
  \citenamefont {Berry}, \citenamefont {Gidney}, \citenamefont {Huggins},
  \citenamefont {McClean}, \citenamefont {Wiebe},\ and\ \citenamefont
  {Babbush}}]{PRXQuantum.2.030305}%
  \BibitemOpen
  \bibfield  {author} {\bibinfo {author} {\bibfnamefont {J.}~\bibnamefont
  {Lee}}, \bibinfo {author} {\bibfnamefont {D.~W.}\ \bibnamefont {Berry}},
  \bibinfo {author} {\bibfnamefont {C.}~\bibnamefont {Gidney}}, \bibinfo
  {author} {\bibfnamefont {W.~J.}\ \bibnamefont {Huggins}}, \bibinfo {author}
  {\bibfnamefont {J.~R.}\ \bibnamefont {McClean}}, \bibinfo {author}
  {\bibfnamefont {N.}~\bibnamefont {Wiebe}},\ and\ \bibinfo {author}
  {\bibfnamefont {R.}~\bibnamefont {Babbush}},\ }\bibfield  {title} {\bibinfo
  {title} {Even more efficient quantum computations of chemistry through tensor
  hypercontraction},\ }\href {https://doi.org/10.1103/PRXQuantum.2.030305}
  {\bibfield  {journal} {\bibinfo  {journal} {PRX Quantum}\ }\textbf {\bibinfo
  {volume} {2}},\ \bibinfo {pages} {030305} (\bibinfo {year}
  {2021})}\BibitemShut {NoStop}%
\bibitem [{QCM()}]{QCMaterialNew}%
  \BibitemOpen
  \href@noop {} {}\bibinfo {note}
  {\url{https://github.com/sakurairihito/QCMaterialNew}}\BibitemShut {NoStop}%
\bibitem [{\citenamefont {Suzuki}\ \emph {et~al.}(2021)\citenamefont {Suzuki},
  \citenamefont {Kawase}, \citenamefont {Masumura}, \citenamefont {Hiraga},
  \citenamefont {Nakadai}, \citenamefont {Chen}, \citenamefont {Nakanishi},
  \citenamefont {Mitarai}, \citenamefont {Imai}, \citenamefont {Tamiya} \emph
  {et~al.}}]{suzuki2021qulacs}%
  \BibitemOpen
  \bibfield  {author} {\bibinfo {author} {\bibfnamefont {Y.}~\bibnamefont
  {Suzuki}}, \bibinfo {author} {\bibfnamefont {Y.}~\bibnamefont {Kawase}},
  \bibinfo {author} {\bibfnamefont {Y.}~\bibnamefont {Masumura}}, \bibinfo
  {author} {\bibfnamefont {Y.}~\bibnamefont {Hiraga}}, \bibinfo {author}
  {\bibfnamefont {M.}~\bibnamefont {Nakadai}}, \bibinfo {author} {\bibfnamefont
  {J.}~\bibnamefont {Chen}}, \bibinfo {author} {\bibfnamefont {K.~M.}\
  \bibnamefont {Nakanishi}}, \bibinfo {author} {\bibfnamefont {K.}~\bibnamefont
  {Mitarai}}, \bibinfo {author} {\bibfnamefont {R.}~\bibnamefont {Imai}},
  \bibinfo {author} {\bibfnamefont {S.}~\bibnamefont {Tamiya}}, \emph
  {et~al.},\ }\bibfield  {title} {\bibinfo {title} {Qulacs: a fast and
  versatile quantum circuit simulator for research purpose},\ }\href@noop {}
  {\bibfield  {journal} {\bibinfo  {journal} {Quantum}\ }\textbf {\bibinfo
  {volume} {5}},\ \bibinfo {pages} {559} (\bibinfo {year} {2021})}\BibitemShut
  {NoStop}%
\bibitem [{\citenamefont {McClean}\ \emph {et~al.}(2020)\citenamefont
  {McClean}, \citenamefont {Rubin}, \citenamefont {Sung}, \citenamefont
  {Kivlichan}, \citenamefont {Bonet-Monroig}, \citenamefont {Cao},
  \citenamefont {Dai}, \citenamefont {Fried}, \citenamefont {Gidney},
  \citenamefont {Gimby} \emph {et~al.}}]{mcclean2020openfermion}%
  \BibitemOpen
  \bibfield  {author} {\bibinfo {author} {\bibfnamefont {J.~R.}\ \bibnamefont
  {McClean}}, \bibinfo {author} {\bibfnamefont {N.~C.}\ \bibnamefont {Rubin}},
  \bibinfo {author} {\bibfnamefont {K.~J.}\ \bibnamefont {Sung}}, \bibinfo
  {author} {\bibfnamefont {I.~D.}\ \bibnamefont {Kivlichan}}, \bibinfo {author}
  {\bibfnamefont {X.}~\bibnamefont {Bonet-Monroig}}, \bibinfo {author}
  {\bibfnamefont {Y.}~\bibnamefont {Cao}}, \bibinfo {author} {\bibfnamefont
  {C.}~\bibnamefont {Dai}}, \bibinfo {author} {\bibfnamefont {E.~S.}\
  \bibnamefont {Fried}}, \bibinfo {author} {\bibfnamefont {C.}~\bibnamefont
  {Gidney}}, \bibinfo {author} {\bibfnamefont {B.}~\bibnamefont {Gimby}}, \emph
  {et~al.},\ }\bibfield  {title} {\bibinfo {title} {Openfermion: the electronic
  structure package for quantum computers},\ }\href@noop {} {\bibfield
  {journal} {\bibinfo  {journal} {Quantum Science and Technology}\ }\textbf
  {\bibinfo {volume} {5}},\ \bibinfo {pages} {034014} (\bibinfo {year}
  {2020})}\BibitemShut {NoStop}%
\bibitem [{\citenamefont {Shinaoka}\ \emph {et~al.}(2021)\citenamefont
  {Shinaoka}, \citenamefont {Otsuki}, \citenamefont {Kawamura}, \citenamefont
  {Takemori},\ and\ \citenamefont {Yoshimi}}]{shinaoka2021dcore}%
  \BibitemOpen
  \bibfield  {author} {\bibinfo {author} {\bibfnamefont {H.}~\bibnamefont
  {Shinaoka}}, \bibinfo {author} {\bibfnamefont {J.}~\bibnamefont {Otsuki}},
  \bibinfo {author} {\bibfnamefont {M.}~\bibnamefont {Kawamura}}, \bibinfo
  {author} {\bibfnamefont {N.}~\bibnamefont {Takemori}},\ and\ \bibinfo
  {author} {\bibfnamefont {K.}~\bibnamefont {Yoshimi}},\ }\bibfield  {title}
  {\bibinfo {title} {Dcore: Integrated dmft software for correlated
  electrons},\ }\href@noop {} {\bibfield  {journal} {\bibinfo  {journal}
  {SciPost Physics}\ }\textbf {\bibinfo {volume} {10}},\ \bibinfo {pages} {117}
  (\bibinfo {year} {2021})}\BibitemShut {NoStop}%
\bibitem [{dys()}]{dyson}%
  \BibitemOpen
  \href@noop {} {}\bibinfo {note}
  {\url{(https://github.com/BoothGroup/dyson}}\BibitemShut {NoStop}%
\bibitem [{\citenamefont {Kantorovich}(1960)}]{10.1287/mnsc.6.4.366}%
  \BibitemOpen
  \bibfield  {author} {\bibinfo {author} {\bibfnamefont {L.~V.}\ \bibnamefont
  {Kantorovich}},\ }\bibfield  {title} {\bibinfo {title} {Mathematical methods
  of organizing and planning production},\ }\href
  {https://doi.org/10.1287/mnsc.6.4.366} {\bibfield  {journal} {\bibinfo
  {journal} {Manage. Sci.}\ }\textbf {\bibinfo {volume} {6}},\ \bibinfo {pages}
  {366–422} (\bibinfo {year} {1960})}\BibitemShut {NoStop}%
\bibitem [{ctx(1969)}]{ctx51223882530003681}%
  \BibitemOpen
  \bibfield  {title} {\bibinfo {title} {Markov processes over denumerable
  products of spaces, describing large systems of automata},\ }\href@noop {}
  {\bibfield  {journal} {\bibinfo  {journal} {Probl. Peredachi Inf.}\ }\textbf
  {\bibinfo {volume} {5}} (\bibinfo {year} {1969})}\BibitemShut {NoStop}%
\bibitem [{\citenamefont {Gomes}\ \emph {et~al.}(2021)\citenamefont {Gomes},
  \citenamefont {Mukherjee}, \citenamefont {Zhang}, \citenamefont {Iadecola},
  \citenamefont {Wang}, \citenamefont {Ho}, \citenamefont {Orth},\ and\
  \citenamefont {Yao}}]{gomes2021adaptive}%
  \BibitemOpen
  \bibfield  {author} {\bibinfo {author} {\bibfnamefont {N.}~\bibnamefont
  {Gomes}}, \bibinfo {author} {\bibfnamefont {A.}~\bibnamefont {Mukherjee}},
  \bibinfo {author} {\bibfnamefont {F.}~\bibnamefont {Zhang}}, \bibinfo
  {author} {\bibfnamefont {T.}~\bibnamefont {Iadecola}}, \bibinfo {author}
  {\bibfnamefont {C.-Z.}\ \bibnamefont {Wang}}, \bibinfo {author}
  {\bibfnamefont {K.-M.}\ \bibnamefont {Ho}}, \bibinfo {author} {\bibfnamefont
  {P.~P.}\ \bibnamefont {Orth}},\ and\ \bibinfo {author} {\bibfnamefont
  {Y.-X.}\ \bibnamefont {Yao}},\ }\bibfield  {title} {\bibinfo {title}
  {Adaptive variational quantum imaginary time evolution approach for ground
  state preparation},\ }\href@noop {} {\bibfield  {journal} {\bibinfo
  {journal} {Advanced Quantum Technologies}\ }\textbf {\bibinfo {volume} {4}},\
  \bibinfo {pages} {2100114} (\bibinfo {year} {2021})}\BibitemShut {NoStop}%
\bibitem [{\citenamefont {Mukherjee}\ \emph {et~al.}(2022)\citenamefont
  {Mukherjee}, \citenamefont {Berthusen}, \citenamefont {Getelina},
  \citenamefont {Orth},\ and\ \citenamefont {Yao}}]{mukherjee2022comparative}%
  \BibitemOpen
  \bibfield  {author} {\bibinfo {author} {\bibfnamefont {A.}~\bibnamefont
  {Mukherjee}}, \bibinfo {author} {\bibfnamefont {N.~F.}\ \bibnamefont
  {Berthusen}}, \bibinfo {author} {\bibfnamefont {J.~C.}\ \bibnamefont
  {Getelina}}, \bibinfo {author} {\bibfnamefont {P.~P.}\ \bibnamefont {Orth}},\
  and\ \bibinfo {author} {\bibfnamefont {Y.-X.}\ \bibnamefont {Yao}},\
  }\bibfield  {title} {\bibinfo {title} {Comparative study of adaptive
  variational quantum eigensolvers for multi-orbital impurity models},\
  }\href@noop {} {\bibfield  {journal} {\bibinfo  {journal} {arXiv preprint
  arXiv:2203.06745}\ } (\bibinfo {year} {2022})}\BibitemShut {NoStop}%
\bibitem [{\citenamefont {Gokhale}\ \emph {et~al.}(2019)\citenamefont
  {Gokhale}, \citenamefont {Angiuli}, \citenamefont {Ding}, \citenamefont
  {Gui}, \citenamefont {Tomesh}, \citenamefont {Suchara}, \citenamefont
  {Martonosi},\ and\ \citenamefont {Chong}}]{Gokhale2019MinimizingSP}%
  \BibitemOpen
  \bibfield  {author} {\bibinfo {author} {\bibfnamefont {P.}~\bibnamefont
  {Gokhale}}, \bibinfo {author} {\bibfnamefont {O.}~\bibnamefont {Angiuli}},
  \bibinfo {author} {\bibfnamefont {Y.}~\bibnamefont {Ding}}, \bibinfo {author}
  {\bibfnamefont {K.}~\bibnamefont {Gui}}, \bibinfo {author} {\bibfnamefont
  {T.}~\bibnamefont {Tomesh}}, \bibinfo {author} {\bibfnamefont
  {M.}~\bibnamefont {Suchara}}, \bibinfo {author} {\bibfnamefont
  {M.}~\bibnamefont {Martonosi}},\ and\ \bibinfo {author} {\bibfnamefont
  {F.~T.}\ \bibnamefont {Chong}},\ }\bibfield  {title} {\bibinfo {title}
  {Minimizing state preparations in variational quantum eigensolver by
  partitioning into commuting families},\ }\href
  {https://api.semanticscholar.org/CorpusID:199000717} {\bibfield  {journal}
  {\bibinfo  {journal} {arXiv: Quantum Physics}\ } (\bibinfo {year}
  {2019})}\BibitemShut {NoStop}%
\bibitem [{\citenamefont {Kingma}\ and\ \citenamefont
  {Ba}(2017)}]{kingma2017adam}%
  \BibitemOpen
  \bibfield  {author} {\bibinfo {author} {\bibfnamefont {D.~P.}\ \bibnamefont
  {Kingma}}\ and\ \bibinfo {author} {\bibfnamefont {J.}~\bibnamefont {Ba}},\
  }\href@noop {} {\bibinfo {title} {Adam: A method for stochastic
  optimization}} (\bibinfo {year} {2017}),\ \Eprint
  {https://arxiv.org/abs/1412.6980} {arXiv:1412.6980 [cs.LG]} \BibitemShut
  {NoStop}%
\bibitem [{\citenamefont {Nakanishi}\ \emph {et~al.}(2020)\citenamefont
  {Nakanishi}, \citenamefont {Fujii},\ and\ \citenamefont
  {Todo}}]{PhysRevResearch.2.043158}%
  \BibitemOpen
  \bibfield  {author} {\bibinfo {author} {\bibfnamefont {K.~M.}\ \bibnamefont
  {Nakanishi}}, \bibinfo {author} {\bibfnamefont {K.}~\bibnamefont {Fujii}},\
  and\ \bibinfo {author} {\bibfnamefont {S.}~\bibnamefont {Todo}},\ }\bibfield
  {title} {\bibinfo {title} {Sequential minimal optimization for
  quantum-classical hybrid algorithms},\ }\href
  {https://doi.org/10.1103/PhysRevResearch.2.043158} {\bibfield  {journal}
  {\bibinfo  {journal} {Phys. Rev. Res.}\ }\textbf {\bibinfo {volume} {2}},\
  \bibinfo {pages} {043158} (\bibinfo {year} {2020})}\BibitemShut {NoStop}%
\bibitem [{\citenamefont {Rogers}\ \emph {et~al.}(2021)\citenamefont {Rogers},
  \citenamefont {Bhattacharyya}, \citenamefont {Frank}, \citenamefont {Jiang},
  \citenamefont {Christiansen}, \citenamefont {Yao},\ and\ \citenamefont
  {Lanatà}}]{rogers2021error}%
  \BibitemOpen
  \bibfield  {author} {\bibinfo {author} {\bibfnamefont {J.}~\bibnamefont
  {Rogers}}, \bibinfo {author} {\bibfnamefont {G.}~\bibnamefont
  {Bhattacharyya}}, \bibinfo {author} {\bibfnamefont {M.~S.}\ \bibnamefont
  {Frank}}, \bibinfo {author} {\bibfnamefont {T.}~\bibnamefont {Jiang}},
  \bibinfo {author} {\bibfnamefont {O.}~\bibnamefont {Christiansen}}, \bibinfo
  {author} {\bibfnamefont {Y.-X.}\ \bibnamefont {Yao}},\ and\ \bibinfo {author}
  {\bibfnamefont {N.}~\bibnamefont {Lanatà}},\ }\href@noop {} {\bibinfo
  {title} {Error mitigation in variational quantum eigensolvers using
  probabilistic machine learning}} (\bibinfo {year} {2021}),\ \Eprint
  {https://arxiv.org/abs/2111.08814} {arXiv:2111.08814 [quant-ph]} \BibitemShut
  {NoStop}%
\bibitem [{\citenamefont {Endo}\ \emph {et~al.}(2018)\citenamefont {Endo},
  \citenamefont {Benjamin},\ and\ \citenamefont {Li}}]{endo2018practical}%
  \BibitemOpen
  \bibfield  {author} {\bibinfo {author} {\bibfnamefont {S.}~\bibnamefont
  {Endo}}, \bibinfo {author} {\bibfnamefont {S.~C.}\ \bibnamefont {Benjamin}},\
  and\ \bibinfo {author} {\bibfnamefont {Y.}~\bibnamefont {Li}},\ }\bibfield
  {title} {\bibinfo {title} {Practical quantum error mitigation for near-future
  applications},\ }\href@noop {} {\bibfield  {journal} {\bibinfo  {journal}
  {Physical Review X}\ }\textbf {\bibinfo {volume} {8}},\ \bibinfo {pages}
  {031027} (\bibinfo {year} {2018})}\BibitemShut {NoStop}%
\bibitem [{\citenamefont {Carleo}\ and\ \citenamefont
  {Troyer}(2017)}]{PMID:28183973}%
  \BibitemOpen
  \bibfield  {author} {\bibinfo {author} {\bibfnamefont {G.}~\bibnamefont
  {Carleo}}\ and\ \bibinfo {author} {\bibfnamefont {M.}~\bibnamefont
  {Troyer}},\ }\bibfield  {title} {\bibinfo {title} {Solving the quantum
  many-body problem with artificial neural networks},\ }\href
  {https://doi.org/10.1126/science.aag2302} {\bibfield  {journal} {\bibinfo
  {journal} {Science (New York, N.Y.)}\ }\textbf {\bibinfo {volume} {355}},\
  \bibinfo {pages} {602—606} (\bibinfo {year} {2017})}\BibitemShut {NoStop}%
\bibitem [{\citenamefont {Carleo}\ \emph {et~al.}(2019)\citenamefont {Carleo},
  \citenamefont {Cirac}, \citenamefont {Cranmer}, \citenamefont {Daudet},
  \citenamefont {Schuld}, \citenamefont {Tishby}, \citenamefont
  {Vogt-Maranto},\ and\ \citenamefont {Zdeborov\'a}}]{RevModPhys.91.045002}%
  \BibitemOpen
  \bibfield  {author} {\bibinfo {author} {\bibfnamefont {G.}~\bibnamefont
  {Carleo}}, \bibinfo {author} {\bibfnamefont {I.}~\bibnamefont {Cirac}},
  \bibinfo {author} {\bibfnamefont {K.}~\bibnamefont {Cranmer}}, \bibinfo
  {author} {\bibfnamefont {L.}~\bibnamefont {Daudet}}, \bibinfo {author}
  {\bibfnamefont {M.}~\bibnamefont {Schuld}}, \bibinfo {author} {\bibfnamefont
  {N.}~\bibnamefont {Tishby}}, \bibinfo {author} {\bibfnamefont
  {L.}~\bibnamefont {Vogt-Maranto}},\ and\ \bibinfo {author} {\bibfnamefont
  {L.}~\bibnamefont {Zdeborov\'a}},\ }\bibfield  {title} {\bibinfo {title}
  {Machine learning and the physical sciences},\ }\href
  {https://doi.org/10.1103/RevModPhys.91.045002} {\bibfield  {journal}
  {\bibinfo  {journal} {Rev. Mod. Phys.}\ }\textbf {\bibinfo {volume} {91}},\
  \bibinfo {pages} {045002} (\bibinfo {year} {2019})}\BibitemShut {NoStop}%
\bibitem [{\citenamefont {Endo}\ \emph {et~al.}(2020)\citenamefont {Endo},
  \citenamefont {Kurata},\ and\ \citenamefont
  {Nakagawa}}]{endo2020calculation}%
  \BibitemOpen
  \bibfield  {author} {\bibinfo {author} {\bibfnamefont {S.}~\bibnamefont
  {Endo}}, \bibinfo {author} {\bibfnamefont {I.}~\bibnamefont {Kurata}},\ and\
  \bibinfo {author} {\bibfnamefont {Y.~O.}\ \bibnamefont {Nakagawa}},\
  }\bibfield  {title} {\bibinfo {title} {Calculation of the green's function on
  near-term quantum computers},\ }\href@noop {} {\bibfield  {journal} {\bibinfo
   {journal} {Physical Review Research}\ }\textbf {\bibinfo {volume} {2}},\
  \bibinfo {pages} {033281} (\bibinfo {year} {2020})}\BibitemShut {NoStop}%
\end{thebibliography}%

\appendix{}

\section{A quantum circuit to compute transition amplitude}
\label{sec:circuit_transition_am}

\begin{figure}[h]
    \centering
    \vspace{-20mm} 
    \includegraphics[width=1.00\linewidth]{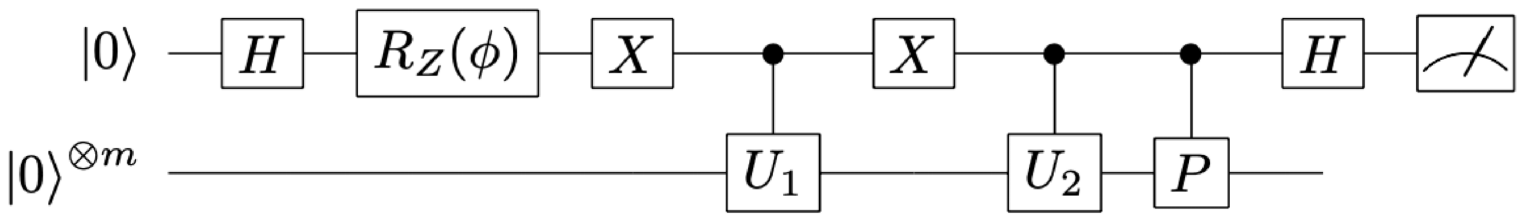}
    \vspace{-30mm} 
    \caption{Quantum circuit for computing the transition amplitude in Eq.~(\ref{general_transition_amplitude}).
    The quantum circuit employs $m$ qubits (on the bottom line) and one additional qubit as an ancilla (on the top line). The transition amplitude can be obtained by summing the measurement outcomes of the ancilla qubit for the $Z$ basis.
    } 
    \label{fig:transition_amplitude}
\end{figure}

We evaluate the transition amplitude on a quantum computer by measuring the Hermitian and anti-Hermitian parts of the following form:
\begin{align}
    \label{general_transition_amplitude}
    \mel{0} {U_{1}^{\dagger}{P}_{}U_{2}}{0},
\end{align}where $P$ are Pauli operators with $m$ qubits, and $U_1$ and $U_{2}$ are unitary operators with $m$ qubits. Equation~\eqref{general_transition_amplitude} can be measured using the quantum circuit in Fig.~\ref{fig:transition_amplitude}~\cite{ibe2022calculating, endo2020calculation,PhysRevA.104.032405},
which requires one ancilla qubit.

Let $p_{0}$/$p_{1}$ be the probability of measuring 0/1 in the ancilla qubit.
The real and imaginary parts of the transition amplitude can be measured separately by setting $\phi=0$ and $\pi/2$ in the $R_z$ gate, respectively, as
\begin{align}
p_{0} - p_{1} &=
\begin{cases}
\Re\mel{0} {U_{1}^{\dagger}(\vec{\theta_1}) P U_{2}(\boldsymbol{\theta}_2)}{0}  & \phi=0,\\
-\Im\mel{0}{U_{1}^{\dagger}(\vec{\theta_1}) P U_{2}(\boldsymbol{\theta}_2)}{0} & \phi=\pi/2.
\end{cases}
\end{align}
As this method is based on a single ancilla qubit, we need complex quantum circuits for NISQ devices because of the control unitary operators.

\section{moment calculations via the recursive VQE} \label{appendix:moment}
This appendix shows the computed spectral moments via VQE and recursive VQE for the single-site impurity model with $N_{\mathrm{imp}}=3$, with or without shot noise.

\subsection{State vector simulation}
\begin{figure}
    \centering
\includegraphics[width=0.95\linewidth]{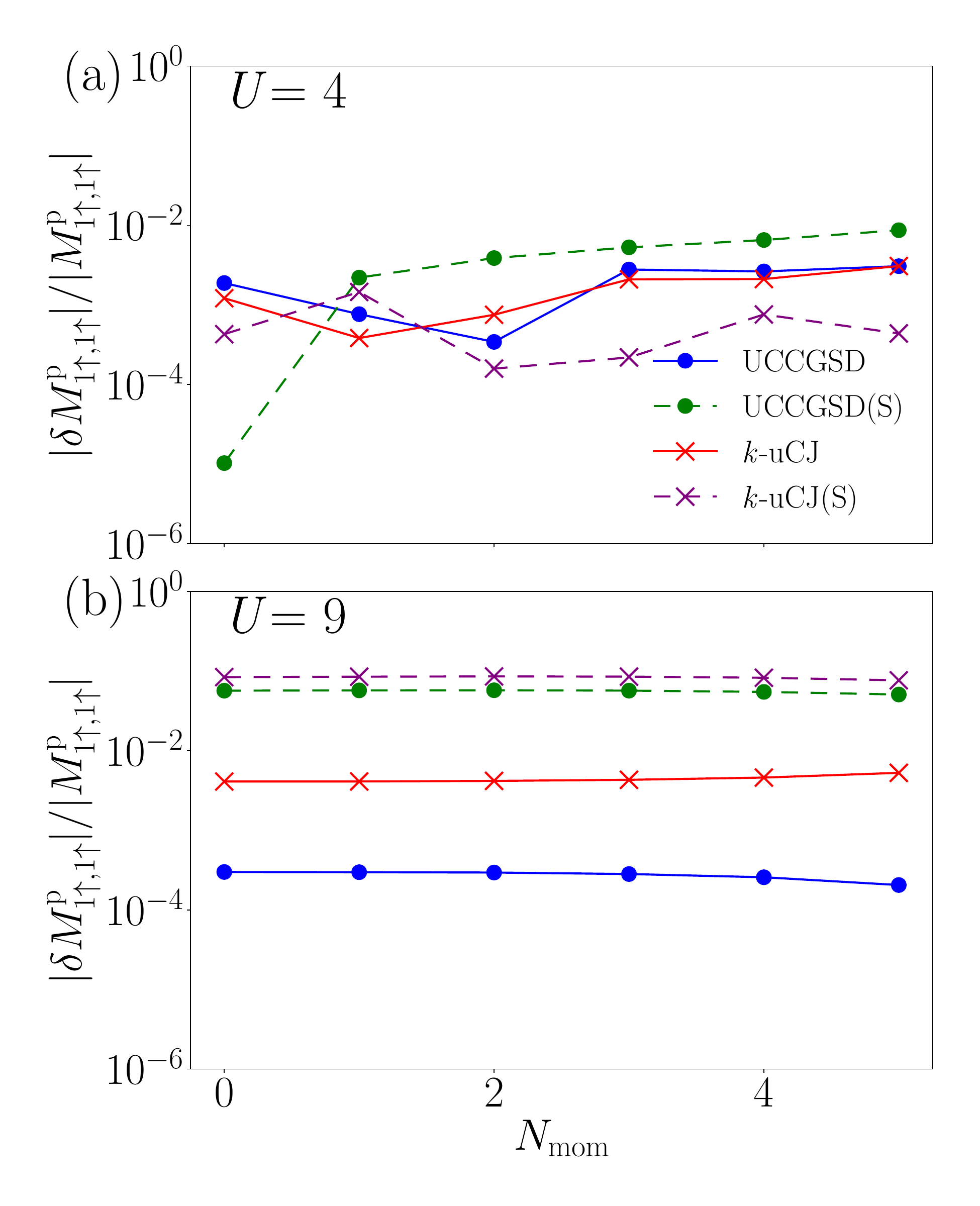}
    \caption{
    Computed $|\delta M_{r s}^{\mathrm{p}}| / M_{r s}^{\mathrm{p}}$ for the single-site impurity model.
    Panels (a) and (b) show the results for $U=4$ and $U=9$, respectively.
    In the $k$-uCJ and the $k$-uCJ(S), we set $k=5$. 
    } 
\label{fig:singlesite_moments}
\end{figure}

Figures~\ref{fig:singlesite_moments}(a) and (b) show the relative error of the spectral moments $ |\delta M_{r s}^{\mathrm{p}}|/|M_{r s}^{\mathrm{p}}|$ for $U=4$ and $U=9$, respectively.
$ |\delta M_{r s}^{\mathrm{p}}|/|M_{r s}^{\mathrm{p}}|$ are calculated via VQE/recursive VQE and exact diagonalization.
The relative error for each ansatz remains nearly constant.
In Fig.~\ref{fig:singlesite_moments}(a), for $U=4$, the $k$-uCJ(S) has the highest accuracy at $N_{\mathrm{mom}}=5$, followed by the $k$-uCJ, UCCGSD. 
In Fig.~\ref{fig:singlesite_moments}(b), for $U=9$, the UCCGSD has the highest accuracy, followed by the $k$-uCJ. 

\subsection{Shot noise}
\begin{figure}
    \centering
    \includegraphics[width=0.95\linewidth]{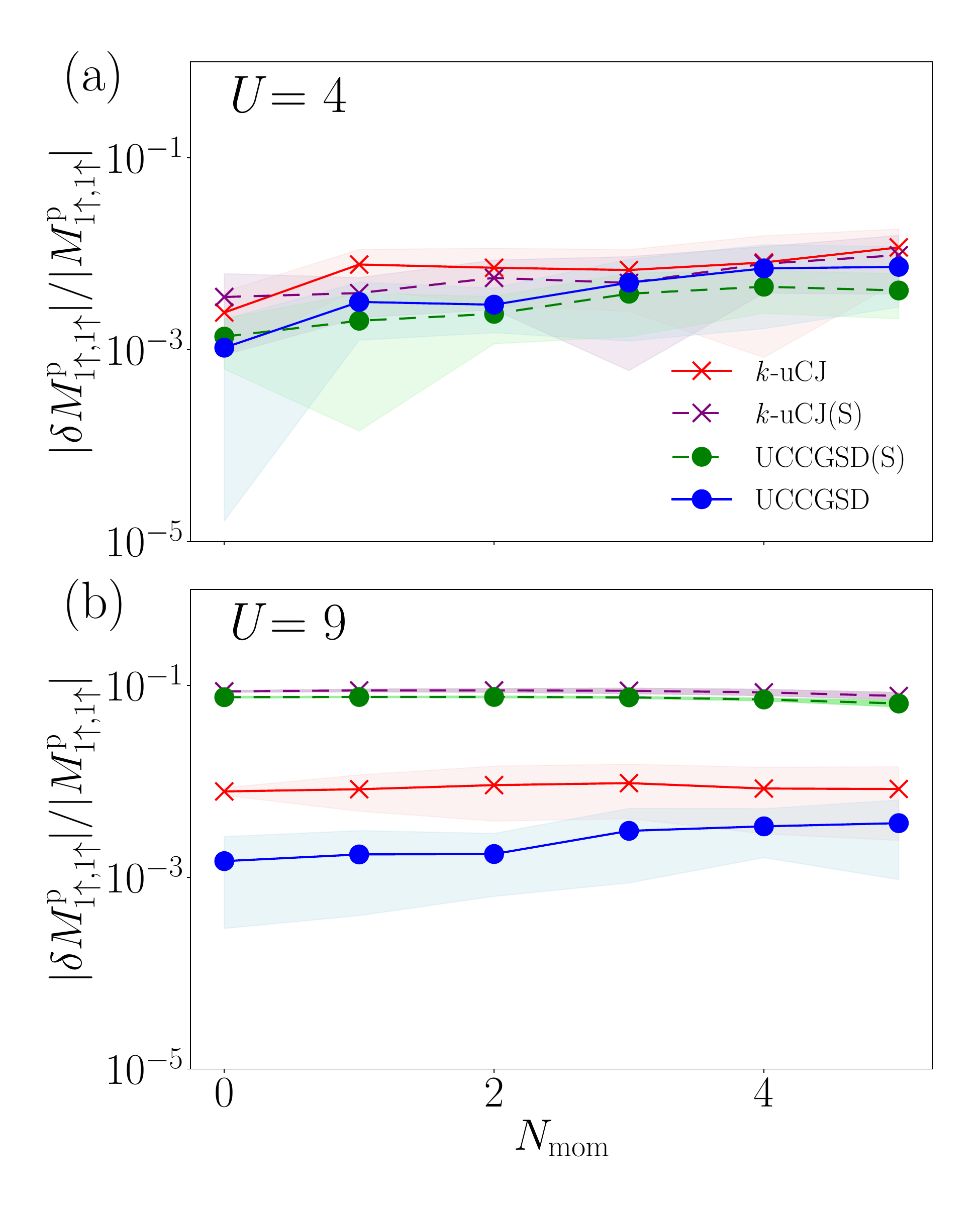}
    \caption{
     Computed $|\delta M_{r s}^{\mathrm{p}}| / M_{r s}^{\mathrm{p}}$ for the single-site impurity model with shot noise.
    Panels (a) and (b) show the results for $U=4$ and $U=9$, respectively.
    In the $k$-uCJ and the $k$-uCJ(S), we set $k=5$.
    The markers indicate the mean, while the lightly shaded areas represent the standard deviation. 
    } 
    \label{fig:singlesite_moment_noise}
\end{figure}

Figures~\ref{fig:singlesite_moment_noise}(a) and (b) show the relative errors of the spectral moments $ |\delta M_{r s}^{\mathrm{p}}|/|M_{r s}^{\mathrm{p}}|$ with a finite number of measurements, 30000 for $U=4$ and $U=9$, respectively.
The markers in the figure denote the mean, and the lightly shaded areas indicate the standard deviation derived from the calculation repeated ten times with shot noise for each ansatz.
The sparse ansatz is generally less accurate than the original ansatz due to the shot noise. 
In Fig.~\ref{fig:singlesite_moments}(a), for $U=4$, no significant difference in relative error between ansatzes was observed due to the shot noise.
Still, the relative error for each ansatz remains nearly constant.
\end{document}